\documentclass[seceq]{ptptex}
\usepackage{graphicx}

\def\img{\mathrm{i}}
\newcommand{\bm}[1]{\boldsymbol{#1}}
\usepackage{macro}
\usepackage{amsthm}

\newtheorem{example}{Example}[section]

\title{
Symmetry Classes of Spin and Orbital Ordered States in 
a $\mathrm{t_{2g}}$ Hubbard Model on a Two-dimensional Squar Lattice
}

\author{
Masanori \textsc{Hamada}, Akira \textsc{Nakanishi}$^1$, Akira \textsc{Goto}$^2$ and Masa-aki \textsc{Ozaki}$^3$
}

\inst{
Graduate School of Materials Science, Nara 
Institute of Science and Technology,
8916-5, Ikoma 630-0192, Japan\\
$^1$Department of
General Education, Osaka Institute of Technology, 
 Osaka 535-8585,Japan \\
$^2$Department of General Education, Kochi National College of Technology, 
 Nangoku 599-8531, Japan \\
$^3$Uji 611-0002,Japan 
}

\abst{
This paper presents  symmetry classes  of the
Hartree-Fock (HF)
solutions of spin and orbital ordered states in a 
 $\mathrm{t_{2g}}$ Hubbard model on a two-dimensional square lattice.
Using a group theoretical bifurcation theory of the Hartree-Fock equation, 
we obtained many types of broken symmetry solutions which bifurcate 
from the normal state through  one step transition 
in cases of commensurate 
ordering 
vectors $\Qb_0=(0,0)$, $\Qb_1=(\pi,\pi)$, $\Qb_2=(\pi,0)$ and $\Qb_3=(0,\pi)$.
Each broken symmetry state is characterized by the presence of 
local order parameters(LOP) at each lattice  site:
quadrupole moment $\Qb=(Q_2^2,Q_{12},Q_{23},Q_{31})$, 
orbital angular momentum $\lb=(l_1,l_2,l_3)$, spin density
 $\ssb=(s^1,s^2,s^3)$, 
 spin quadrupole moment $\Qb^{\lambda}=(Q_2^{2\lambda},
 Q_{12}^{\lambda},Q_{23}^{\lambda},Q_{31}^{\lambda})$ and spin orbital angular 
 momentum $\lb^{\lambda}=(l_1^{\lambda},l_2^{\lambda},l_3^{\lambda})$ where 
 $\lambda=1,2,3$.
We performed numerical calculations for some parameter sets. 
Then we have found that many types of  non-collinear magnetic 
orbital ordered states having LOP:$\Qb^{\lambda}$ and $\lb^{\lambda}$ 
can be the ground state for these parameter sets.
}
\begin{document}

\maketitle

\section{Introduction}
Perovskite-type \textit{R}TiO$_3$ and \textit{R}VO$_3$\cite{Mizokawa, Ren, Mochizuki} 
(\textit{R} being a rare-earth ion or Y)
and quasi two-dimensional ruthenate compounds Ca$_{2-x}$Sr$_{x}$RuO$_{4}$
\cite{Kubota,Hotta,Kurokawa,Anisimov,Fang}
have gained considerable interest 
because of their plentiful phases including various orbital orders.
In a previous paper\cite{Nakanishi}
 (hereafter we cite it as I), in order to study these perovskite-type
  compounds, 
we have considered a triply degenerate $\mathrm{t_{2g}}$ Hubbard model 
on a three-dimensional cubic lattice. 

In the paper we have presented a brief review of the general group theoretical
bifurcation theory of the Hartree-Fock (HF) equation. 
By listing axial isotropy subgroups of the R-reps(irreducible 
representation over the real number field) of the group $G_0$ of
 symmetry of the system,
 we have obtained various types of non magnetic
orbital ordered states and collinear
\footnote{All spins are along the $z$ axis.}
magnetic
\footnote{The phrase "magnetic" means magnetic due to the spin of electrons.} 
orbital ordered states, which bifrurcate from the normal 
state through one step 
 transition in the cases of ordering vectors $\Qb_{\rm F}=(0,0,0)$ and 
$\Qb_{\rm G}=(\pi,\pi,\pi)$. 

In the present paper, to study quasi two-dimensional 
ruthenate compounds, 
we consider  a three-orbital 
$\mathrm{t_{2g}}$ Hubbard model
on a two-dimensional square lattice\cite{Hotta}. 
We apply the group theoretical bifurcation theory 
to this model. 

In this paper we consider broken symmetry states 
with ordering vectors 
$\mathrm{\Gamma\ point}:\Qb_0=(0,0)$, $\mathrm{M\ point}:\Qb_1=(\pi,\pi)$
and $\mathrm{X\ point}:
\Qb_2=(\pi,0)$ and $\Qb_3=(0,\pi)$ which can allow 
states with double $\Qb$. Also we consider magnetic states 
with non-collinear spin structure, which were not
treated in I.  

This paper is organized as follows.
In \S 2 we give a model Hamiltonian and its symmetry group $G_0$. 
In \S 3 we present a general HF Hamiltonian with ordering 
vectors $\Qb_0$, $\Qb_1$, $\Qb_2$ and $\Qb_3$, and 
define the isotropy subgroup of  
the HF Hamiltonian. We present 
general formulae for the local order parameters (LOP) at the lattice site 
$\mb$: the charge density, the spin density, the quadrupole moment, 
the orbital angular momentum, the spin quadrupole moment, 
and the spin orbital angular momentum. 
In \S 4 we present R-reps of the symmetry group $G_0$ of 
the system over the HF Hamiltonian space.

In \S 5 we present symmetry classes of non magnetic 
orbital ordered states by listing axial isotropy subgroups of 
R-reps which do not break the spin rotation symmetry. 
In three examples,  
we show how to derive the canonical form of the HF Hamiltonian 
and occupied orbitals from the isotropy subgroup of a state.
We show that all states, which break the time reversal symmetry, 
have  orbital angular momentum and 
almost of all states, which do not break the 
time reversal symmetry, have the  quadrupole moment.
 
In \S 6 we  present symmetry classes of  magnetic 
orbital ordered states by listing axial isotropy subgroups of 
R-reps which break the spin rotation symmetry.
From R-reps which break time reversal symmetry,  we obtain states
 which have  spin densities or  spin quadrupole moments as LOP.
From R-reps which do not break time reversal symmetry,  we obtain states
 which have spin orbital angular moment as LOP.
 
In two  collinear examples, in which 
hold the spin rotation symmetry around the $z$ axis, and 
the time reversal symmetry is broken, 
we show how to derive the canonical form of occupied orbital 
and explicit form of the spin density or the spin quadrupole moment as LOP.
In three non-collinear examples, 
in which both of  the spin rotation symmetry around the $z$ axis and 
the time reversal symmetry are broken, 
we show how to derive the canonical form of occupied orbital 
and explicit form of the non-collinear spin density or 
the spin quadrupole moment as LOP.

In \S 7 we report numerical results for some parameter sets.
There we show that various  non-collinear magnetic states can be 
the most stable  in all states described in \S 5 and \S 6 for 
these parameter sets. 
In \S 8 a summary and discussion are given. In this section we
discuss two examples of states bifurcating through two step transitions 
from the normal state. It is shown that these are 
the coexistent states of the spin density and the quadrupole moment. 
The notation used in this paper follows the one in I.

\section{The model Hamiltonian and its symmetry}
We consider the three-orbital $\mathrm{t_{2g}}$ Hubbard 
Hamiltonian on the two-dimensional 
square lattice (lattice constant$=1$):
\begin{align}
\label{site-hamiltonian}
\mathcal{H} =&\sum_{\snb}\sum_{i=1}^3\sum_{s}\epsilon_i
a^{\dag}_{\bm{n}is}a^{}_{\bm{n}is}
- \sum^{}_{\bm{n},\bm{a}} \sum^{3}_{i,j=1} \sum^{}_{s} t^{\bm{a}}_{ij}
 a^{\dag}_{\bm{n}is} a^{}_{(\bm{n+a})js} 
-\mu \sum^{}_{\bm{n},i,s}n^{}_{\bm{n}is}\nonumber  \\
& + U \sum^{}_{\bm{n},i} n^{}_{\bm{n}i\uparrow} n^{}_{\bm{n}i\downarrow}
+ \dfrac{U'}{2} \sum^{}_{\bm{n}} \sum^{}_{i \neq j }
 \sum^{}_{s,s'} n^{}_{\bm{n} is} n^{}_{\bm{n}js'}
\nonumber\\
&+ \dfrac{J}{2} \sum^{}_{\bm{n}} \sum^{}_{i \neq j } \sum^{}_{s,s'}
a^{\dag}_{\bm{n}is} a^{\dag}_{\bm{n}js'} a^{}_{\bm{n}is'} a^{}_{\bm{n}js}
+
\dfrac{J'}{2} \sum^{}_{\bm{n}} \sum^{}_{i \neq j } \sum^{}_{s,s'}
a^{\dag}_{\bm{n}is} a^{\dag}_{\bm{n}is'} a^{}_{\bm{n}js'}
a^{}_{\bm{n}js},
\end{align}
where $a^{\dag}_{\bm{n}is}$ is the creation operator for 
a $\mathrm{t_{2g}}$ electron with spin $s$ in the $i$-th 
$(i=1\text{ for }d_{yz},i=2\text{ for }d_{zx}\text{ and }i=3\text{ for }d_{xy})$ orbital
at site $\bm{n}$  
and $n^{}_{\bm{n}is}=a^{\dag}_{\bm{n}is}a^{}_{\bm{n}is}$, 
$\epsilon_i$ is the crystal field for the $i$-th orbital,
$\bm{a}$ is the vector connecting nearest neighbor sites  
, that is, $\bm{a}=(1,0), (0,1), (-1,0), (0,-1)$
and $t^{\bm{a}}_{ij}$ is the nearest neighbor hopping integral 
between $i$ and $j$ orbitals along $\ab$ direction, $U$ is 
the intra-orbital Coulomb interaction,
$U'$ is the inter-orbital Coulomb interaction, $J$ is the exchange integral 
and $J'$ is the pair hopping interaction. 
$U=U^{\prime}+J+J'$ is derived from 
rotational
invariance in orbital space and $J=J'$ from the evaluation of Coulomb 
integrals. 
Three $\mathrm{t_{2g}}$ atomic orbitals $\phi_1(\rb)$,
$\phi_2(\rb)$ and $\phi_3(\rb)$ of an atom located at the origin are defined by
\begin{align}
\phi_1(\rb)&=d_{yz}=f(r)yz,\nonumber \\
\phi_2(\rb)&=d_{zx}=f(r)zx,\nonumber \\
\phi_3(\rb)&=d_{xy}=f(r)xy,
\end{align}
where $\rb=(x,y,z)$, $r=|\rb|$ and $f(r)$ is a spherical symmetric function. 
Three $\mathrm{t_{2g}}$ atomic orbitals of an atom located at a site 
$\nb$
are given by $\phi_1(\rb-\nb)$, $\phi_2(\rb-\nb)$ and $\phi_3(\rb-\nb)$.
\begin{table}[h]
{\footnotesize
\caption{R-rep matrices $D^{(\gamma)}$ of $\Db_{4h}$}
$$
\begin{array}{c|ccccc}
{\rm Representation}&\ A_{1g}\ &\  A_{2g} \ &\ B_{1g}\ & \ 
B_{2g}\ &\ E_{g} \ \\
\hline
\\[-3mm]
E& 1&1& 1& 1& \left(
\begin{array}{cc}
\,\, 1& \,\,0\\
\,\,0&\,\, 1 
\end{array}
\right)\\
\\[-3mm]
\hline
\\[-3mm]
C_{4z}^+& 1& 1& -1 &-1& \left(
\begin{array}{cc}
0& 1\\
-1& 0
\end{array}
\right)\\
\\[-3mm]
\hline
\\[-3mm]
C_{4z}^-& 1& 1& -1 &-1& \left(
\begin{array}{cc}
0& -1\\
1& 0
\end{array}
\right)\\
\\[-3mm]
\hline
\\[-3mm]
C_{2z}& 1& 1& 1 &1& \left(
\begin{array}{cc}
-1& 0\\
0& -1
\end{array}
\right)\\
\\[-3mm]
\hline
\\[-3mm]
C_{2x}& 1& -1& 1 &-1& \left(
\begin{array}{cc}
1& 0\\
0& -1
\end{array}
\right)\\
\\[-3mm]
\hline
\\[-3mm]
C_{2y}& 1& -1& 1 &-1& \left(
\begin{array}{cc}
-1& 0\\
0& 1
\end{array}
\right)\\
\\[-3mm]
\hline
\\[-3mm]
C_{2a}& 1& -1& -1 &1& \left(
\begin{array}{cc}
0& -1\\
-1& 0
\end{array}
\right)\\
\\[-3mm]
\hline
\\[-3mm]
C_{2b}& 1& -1& -1 &1& \left(
\begin{array}{cc}
0& \,1\\
\,1& 0
\end{array}
\right)
\\[-3mm]
&&&&&\\
\hline
\multicolumn{6}{l}{D^{(\gamma)}(Ip)=D^{(\gamma)}(p), \ {\rm for }\ p\in \Db_{4},
\ I:{\rm inversion}.
}\\
\hline
\hline
\end{array}
$$
\label{D4reptab}
}
\end{table}

The symmetry group $G_0$ of the Hamiltonian $\mathcal{H}$ in  
\eqref{site-hamiltonian}
is given by 
\begin{align}
G_0&=\Pb\times \bm{S}\times \Rbb,
\end{align}
where $\Pb=\Lb(\eb_1,\eb_2)\wedge \Db_{4h}$ is the space group of the 
square lattice, $\Lb(\eb_1,\eb_2)$ is a two-dimensional translation group
with basis vectors $\eb_1$ and $\eb_2$, and $\wedge$ denotes the semidirect 
product. $\Db_{4h}$ is the point group of the square lattice. 
$\SSb$ is the group of spin rotation, ${\Rbb}=\{ E,t\}$ is the
 group of time reversal. 

Three orbitals $\phi_1(\rb), \phi_2(\rb)$ and 
$\phi_3(\rb)$ have following symmetry properties for $p\in \Db_{4h}$. 
\begin{align}
p\cdot \phi_i(\rb)&\equiv \phi_i(p^{-1}\cdot \rb)=
\sum _{j=1}^2\phi_j(\rb)D_{ji}^{(E_g)}(p),\quad  {\rm for}\quad i=1,2
\nonumber\\
p\cdot \phi_3(\rb)&\equiv \phi_3(p^{-1}\cdot \rb)=
\chi^{(B_{2g})}(p)\phi_3(\rb),
\label{pactphi-i-1}
\end{align}
where $D_{ji}^{(E_g)}(p)$($\chi^{(B_{2g})}(p)$) are R-rep matrices
of $E_{g}$($B_{2g}$). The equations \eqref{pactphi-i-1} can be expressed 
in a more 
compact form 
\begin{align}
p\cdot\phi_i(\rb)&=\sum_{j=1}^3\phi_j(\rb)D_{ji}(p),\ i=1,2,3
\end{align}
where 
\begin{align}
D(p)&=\left(
\begin{array}{ccc}
D^{(E_g)}_{11}(p)& D^{(E_g)}_{12}(p)&0\\
D^{(E_g)}_{21}(p)& D^{(E_g)}_{22}(p)&0\\
0& 0& \chi^{(B_{2g})}(p)
\end{array}
\right).
\end{align}
Action of $p\in \Db_{4h}$ on $\phi_i(\rb-\nb)$ are given by
\begin{align}
p\cdot \phi_i(\rb-\nb)&\equiv \phi_i(p^{-1}\cdot \rb -\nb)=
\phi_i(p^{-1}\cdot(\rb-p\cdot \nb))\nonumber \\
&=\sum_{j=1}^3\phi_j(\rb-p\cdot\nb)D_{ji}(p).
\end{align} 
Then actions of $p\in \Db_{4h}$ on $\{a^{\dag}_{\bm{n}is},a^{}_{\bm{n}is}\}$ are
given by 
\begin{align}
p\cdot a^{\dag}_{\bm{n}is}&=\sum _{j=1}^3a^{\dag}_{(p\cdot\bm{n})js}
D_{ji}(p),\quad p\cdot a^{}_{\bm{n}is}
=\sum _{j=1}^3a^{}_{(p\cdot\bm{n})js}
D_{ji}(p).
\label{D4han}
\end{align}

Actions of translation $T(\nb)\in \Lb(\eb_1,\eb_2)$
with a vector 
$\nb=n_1\eb_1+n_2\eb_2$($n_1$ and $n_2$ are integers)
on $\{a^{\dag}_{\bm{m}is}, a^{}_{\bm{m}is}\}$ are given by
\begin{align}
T(\nb)\cdot a^{\dag}_{\bm{m}is}&=a^{\dag}_{(\bm{n+m})is},\qquad
T(\nb)\cdot a^{}_{\bm{m}is}=a^{}_{(\bm{n+m})is}.
\label{Tnam}
\end{align}
Actions of spin rotation $u(\nb,\theta)\in \SSb$, time 
reversal $t\in \Rbb$ on 
$\{a^{\dag}_{\smb is}, 
a^{}_{\smb is}\}$ are defined 
by\cite{ozaki1,Masago} 
\begin{align}
u(\nb,\theta)\cdot a^{\dag}_{\smb i s}&=
\sum_{s^{\prime}}u(\nb,\theta)_{s^{\prime} s} 
a^{\dag}_{\smb i s^{\prime}},& 
u(\nb,\theta)\cdot a^{}_{\smb i s}&=
\sum_{s^{\prime}}\{u(\nb,\theta)_{s^{\prime} s}\}^* 
a^{}_{\smb i s^{\prime}},\nonumber \\
t\cdot (za^{\dag}_{\smb i \uparrow})
&=-z^{*}a^{\dag}_{\smb i \downarrow},& 
t\cdot (za^{}_{\smb i \uparrow})&
=-z^{*}a^{}_{\smb i \downarrow},\nonumber \\
t\cdot (za^{\dag}_{\smb i \downarrow})
&=z^{*}a^{\dag}_{\smb i \uparrow},& 
t\cdot (za^{}_{\smb i \downarrow})
&=z^{*}a^{}_{\smb i \uparrow},
\label{uTam}
\end{align}
where $u(\nb,\theta)$ is a spin rotation by $\theta$ around the $\nb$ axis 
and is given by $2\times 2$ unitary matrix;
\begin{align}
u(\nb,\theta)&=\cos (\dfrac{\theta}{2}){\bf 1}_2-i(\sigmab \cdot \nb)
\sin (\dfrac{\theta}{2}),\nonumber\\
\sigmab&=(\sigma^1,\sigma^2,\sigma^3):\ {\rm Pauli\  matrices},\nonumber \\
{\bf 1}_2&=2\times 2\ \text{unit matrix}, 
\end{align}
and 
$z$ is a complex number and $z^{\ast}$ is a complex conjugate of $z$.

From $\Db_{4h}$ invariance of the Hamiltonian $\mathcal{H}$, we 
obtain following conditions 
for $\epsilon_i$ and  $t^{\sab}_{ij}$
\begin{align}
\epsilon_1&=\epsilon_2=\delta,\ \epsilon_3=0  \nonumber\\
t^{(1,0)}_{11}&=t^{(-1,0)}_{11}=t^{(0,1)}_{22}=t^{(0,-1)}_{22}=t_1,\nonumber\\
t^{(1,0)}_{22}&=t^{(-1,0)}_{22}=t^{(0,1)}_{11}=t^{(0,-1)}_{11}=t_2,\nonumber\\
t^{(1,0)}_{33}&=t^{(-1,0)}_{33}=t^{(0,1)}_{33}=t^{(0,-1)}_{33}=t_3,\nonumber\\
t^{\sab}_{ij}&=0, \quad{\rm in\ other\ cases}
\end{align}
where $\delta$ is the level splitting between $\phi_3$  and 
$(\phi_1,\phi_2)$ orbitals. 

Using Fourier transformations
\begin{align}\label{fourier}
a^{\dag}_{i \bm{k} s } &= \dfrac{1}{\sqrt{N}} 
\sum_{\bm{n}}^{} \mathrm{e}^{\img \bm{k} \cdot \bm{n} } 
a^{\dag}_{\bm{n}is},
\quad 
a^{}_{i \bm{k} s }= 
\dfrac{1}{\sqrt{N}} 
\sum_{\bm{n}}^{} \mathrm{e}^{-\img \bm{k} \cdot \bm{n} } a^{}_{\bm{n}is},
\end{align}
we obtain $\mathcal{H}$ in the momentum representation
\begin{align}
\mathcal{H} =&
\sum_{\bm{k},s} \sum_{i=1}^{3} \left\{-t^{}_{ii}(\bm{k}) - \mu 
\right\}
a^{\dag}_{i\bm{k}s} a^{}_{i\bm{k}s}\nonumber \\
& + \dfrac{1}{2} \sum^{}_{\bm{k},\bm{k'},\bm{q}}\quad
 \sum^{3}_{i,j,n,m=1} 
\sum^{}_{s ,s'}
\langle i(\kb+\qb)s,n\kb's'\mid V\mid j\kb s,m(\kb'+\qb)s'
\rangle \nonumber\\
&\qquad \times a^{\dag}_{i(\bm{k+q})s} 
a^{\dag}_{n\bm{k'}s '}
a^{}_{m(\bm{k'+q})s'} a^{}_{j\bm{k}s}.
\label{Hamk}
\end{align}
where $N$ is the number of lattice sites and 
$\kb$ runs in the first
Brillouin zone:\\ $\{\kb=(k_1,k_2)\mid -\pi 
\leq k_1,k_2 \leq \pi\}$, and\\
$t_{ii}(\bm{k})$ and $
\langle i(\kb+\qb)s,n\kb's'\mid V\mid j\kb s,m(\kb'+\qb)s'
\rangle$  are
given by
\begin{align}
t_{11}(\bm{k})&=-\delta+2(t_1\cos k_1+t_2\cos k_2),\nonumber\\
t_{22}(\bm{k})&=-\delta+2(t_2\cos k_1+t_1\cos k_2),\nonumber\\
t_{33}(\bm{k})&=2t_3(\cos k_1+\cos k_2),
\end{align}
\\[-1.0cm]
\begin{align}
\langle i(\kb+\qb)s,i\kb's'\mid V\mid i\kb s,i(\kb'+\qb)s'
\rangle&= \dfrac{U}{N},\nonumber\\
\langle i(\kb+\qb)s,j\kb's'\mid V\mid i\kb s,j(\kb'+\qb)s'
\rangle&= \dfrac{U'}{N},\quad i\neq j,\nonumber\\
\langle i(\kb+\qb)s,j\kb's'\mid V\mid j\kb s,i(\kb'+\qb)s'
\rangle&= \dfrac{J}{N},\quad i\neq j,\nonumber\\
\langle i(\kb+\qb)s,i\kb's'\mid V\mid j\kb s,j(\kb'+\qb)s'
\rangle&= \dfrac{J'}{N},\quad i\neq j,\nonumber\\
\langle i(\kb+\qb)s,n\kb's'\mid V\mid j\kb s,m(\kb'+\qb)s'
\rangle&=0,\quad {\rm otherwise}.
\label{intK}
\end{align}
From \eqref{D4han}, \eqref{Tnam}, \eqref{uTam} and 
\eqref{fourier}, 
actions of $p\in \Db_{4h}, T(\nb)\in \Lb(\eb_1,\eb_2),\\ 
u(\nb,\theta)\in \SSb, t\in \Rbb$ on $\{a^{\dag}_{i\skb s}, 
a^{}_{i\skb s}\}$ are given by
\begin{align}
p\cdot a^{\dag}_{i\skb s}&=\sum_{j=1}^3D_{ji}(p)
a^{\dag}_{j(p\cdot \skb)s},&
p\cdot a^{}_{i\skb s}&=\sum_{j=1}^3D_{ji}(p)
a^{}_{j(p\cdot \skb)s},\nonumber \\
T(\nb)\cdot a^{\dag}_{i\skb s}&=
e^{-i\skb\cdot\snb}a^{\dag}_{i\skb s},&
T(\nb)\cdot a^{}_{i\skb s}&=
e^{i\skb\cdot\snb}a^{}_{i\skb s},\nonumber \\
u(\nb,\theta)\cdot a^{\dag}_{i\skb s}&=\sum_{s'=1}^2
u(\nb,\theta)_{s's}a^{\dag}_{i\skb s'},&
u(\nb,\theta)\cdot a^{}_{i\skb s}&=\sum_{s'=1}^2
u(\nb,\theta)^*_{s's}a^{}_{i\skb s'},\nonumber \\
t\cdot (za^{\dag}_{i\skb \uparrow})&=
-z^*a^{\dag}_{i(-\skb)\downarrow},&
t\cdot (za^{}_{i\skb \uparrow})&=
-z^*a^{}_{i(-\skb)\downarrow},\nonumber \\
t\cdot (za^{\dag}_{i\skb \downarrow})&=
z^*a^{\dag}_{i(-\skb)\uparrow},&
t\cdot (za^{}_{i\skb \downarrow})&=
z^*a^{}_{i(-\skb)\uparrow}.
\end{align}

\section{Hartree-Fock Hamiltonian and its isotropy subgroup}
 In this paper we consider 
HF solutions with four types 
of ordering vectors $\Qb_0=(0,0), \Qb_1=(\pi,\pi), \Qb_2=(\pi,0),
\Qb_3=(0,\pi)$. 
Thus a general HF Hamiltonian $H_m$ is written as
\begin{align}
H_m&=H_K+\sum_{l=0}^3\sum_{i,j=1}^3\sum_{\lambda=0}^3\sum_{s,s'=1}^2
\sum_{\skb}x^{l\lambda}_{ij}(\kb)a^{\dag}_{i(\skb+\sQb_l)s}
\sigma^{\lambda}_{ss'}a^{}_{j\skb s'},
\label{Hmdef}
\end{align}
where 
$H_K$ is the kinetic energy written as
\begin{align}
H_K &= \sum_{\skb}\sum_{i=1}^3\sum _{s=1}^2 \left\{ -t_{ii}(\kb)-\mu \right\}
a^{\dag}_{i\skb s}a^{}_{i\skb s}.
\end{align}
From the Hermite condition of $H_m$ we have 
\begin{align}
x^{l\lambda}_{ij}(\kb+\Qb_l)^*&=x^{l\lambda}_{ji}(\kb).
\label{Her-x-1}
\end{align}
$x^{l\lambda}_{ij}(\kb)$ in \eqref{Hmdef} satisfy following 
SCF conditions
\begin{align}
x^{l0}_{ij}(\kb)&=\sum_{\skb'}\sum_{m,n=1}^3W^l_{injm}(\kb,\kb')
\rho^{l0}_{mn}(\kb'),\quad l=0,1,2,3\nonumber\\
x^{l\lambda}_{ij}(\kb)&=\sum_{\skb'}\sum_{m,n=1}^3Y^l_{injm}(\kb,\kb')
\rho^{l\lambda}_{mn}(\kb'),\quad l=0,1,2,3,\ \lambda=1,2,3
\label{SCF1}
\end{align}
where
\begin{align}
\rho^{l\lambda}_{ij}(\kb)&=\dfrac{1}{2}\sum_{s,s'=1}^2
\langle a^{\dag}_{j(\skb+\sQb_l)s}a^{}_{i\skb s'}
\rangle \sigma^{\lambda}_{ss'},\nonumber\\
W^l_{injm}(\kb,\kb')&=2\langle i(\kb+\Qb_l)s,n\kb's\mid V\mid
j\kb s,m(\kb'+\Qb_l)s\rangle\nonumber\\
&\quad -\langle i(\kb+\Qb_l)s,n\kb's\mid V\mid m(\kb'+\Qb_l)s,
j\kb s\rangle,\nonumber\\
Y^l_{injm}(\kb,\kb')&=
-\langle i(\kb+\Qb_l)s,n\kb's\mid V\mid m(\kb'+\Qb_l)s,
j\kb s\rangle,
\label{RoWY}
\end{align}
where $\langle \hat{A} \rangle$ denotes the expectation value of $\hat{A}$
in the ground state of the HF Hamiltonian $H_m$ and 
$\rho^{l\lambda}_{ij}(\kb)$ satisfies 
\begin{align}
\rho^{l\lambda}_{ji}(\kb)^*&=\rho^{l\lambda}_{ij}(\kb+\Qb_l),\nonumber\\
\langle a^{\dag}_{j(\skb+\sQb_l)s'}a^{}_{i\skb s}\rangle
&=\sum_{\lambda=0}^3\sigma ^{\lambda}_{s s'}\rho^{l\lambda}_{ij}(\kb).
\label{roHermite-1}
\end{align}
From \eqref{intK} we have for $i\neq j$
\begin{align}
W^l_{iiii}(\kb,\kb')&\equiv W_{iiii}=\dfrac{U}{N},& 
Y^l_{iiii}(\kb,\kb')&\equiv Y_{iiii}=-\dfrac{U}{N},\nonumber \\
W^l_{ijij}(\kb,\kb')&\equiv W_{ijij}=\dfrac{2U'-J}{N},&
Y^l_{ijij}(\kb,\kb')&\equiv Y_{ijij}=-\dfrac{J}{N},\nonumber \\
W^l_{iijj}(\kb,\kb')&\equiv W_{iijj}=\dfrac{J'}{N},&
Y^l_{iijj}(\kb,\kb')&\equiv Y_{iijj}=-\dfrac{J'}{N},\nonumber \\
W^l_{ijji}(\kb,\kb')&\equiv W_{ijji}=\dfrac{2J-U'}{N},&
Y^l_{ijji}(\kb,\kb')&\equiv Y_{ijji}=-\dfrac{U'}{N},\nonumber \\
W^l_{injm}(\kb,\kb')&\equiv W_{injm}=0,& 
Y^l_{injm}(\kb,\kb')&\equiv Y_{injm}=0,\  {\rm otherwise}
\label{WY}
\end{align}
From \eqref{SCF1} and \eqref{WY} we see that $x^{l\lambda}_{ij}(\kb)$ are 
 independent of $\kb $ and given by
\begin{align}
x^{l0}_{ij}(\kb)&\equiv x^{l0}_{ij}=N\sum_{m,n=1}^3W_{injm}R^{l0}_{mn},
\quad l=0,1,2,3\nonumber\\
x^{l\lambda}_{ij}(\kb)&\equiv x^{l0}_{ij}=N\sum_{m,n=1}^3
Y_{injm}R^{l\lambda}_{mn},\quad l=0,1,2,3,\quad \lambda=1,2,3
\label{xlmuconst}
\end{align}
where 
\begin{align}
R^{l\lambda}_{ij}&=\dfrac{1}{N}\sum_{\skb}\rho^{l\lambda}_{ij}(\kb).
\label{Rlmuij-def}
\end{align}
We denote a $3\times 3$ matrix whose $ (i,j)$ component is $x^{l\lambda}_{ij}$
($R^{l\lambda}_{ij}$) by $\xb^{l\lambda}$($\Rb^{l\lambda}$).
From \eqref{Her-x-1}, \eqref{roHermite-1}, \eqref{xlmuconst}, 
\eqref{Rlmuij-def}, we see that 
$\xb^{l\lambda}$ and $\Rb^{l\lambda}$ are Hermite matrices.

From \eqref{xlmuconst} the HF Hamiltonian $H_m$ 
in \eqref{Hmdef} can be written as 
\begin{align}
H_m &= H_K + \sum_{l=0}^3\sum_{i,j=1}^3\sum_{\lambda=0}^3
\sum_{s,s'=1}^2
\sum_{\skb}
x^{l\lambda}_{ij}
a^{\dag}_{i(\skb+\sQb_l)s}
\sigma^{\lambda}_{ss'}a^{}_{j\skb s'}.
\label{Hmdef2}
\end{align}
From \eqref{Hmdef2} and the Hermiticity of $\xb^{l\lambda}$,
$H_m$ is characterized by $3\times 3$ Hermitian matrices 
$\xb^{l\lambda}$ ($16$ matrices in all, corresponding $l=0,1,2,3,\ 
\lambda=0,1,2,3)$. Thus $H_m$ is specified by a vector in the
{\bf HF Hamiltonian space} $W_{\rm HF}$ over real
number field ${\rm R}$:
\begin{align}  
W_{\rm HF} &= \left\{ \sum_{\skb}(a^{\dag}_{i(\skb+\sQb_l)s}a^{}_{j\skb s'}
+a^{\dag}_{j\skb s'}a^{}_{i(\skb+\sQb_l)s}),\sum_{\skb}
(ia^{\dag}_{i(\skb+\sQb_l)s}a^{}_{j\skb s'}
-ia^{\dag}_{j\skb s'}a^{}_{i(\skb+\sQb_l)s})
\right\}_{\rm R},
\end{align}
where $l=0,1,2,3,\  i,j=1,2,3$ and 
$\{A,B,\cdots,\}_{\rm R}$ denotes a vector 
space with bases $A,B,\cdots$
over the real number field.

The HF energy is expressed in terms of $\Rb^{l\lambda}$ 
\begin{align}
E_{\rm FH}&=\langle H_{\rm K} \rangle' 
+N^2\sum_{i,n,j,m=1}^3
W_{injm}R^{l0}_{ji}R^{l0}_{mn}\nonumber \\
&\ +N^2\sum_{i,n,j,m=1}^3\sum_{\lambda=1}^3
Y_{injm}R^{l\lambda}_{ji}R^{l\lambda}_{mn},\nonumber\\
\langle H_{\rm K} \rangle' &=
-2\sum_{\skb}\sum_{i=1}^3t_{ii}(\kb)\rho^{00}_{ii}(\kb).
\label{HF-energy}
\end{align}
The SCF condition \eqref{xlmuconst} corresponds to the extremum condition
of $E_{\rm FH}$\cite{ozakim}.

Actions $g\in G_0$ on $H_m$ are defined by
\begin{align}
g\cdot H_m&=H_K\nonumber \\
&\quad +\sum_{l=0}^3\ \sum_{i,j=1}^3\sum_{\lambda=0}^3\ \sum_{s,s'=1}^2
\sum_{\skb}(x^{l\lambda}_{ij})^{(*)}
(g\cdot a^{\dag}_{i(\skb+\sQb_l)s})
(\sigma^{\lambda}_{ss'})^{(*)}(g\cdot a^{}_{j\skb s'}),
\end{align}
where $A^{(*)}$ denotes the complex conjugation of 
a complex number $A$ in the case of anti-unitary $g$ which contains
time reversal $t$.
Note that 
\begin{align}
g\cdot H_K&\equiv\sum_{\skb}\sum_{i=1}^3\sum _{s=1}^2
\left\{-t_{ii}(\kb)-\mu\right\}
(g\cdot a^{\dag}_{i\skb s})(g\cdot a^{}_{i\skb s})=H_K.
\end{align}
We define {\it the isotropy subgroup} $G(H_m)$ of $H_m$ by 
\begin{align}
G(H_m)&\equiv \{g\in G_0 \mid g\cdot H_m=H_m\}.
\end{align}
Actions  of $g\in G_0$ on $\Rb^{l\lambda}$ are defined by
\begin{align}
(g\cdot \Rb^{l\lambda})_{ij}&=\dfrac{1}{2N}\sum_{\skb}
\sum_{s,s'=1}^2\langle (g^{-1}\cdot a^{\dag}_{j(\skb+\sQb_l)s})
(g^{-1}\cdot a_{i\skb s'})\rangle ^{(*)}\sigma ^{\lambda}_{ss'}.
\end{align}
In previous papers\cite{ozakim,ozaki2},
we have shown that for $g\in G(H_m)$ 
\begin{align}
\langle (g\cdot a^{\dag}_{i\skb s})
(g\cdot a^{}_{j\skb' s'})\rangle ^{(*)}
&=\langle a^{\dag}_{i\skb s}a^{}_{j\skb' s'}\rangle.
\end{align}
Then we obtain for $g\in G(H_m)$
\begin{align}
g\cdot \Rb^{l\lambda}&=\Rb^{l\lambda}.
\label{gRR}
\end{align}

For subsequent uses we list explicit forms of $G_0$ actions.
For $p\in \Db_{4h}, T(\mb)\in \Lb_0, u(\nb,\theta)\in \SSb,
t\in \Rbb$
\begin{align}
p\cdot \Rb^{l\lambda}&=D(p)\Rb^{(p^{-1}\cdot l)\lambda}D^{\dag}(p),\nonumber\\
T(\mb)\cdot \Rb^{l\lambda}&=e^{i\sQb_l\cdot \smb}\Rb^{\l\lambda},\nonumber\\
u(\nb,\theta)\cdot \Rb^{l0}&=\Rb^{l0},\nonumber\\
u(\nb,\theta)\cdot \Rb^{l\lambda}&=\sum_{\lambda'=1}^3
R(\nb,-\theta)_{\lambda'\lambda}
 \Rb^{l\lambda'},\quad \lambda=1,2,3,\nonumber\\
t\cdot\Rb^{l0}&=(\Rb^{l0})^*,\nonumber\\
t\cdot\Rb^{l\lambda}&=-(\Rb^{l\lambda})^*,\ \lambda=1,2,3, 
\label{gaction-R}
\end{align}
where $(p^{-1}\cdot l)$ is defined such that 
\begin{align}
\Qb_{p^{-1}\cdot l}&\equiv p^{-1}\cdot \Qb_l,
\end{align}
and $R(\nb,\theta)$ is the rotation matrix by $\theta$ radian around 
$\nb=(n_1,n_2,n_3)$ axis in the three dimensional Euclid space and is
given by\cite{Cornwell,ozaki2}
\begin{equation}
\begin{split}
R&(u(\nb,\theta))= R(\nb, -\theta)=\\[2mm]
&\left(
\begin{array}{ccc}
\cos \theta +(1-\cos \theta)n_1^2& (1-\cos \theta)n_1n_2-n_3\sin \theta &
(1-\cos \theta)n_1n_3+n_2\sin \theta\\
(1-\cos \theta)n_1n_2+n_3\sin\theta&\cos \theta+(1-\cos \theta)n_2^2&
(1-\cos \theta)n_2n_3-n_1\sin\theta\\
(1-\cos \theta)n_1n_3-n_2\sin\theta & (1-\cos \theta)n_2n_3+n_1\sin\theta
&\cos \theta +(1-\cos \theta)n_3^2
\end{array}
\right).
\end{split}
\label{Rotation-matrix}
\end{equation}

We define 
density matrices $\Db^{s s'}(\mb)=\{D^{ss'}_{ij} (i,j=1,2,3, s,s'=1,2)$\}
at a site $\mb$ as follows:
\begin{align}
D^{ss'}_{ij}(\mb)&=\langle a^{\dag}_{\smb j s'}a~{}_{\smb i s}\rangle. 
\label{Dssdij}
\end{align}
Since for all states with ordering vectors $\Qb_l\ (l=0,1,2,3)$ 
\begin{align}
\langle a^{\dag}_{j\skb' s'}a^{}_{i\skb s}\rangle &=0, \ {\rm for}\ 
\kb'\neq \kb+\Qb_l,
\end{align}
using \eqref{roHermite-1}
we obtain 
\begin{align}
D^{ss'}_{ij}(\mb) &= \dfrac{1}{N}\sum_{\skb, \skb'}e^{-i\skb'\cdot \smb}
e^{i\skb\cdot \smb}\langle a^{\dag}_{j\skb's'}a^{}_{i\skb s}\rangle
 \nonumber \\
&=\dfrac{1}{N}\sum_{\skb }\sum_{l=0}^3e^{-i\sQb_l\cdot \smb}
\langle a^{\dag}_{j(\skb+\sQb_l)s'}a^{}_{i\skb s}\rangle\nonumber \\
&=\dfrac{1}{N}\sum_{\skb }\sum_{l=0}^3\sum_{\lambda=0}^3
e^{-i\sQb_l\cdot \smb}\sigma^{\lambda}_{ss'}\rho^{l\lambda}_{ij}(\kb)
\nonumber \\
&=\sum_{l=0}^3\sum_{\lambda=0}^3
e^{-i\sQb_l\cdot \smb}\sigma^{\lambda}_{ss'}R^{l\lambda}_{ij}.
\end{align}
Thus we obtain 
\begin{align}
\Db^{ss'}(\mb)&=\sum_{l=0}^3\sum_{\lambda=0}^3
e^{-i\sQb_l\cdot \smb}\sigma^{\lambda}_{ss'}\Rb^{l\lambda}.
\label{DssdijRij}
\end{align}
Then we obtain explicit expression of density matrices 
as follows:
\begin{align}
\Db^{\uparrow\uparrow}(\mb)&=\sum_{l=0}^3e^{-i\sQb_l\cdot \smb}
(\Rb^{l0}+\Rb^{l3}),\nonumber\\
\Db^{\downarrow\downarrow}(\mb)&=\sum_{l=0}^3e^{-i\sQb_l\cdot \smb}
(\Rb^{l0}-\Rb^{l3}),\nonumber\\
\Db^{\uparrow\downarrow}(\mb)&=\sum_{l=0}^3e^{-i\sQb_l\cdot \smb}
(\Rb^{l1}-i\Rb^{l2}),\nonumber\\
\Db^{\downarrow\uparrow}(\mb)&=\sum_{l=0}^3e^{-i\sQb_l\cdot \smb}
(\Rb^{l1}+i\Rb^{l2}).
\label{Dmupdown}
\end{align}
From \eqref{gRR} we can see that symmetry properties 
of density matrices $\Db^{ss'}(\mb)$ are determined by $G(H_m)$.

Using notations 
\begin{align}
\Ab^{\dag}(\mb)&=\left(A^{\dag}_1(\mb),A^{\dag}_2(\mb),A^{\dag}_3(\mb),
A^{\dag}_4(\mb),A^{\dag}_5(\mb),A^{\dag}_6(\mb)\right)\nonumber\\
&\equiv \left(a^{\dag}_{\smb 1\uparrow},a^{\dag}_{\smb 2\uparrow},
a^{\dag}_{\smb 3\uparrow},a^{\dag}_{\smb 1\downarrow},
a^{\dag}_{\smb 2\downarrow},a^{\dag}_{\smb 3\downarrow}\right),
\end{align}
we define  generalized density matrix $\Bf{\cal{D}}(\mb)$ whose $(i,j)$ component is given by
\begin{align}
{\cal{D}}_{ij}(\mb)&=\left\langle A^{\dag}_{j}(\mb)A_{i}(\mb)\right\rangle.
\end{align}
From \eqref{Dssdij} we obtain 
\begin{align}
\Bf{\cal{D}}(\mb)&=\left(
\begin{array}{cc}
\Db^{\uparrow \uparrow}(\mb) & \Db^{\uparrow \downarrow}(\mb)\\
\Db^{\downarrow \uparrow}(\mb)& \Db^{\downarrow \downarrow}(\mb)
\end{array}
\right).
\label{calDb-def}
\end{align}
The $6\times 6$ Hermitian matrix $\Bf{\cal{D}}$(\mb) is diagonalized by
a unitary matrix $\cal{U}$ as follows:
\begin{align}
\cal{U}^{\dag}\calDb \cal{U}&=\left(
\begin{array}{cccccc}
\lambda_1& 0 & 0 & 0 & 0 & 0 \\
0 & \lambda _2 & 0 & 0 & 0 & 0\\
0 & 0 & \lambda_3 & 0 & 0 & 0\\
0 & 0 & 0 &  \lambda_4 & 0 & 0 \\
0 & 0 & 0 & 0 & \lambda_5 & 0 \\
0 & 0 & 0 & 0 & 0&\lambda_6 
\end{array}
\right).
\end{align}
Defining $\Bb^{\dag}(\mb)=(B^{\dag}_1(\mb),B^{\dag}_2(\mb),
B^{\dag}_3(\mb),B^{\dag}_4(\mb),
B^{\dag}_5(\mb),B^{\dag}_6(\mb))$ by 
\begin{align}
\Bb^{\dag}(\mb)&=\Ab^{\dag}(\mb)\calU,\ \ \ \Bb(\mb)=\calU^{\dag}\Ab(\mb),
\end{align}
we obtain 
\begin{align}
\langle B_l^{\dag}(\mb)B_k(\mb)\rangle &=\sum_{i,j=1}^6
\langle A^{\dag}_{i}(\mb){\calU}_{i l}{\calU}^{\dag}_{kj}
A_{j}(\mb)\rangle 
\nonumber \\
&=\sum_{i,j=1}^6{\calU}_{i l}\langle A^{\dag}_{i}(\mb)
A_{j}(\mb)\rangle {\calU}^{\dag}_{kj} \nonumber \\
&=\sum_{i,j=1}^6{\calU}^{\dag}_{k j}{\calDb}(\mb)_{ji}
{\calU}_{i l}=\delta_{kl}\lambda_{l}.
\end{align}
Thus occupied spin orbitals $\psi_l$ and 
their occupation numbers $n(\psi_l)$ are given by
\begin{align}
\psi_l&=\sum_{i=1}^3\left(\phi_i\mid \uparrow\rangle {\calU}_{il}+
\phi_i\mid \downarrow\rangle {\calU}_{(i+3)l}\right),
\nonumber\\
n(\psi_l)&=\lambda_l.
\end{align}
where $l=1,\dots,6$.

 Here we present formulae for the local order parameter(LOP) at a site $\mb$.
The charge density at a site $\mb$: $d(\mb)$  is expressed by
\begin{align}
d(\mb)&= \sum_{j=1}^3\sum_{ss'}
\langle a^{\dag}_{\smb js}a^{}_{\smb j s'}
\rangle 
\sigma^{0}_{ss'}.
\end{align}
The $\lambda$ th component of the spin density at 
the $\mb$: $s^{\lambda}(\mb)$ 
is written as 
\begin{align}
s^{\lambda}(\mb)&=\dfrac{1}{2}\sum_{j=1}^3\sum_{ss'}\langle 
a^{\dag}_{\smb js}a^{}_{\smb j s'}
\rangle 
\sigma^{\lambda}_{ss'}, \lambda=1,2,3.
\end{align}
The ($i,j$) component of the  quadrupole moment at 
the site $\mb$: $Q_{ij}$(\mb) are written as 
\begin{align}
Q_{11}(\mb)&=I_1\sum_{s,s'=1}^2\left(2\langle a^{\dag}_{\smb 1 s}
a^{}_{\smb 1 s'}\rangle -\langle a^{\dag}_{\smb 2 s}
a^{}_{\smb 2 s'}\rangle -\langle a^{\dag}_{\smb 3 s}
a^{}_{\smb 3 s'}\rangle \right)\sigma^{0}_{ss'}\nonumber \\
Q_{22}(\mb)&=I_1\sum_{s,s'=1}^2\left(-\langle a^{\dag}_{\smb 1 s}
a^{}_{\smb 1 s'}\rangle +2\langle a^{\dag}_{\smb 2 s}
a^{}_{\smb 2 s'}\rangle -\langle a^{\dag}_{\smb 3 s}
a^{}_{\smb 3 s'}\rangle \right)\sigma^{0}_{ss'}\nonumber \\
Q_{33}(\mb)&=I_1\sum_{s,s'=1}^2\left(-\langle a^{\dag}_{\smb 1 s}
a^{}_{\smb 1 s'}\rangle -\langle a^{\dag}_{\smb 2 s}
a^{}_{\smb 2 s'}\rangle +2\langle a^{\dag}_{\smb 3 s}
a^{}_{\smb 3 s'}\rangle \right)\sigma^{0}_{ss'}\nonumber \\
Q_{ij}(\mb)&=
I_2\sum_{ss'}\langle a^{\dag}_{\smb i s}a^{}_{\smb j s'}
+a^{\dag}_{\smb j s}a^{}_{\smb i s'}\rangle \sigma^{0}_{ss'},\ i\neq j,
\label{quadrupole-1}
\end{align}
where 
\begin{align}
I_1&=\int d\rb \phi_1(\rb)(x^2-y^2)\phi_1(\rb),\nonumber \\
I_2&=3\int d\rb \phi_1(\rb)xy\phi_2(\rb).
\end{align}
The derivation of \eqref{quadrupole-1} is given in Appendix A.

The ($i,j$) component of the  {\it spin-quadrupole moment} at 
the site $\mb$: $Q^{\lambda}_{ij}$(\mb) are defined by, for $\lambda=1,2,3$, 
\begin{align}
Q^{\lambda}_{11}(\mb)&=\dfrac{1}{2}I_1\sum_{s,s'=1}^2
\left(2\langle a^{\dag}_{\smb 1 s}
a^{}_{\smb 1 s'}\rangle -\langle a^{\dag}_{\smb 2 s}
a^{}_{\smb 2 s'}\rangle -\langle a^{\dag}_{\smb 3 s}
a^{}_{\smb 3 s'}\rangle \right)\sigma^{\lambda}_{ss'},\nonumber \\
Q^{\lambda}_{22}(\mb)&=\dfrac{1}{2}I_1\sum_{s,s'=1}^2
\left(-\langle a^{\dag}_{\smb 1 s}
a^{}_{\smb 1 s'}\rangle +2\langle a^{\dag}_{\smb 2 s}
a^{}_{\smb 2 s'}\rangle -\langle a^{\dag}_{\smb 3 s}
a^{}_{\smb 3 s'}\rangle \right)\sigma^{\lambda}_{ss'},\nonumber \\
Q^{\lambda}_{33}(\mb)&=\dfrac{1}{2}I_1\sum_{s,s'=1}^2
\left(-\langle a^{\dag}_{\smb 1 s}
a^{}_{\smb 1 s'}\rangle -\langle a^{\dag}_{\smb 2 s}
a^{}_{\smb 2 s'}\rangle +2\langle a^{\dag}_{\smb 3 s}
a^{}_{\smb 3 s'}\rangle \right)\sigma^{\lambda}_{ss'},\nonumber \\
Q^{\lambda}_{ij}(\mb)&=
\dfrac{I_2}{2}\sum_{ss'}\langle a^{\dag}_{\smb i s}a^{}_{\smb j s'}
+a^{\dag}_{\smb j s}a^{}_{\smb i s'}\rangle 
\sigma^{\lambda}_{ss'},\ i\neq j, \lambda=1,2,3.
\label{spin-Quadru-1}
\end{align}
This type of order parameter has been treated in a paper by
Shiina, Nishitani and Shiba\cite{Shiina} as the  coupled orbital and 
spin morment in the case of the superexchange model 
of $\mathrm{e_g}$ orbital.

In our system with tetragonal symmetry we use $Q_2^2(\mb)$ and 
$Q_2^{2\lambda}(\mb)$ instead of $Q_{ii}$ and $Q_{ii}^{\lambda}(i=1,2,3)$,
where
\begin{align}
Q^2_2(\mb)&=\dfrac{1}{3}(Q_{11}-Q_{22})=I_1\sum_{ss'=1}^2
\left(\langle a^{\dag}_{\smb 1 s}
a^{}_{\smb 1 s'}\rangle -\langle a^{\dag}_{\smb 2 s}
a^{}_{\smb 2 s'}\rangle\right)\sigma^0_{ss'},\nonumber \\
Q^{2\lambda}_2(\mb)&=\dfrac{1}{3}(Q^{\lambda}_{11}-Q^{\lambda}_{22})
=\dfrac{I_1}{2}\sum_{ss'=1}^2
\left(\langle a^{\dag}_{\smb 1 s}
a^{}_{\smb 1 s'}\rangle -\langle a^{\dag}_{\smb 2 s}
a^{}_{\smb 2 s'}\rangle\right)\sigma^{\lambda}_{ss'}.
\end{align}

The $i$-th component$(i=1,2,3)$ of the orbital angular momentum 
at the site $\mb$: $l_i(\mb)$  are written as 
\begin{align}
l_1(\mb)&=\sum_{ss'}\left(
i \langle a^{\dag}_{\smb 2 s}a^{}_{\smb 3s'}\rangle-
i \langle a^{\dag}_{\smb 3 s}a^{}_{\smb 2s'}\rangle\right)
\sigma^{0}_{ss'},\nonumber \\
l_2(\mb)&=\sum_{ss'}\left(
i \langle a^{\dag}_{\smb 3 s}a^{}_{\smb 1s'}\rangle-
i \langle a^{\dag}_{\smb 1 s}a^{}_{\smb 3s'}\rangle\right)
\sigma^{0}_{ss'},\nonumber \\
l_3(\mb)&=\sum_{ss'}\left(
i \langle a^{\dag}_{\smb 1 s}a^{}_{\smb 2s'}\rangle-
i \langle a^{\dag}_{\smb 2 s}a^{}_{\smb 1s'}\rangle\right)
\sigma^{0}_{ss'}.
\label{orbital-angular-1}
\end{align}
The derivation of \eqref{orbital-angular-1} is given in Appendix B.

The $i$-th component($i=1,2,3$) of the  {\it spin-orbital angular momentum} 
at the site $\mb$: $l^{\lambda}_i(\mb)$  are defined by, for $\lambda=1,2,3$,
\begin{align}
l_1^{\lambda}(\mb)&=\dfrac{1}{2}\sum_{ss'}\left(
i \langle a^{\dag}_{\smb 2 s}a^{}_{\smb 3s'}\rangle-
i \langle a^{\dag}_{\smb 3 s}a^{}_{\smb 2s'}\rangle\right)
\sigma^{\lambda}_{ss'},\nonumber \\
l_2^{\lambda}(\mb)&=\dfrac{1}{2}\sum_{ss'}\left(
i \langle a^{\dag}_{\smb 3 s}a^{}_{\smb 1s'}\rangle-
i \langle a^{\dag}_{\smb 1 s}a^{}_{\smb 3s'}\rangle\right)
\sigma^{\lambda}_{ss'},\nonumber \\
l_3^{\lambda}(\mb)&=\dfrac{1}{2}\sum_{ss'}\left(
i \langle a^{\dag}_{\smb 1 s}a^{}_{\smb 2s'}\rangle-
i \langle a^{\dag}_{\smb 2 s}a^{}_{\smb 1s'}\rangle\right)
\sigma^{\lambda}_{ss'}.
\label{spin-orbital-ang}
\end{align}

Here we express these local order parameters in terms of $\Rb^{l\lambda}$.
Using \eqref{Dssdij} and \eqref{DssdijRij} we obtain 
\begin{align}
\sum_{ss'}\langle a^{\dag}_{\smb j s}a^{}_{\smb i s'}\rangle
 \sigma^{\lambda}_{ss'}
&=2\sum_{l=0}^3e^{-i\sQb_l\cdot \smb}R^{l\lambda}_{ij}.
\label{adagRij}
\end{align}
From \eqref{adagRij}, we obtain
\begin{align}
d(\mb)&=2\sum_{l=0}^3\sum_{j=1}^3e^{-i\sQb_l\cdot \smb}R^{l0}_{jj}\nonumber \\
s^{\lambda}(\mb)&=\sum_{l=0}^3\sum_{j=1}^3e^{-i\sQb_l\cdot \smb}
R^{l\lambda}_{jj},\nonumber \\
Q_2^2(\mb)&=2I_1\sum_{l=0}^3e^{-i\sQb_l\cdot \smb}\left(R^{l0}_{11}-
R^{l0}_{22}\right)\nonumber \\
Q_{ij}(\mb)&=2I_2\sum_{l=0}^3e^{-i\sQb_l\cdot \smb}
\left(R^{l0}_{ij}+R^{l0}_{ji}\right),\  i\neq j
,\nonumber \\
Q_2^{2\lambda}(\mb)&=I_1\sum_{l=0}^3e^{-i\sQb_l\cdot \smb}
\left(R^{l\lambda}_{11}-
R^{l\lambda}_{22}\right)\nonumber \\
Q^{\lambda}_{ij}(\mb)&=I_2\sum_{l=0}^3e^{-i\sQb_l\cdot \smb}
(R^{l\lambda}_{ij}
+R^{l\lambda}_{ji}),\ i\neq j 
,\nonumber \\
l_1(\mb)&=2\sum_{l=0}^3e^{-i\sQb_l\cdot \smb}
(-i)(R^{l0}_{23}-R^{l0}_{32}),\quad  
l_1^{\lambda}(\mb)=\sum_{l=0}^3e^{-i\sQb_l\cdot \smb}(-i)
(R^{l\lambda}_{23}-R^{l\lambda}_{32})
,\nonumber \\
l_2(\mb)&=2\sum_{l=0}^3e^{-i\sQb_l\cdot \smb}(-i)
(R^{l0}_{31}-R^{l0}_{13}),\quad
l_2^{\lambda}(\mb)=\sum_{l=0}^3e^{-i\sQb_l\cdot \smb}(-i)
(R^{l\lambda}_{31}-R^{l\lambda}_{13})
,\nonumber \\
l_3(\mb)&=2\sum_{l=0}^3e^{-i\sQb_l\cdot \smb}(-i)
(R^{l0}_{12}-R^{l0}_{21}),\quad
l_3^{\lambda}(\mb)=\sum_{l=0}^3e^{-i\sQb_l\cdot \smb}(-i)
(R^{l\lambda}_{12}-R^{l\lambda}_{21}).
\label{OP-R}
\end{align}

Here we give the physical meanings of 
$Q_{ij}^{\lambda}$ and $l_j^{\lambda}\ (i,j,\lambda=1,2,3)$.
As an example we consider the case of $Q_{12}^{3}$ and $l_1^3$.
The quadrupole moment $Q_{12}^{\uparrow}(\mb)(Q_{12}^{\downarrow}(\mb))$ 
by the up spin (down spin) electrons at 
the site
$\mb$ is written as 
\begin{align}
Q^{\uparrow}_{12}(\mb)&=I_2\left\langle
a^{\dag}_{\smb 1 \uparrow}a^{}_{\smb 2 \uparrow}+
a^{\dag}_{\smb 2 \uparrow}a^{}_{\smb 1 \uparrow}
\right\rangle, \nonumber \\
Q^{\downarrow}_{12}(\mb)&=I_2\left\langle
a^{\dag}_{\smb 1 \downarrow}a^{}_{\smb 2 \downarrow}+
a^{\dag}_{\smb 2 \downarrow}a^{}_{\smb 1 \downarrow}
\right\rangle.
\end{align}
Thus we obtain
\begin{align}
Q_{12}(\mb)&=Q_{12}^{\uparrow}(\mb)+Q_{12}^{\downarrow}(\mb),\nonumber \\
Q_{12}^{3}(\mb)&=\dfrac{1}{2}\left(Q_{12}^{\uparrow}(\mb)-
Q_{12}^{\downarrow}(\mb)\right).
\end{align}
Then we obtain
\begin{align}
Q_{12}^{\uparrow}(\mb)&=\dfrac{1}{2}\left(Q_{12}(\mb)+2Q_{12}^{3}(\mb)\right),
\nonumber \\
Q_{12}^{\downarrow}(\mb)
&=\dfrac{1}{2}\left(Q_{12}(\mb)-2Q_{12}^{3}(\mb)\right).
\label{Q12updown}
\end{align}
From \eqref{Q12updown}  the  existence of $Q_{12}^{3}(\mb)$ 
implies the {\it different quadrupole moment for different spin}.

The orbital angular momentum $l_1^{\uparrow}(\mb) (l_1^{\downarrow}(\mb))$
 by the up spin  
(down spin) electrons at the site $\mb$ is written as
\begin{align}
l_1^{\uparrow}(\mb)&=i\left\langle a^{\dag}_{\smb 2 \uparrow}
a^{}_{\smb 3 \uparrow}-a^{\dag}_{\smb 3 \uparrow}
a^{}_{\smb 2 \uparrow}\right\rangle,\nonumber \\
l_1^{\downarrow}(\mb)&=i\left\langle a^{\dag}_{\smb 2 \downarrow}
a^{}_{\smb 3 \downarrow}-a^{\dag}_{\smb 3 \downarrow}
a^{}_{\smb 2 \downarrow}\right\rangle.
\end{align}
Thus we obtain
\begin{align}
l_1^{\uparrow}(\mb)&=\dfrac{1}{2}\left(l_1(\mb)+2l_1^3(\mb)\right),\nonumber \\
l_1^{\downarrow}(\mb)&=\dfrac{1}{2}\left(l_1(\mb)-2l_1^3(\mb)\right).
\end{align} 
Thus the existence of $\l_1^{3}(\mb)$ implies the 
{\it different orbital angular momentum for different spin}.
We note that other $Q_{ij}^{\lambda}(\mb)$ and $l_j^{\lambda}(\mb)$ 
have physical meanings similar to the
above cases.

\section{R-reps of $G_0$ in $W_{\rm HF}$}
\label{R-repG0}
First we give some notations and definitions in reference to the group
theory.  We denote an R-rep of a group $G$ as $\check{G}^{\gamma}$
where $\gamma$ labels an R-rep.
Let $d^{\gamma}(g)$ be 
the R-rep matrix of $\check{G}^{\gamma}$ corresponding to $g\in G$
and $W(\check{G}^{\gamma})$ be the representation  space of 
$\check{G}^{\gamma}$ spanned by 
$(l^{\gamma}_1,l^{\gamma}_2,\cdots,l^{\gamma}_n)$ over the real number field:
\begin{align}
W(\check{G}^{\gamma})&=\{l^{\gamma}_1,l^{\gamma}_2,\cdots,
l^{\gamma}_n\}^{}_{\rm R}.
\end{align} 
Then for $g\in G$ 
\begin{align}
g\cdot \l_i^{\gamma}&=\sum_{i'=1}^nd^{\gamma}(g)_{i'i}l^{\gamma}_{i'}.
\end{align}
Using  real numbers 
$x_i(i=1,2,\cdots,n)$, a vector $\vb\in W(\check{G}^{\gamma})$ 
is written as follows
\begin{align}
\vb&=\sum _{i=1}^nx_il^{\gamma}_i.
\end{align}
The {\it isotropy subgroup} $\Sigma(\vb)$ of $\vb\in W(\check{G}^{\gamma})$ is 
\begin{align}
\Sigma(\vb)&=\{g\in G\mid g\cdot \vb=\vb\}.
\end{align}
Let $H\subset G_0$ be a subgroup of $G_0$. The {\it fixed-point subspace} 
of $H$ in $W(\check{G}^{\gamma})$
is 
\begin{align}
{\rm Fix}^{\gamma}(H)&=\{\vb\in W(\check{G}^{\gamma})\mid g\cdot \vb=
\vb\ {\rm for \ all}\ g\in H\}.
\end{align}
An isotropy subgroup with one-dimensional 
fixed-point subspace is  called an {\it axial 
isotropy subgroup}\cite{Golubitsky3}. 

According to group theoretical analysis 
of the HF  equation
\cite{ozaki-inst,ozakim,ozaki2,Nakanishi}, 
instabilities of a HF solution (characterized by HF Hamiltonian $H_m$) is 
labeled by 
R-reps of the isotropy subgroup of $H_m$. 

Using the {\it equivariant branching lemma}\cite{Golubitsky2}
in the group theoretical bifurcation theory
\cite{Golubitsky1,Golubitsky2,Sattinger}, 
we can show\cite{ozaki-inst, Nakanishi} that 
if an instability of a HF solution characterized by
 an R-rep $\check{G}_0^{\gamma}$ occurs, there always exists
a branch of a HF solution which bifurcates through the instability
and has the axial isotropic subgroup in $W(\check{G}_0^{\gamma})$.
Thus we can enumerate broken symmetry states bifurcating 
from the normal paramagnetic state  by
listing axial isotropy subgroups of each R-rep $\check{G}_0$ of $G_0$.

Since we consider HF solutions with three types of ordring vectors,
$\Gamma$ point: $\Qb_0=(0,0)$, $M$ point: $\Qb_1(\pi,\pi)$ 
and $X$ point$: \{\Qb_2=(\pi,0),\Qb_3=(0,\pi)\}$,
we present  R-reps of $G_0$ in the 
representation space
$W_{\rm HF}$ with ordering vectors
$\Qb_i, (i=0,1,2,3)$. 
An R-rep of $G_0=\Pb\times \SSb\times \Rbb$ in $W_{\rm HF}$
is written as  Kronecker products of R-reps of $\Pb, \SSb$ and $\Rbb$ 
as follows.
\begin{align}
\check{G}_0&=\check{\Pb}\otimes \check{\SSb}\otimes \check{\Rbb}.
\end{align}
Thus 
relevant R-reps of $G_0$ in $W_{\rm HF}$ are written as
\begin{align}
\check{G}_0^{(\Gamma j,\mu,\nu)}&=\check{\Pb}^{(\Gamma j)}\otimes 
\check{\SSb}^{(\mu)}\otimes \check{\Rbb}^{(\nu)},\nonumber\\
\check{G}_0^{(Mj,\mu,\nu)}&=\check{\Pb}^{(Mj)}\otimes 
\check{\SSb}^{(\mu)}\otimes \check{\Rbb}^{(\nu)},\nonumber\\
\check{G}_0^{(X\gamma,\mu,\nu)}&=\check{\Pb}^{(X\gamma)}\otimes 
\check{\SSb}^{(\mu)}\otimes \check{\Rbb}^{(\nu)},
\label{Ghat-P-S-R}
\end{align}
where $\check{\SSb}^{0}$ is the identity representation and 
$\check{\SSb}^{1}$ is 
a three dimensional R-rep of $\SSb$ written as 
\begin{align}
\check{\SSb}^{(1)}(u(\nb,\theta))&=R(u(\nb,\theta)),
\end{align}
where $R(u(\nb,\theta))$ is the three-dimensional rotation matrix 
defined in \eqref{Rotation-matrix},
$\check{\Rbb}^{(0)}$ is the identity representation and 
$\check{\Rbb}^{(1)}$ is a one-dimensional representation such that
\begin{align}
\check{\Rbb}^{(1)}(E)&=1,\ \check{\Rbb}^{(1)}(t)=-1.
\end{align}
$\check{\Pb}^{(\Gamma j)}$($\check{\Pb}^{(Mj)}$) are R-reps of
 $\Pb$ with ordering vector
$\Qb_0=(0,0)$($\Qb_1=(\pi,\pi))$ and $j$ is the label of R-rep of the
little co-group
 $\Db_{4h}$ of $\Qb_0$($\Qb_1$), that is , 
$j={\rm A_{1g}}, {\rm A_{2g}}, {\rm B_{1g}}, {\rm A_{2g}},\\
{\rm E_{g}}$
\footnote{The HF Hamiltonian space $W_{\rm HF}$ is spanned by only gerade(even)
bases.}.
The Rep
$\check{\Pb}^{(X\gamma)}$ are  R-reps of
 $\Pb$ with ordering vector $\Qb_2=(\pi,0)$ and 
 $\gamma$ is the label of the R-rep of the little co-group $\Db_{2h}$ of 
 $\Qb_2$, that is,
 $\gamma={\rm A_{g}},{\rm B_{1g}},{\rm B_{2g}},{\rm B_{3g}}$.
The R-rep matrices of $\check{\Pb}^{(\Gamma j)}$, 
$\check{\Pb}^{(Mj)}$ and $\check{\Pb}^{(X\gamma)}$ are given in 
Table \ref{PSRrepmat}.
\begin{table}[!]
{\footnotesize
\caption{R-rep matrices of 
$\Pb, \SSb$ and $\Rbb$}
\label{PSRrepmat}
$$
\begin{array}{ccclc}
\hline 
{\rm group }& & & {\rm R-rep\ matrix} & \\
\hline
\\[-2mm]
\Pb& \check{\Pb}^{(\Gamma j)}(pT(\mb))&=&D^{(j)}(p)& p\in \Db_{4h} \\[2mm]
& \check{\Pb}^{(Mj)}(pT(\mb))&=&D^{(j)}(p)e^{-i\sQb_1\cdot \mb}& 
p\in \Db_{4h} \\[2mm]
& \check{\Pb}^{(X\gamma)}(pT(\mb))&=&\left(
\begin{array}{cc}
\chi^{(\gamma)}(p)e^{-i\sQb_2\cdot\mb}& 0\\
0&\chi^{(\gamma)}(C_{2a}pC_{2a})e^{-i\sQb_3\cdot\mb}
\end{array}
\right)&
p\in \Db_{2h} \\[6mm]
& \check{\Pb}^{(X\gamma)}(C_{2a}pT(\mb))&=&\left(
\begin{array}{cc}
0& \chi^{(\gamma)}(C_{2a}pC_{2a})e^{-i\sQb_3\cdot\mb}\\
\chi^{(\gamma)}(p)e^{-i\sQb_2\cdot\mb}& 0
\end{array}
\right)&
p\in \Db_{2h} \\[4mm]
\hline 
\\[-2mm]
\SSb & \check{\SSb}^{(0)}(u(\nb,\theta))&=&1& 
u(\nb,\theta)\in \SSb\\[2mm]
& \check{\SSb}^{(1)}(u(\nb,\theta))&=&R(u(\nb,\theta))& \\[2mm]
\hline
\\[-2mm]
\Rbb& \check{\Rbb}^{(0)}(t)&=&1 & t\in \Rbb\\
& \check{\Rbb}^{(1)}(t)&=&-1 & \\[2mm]
\hline
\\[-4mm]
\multicolumn{5}{l}{(1)T(\mb)\in \Lb_0=\Lb(\eb_1,\eb_2)}\\
\multicolumn{5}{l}{(2)D^{(j)}= \text{an R-rep matrix of}\ \Db_{4h}}\\
\multicolumn{5}{l}{(3)\chi^{(\gamma)}(p)= \text{an R-rep matrix of}\ \Db_{2h}}\\  
\multicolumn{5}{l}{(4)R(u(\nb,\theta))= \text{the matrix defined in}\ \eqref{Rotation-matrix}}\\
\hline
\end{array}
$$
}
\end{table}

In Table \ref{basis-GamM-mu-nu} and Table \ref{X-mu-nu-base}
we list bases in $W_{\rm HF}$ of $\check{G}_0^{(\Gamma j,\mu,\nu)}$, 
$\check{G}_0^{(M j,\mu,\nu)}$
and $\check{G}_0^{(Xj,\mu,\nu)}$.
\begin{table}[!]
{\footnotesize
\caption{Bases of $\check{G}_0^{(\Gamma j,\mu,\nu)}$ and 
$\check{G}_0^{(M j,\mu,\nu)}$
in $W_{\rm HF}$}
\label{basis-GamM-mu-nu}
$$
\begin{array}{cl}
\hline\\[-4mm]
{\rm R-rep} & {\rm Bases\ in}\ W_{\rm HF} \\
\hline 
\\[-4mm]
\check{G}_0^{(\Gamma A_{1g},\mu,\mu)}& 
 h(\Gamma A^1_{1g},\mu,\mu)_{1,n(\lambda),1}
=\sum_{\skb}\sum_{ss'}(a^{\dag}_{1\skb s}a^{}_{1\skb s'}+
a^{\dag}_{2\skb s}a^{}_{2\skb s'})\sigma^{\lambda}_{ss'}\\[2mm]
&h(\Gamma A^2_{1g},\mu,\mu)_{1,n(\lambda),1}
=\sum_{\skb}\sum_{ss'}a^{\dag}_{3\skb s}a^{}_{3\skb s'}\sigma^{\lambda}_{ss'}
\\[2mm]
\hline\\[-3mm]
\check{G}_0^{(\Gamma A_{2g},\mu,1-\mu)}& 
h(\Gamma A_{2g},\mu,1-\mu)_{1,n(\lambda),1}
=\sum_{\skb}\sum_{ss'}(ia^{\dag}_{1\skb s}a^{}_{2\skb s'}-
ia^{\dag}_{2\skb s}a^{}_{1\skb s'})\sigma^{\lambda}_{ss'}\\[2mm]
\hline \\[-3mm]
\check{G}_0^{(\Gamma B_{1g},\mu,\mu)}& 
h(\Gamma B_{1g},\mu,\mu)_{1,n(\lambda),1}
=\sum_{\skb}\sum_{ss'}(a^{\dag}_{1\skb s}a^{}_{1\skb s'}
-a^{\dag}_{2\skb s}a^{}_{2\skb s'})\sigma^{\lambda}_{ss'}\\[2mm]
\hline \\[-3mm]
\check{G}_0^{(\Gamma B_{2g},\mu,\mu)}& 
h(\Gamma B_{2g},\mu,\mu)_{1,n(\lambda),1}
=\sum_{\skb}\sum_{ss'}(a^{\dag}_{1\skb s}a^{}_{2\skb s'}+
a^{\dag}_{2\skb s}a^{}_{1\skb s'})\sigma^{\lambda}_{ss'}\\[2mm]
\hline \\[-3mm]
\check{G}_0^{(\Gamma E_{g},\mu,\mu)}&\left\{
\begin{array}{l}
h(\Gamma E_{g},\mu,\mu)_{1,n(\lambda),1}=\sum_{\skb}\sum_{ss'}
(a^{\dag}_{2\skb s}a^{}_{3\skb s'}+
a^{\dag}_{3\skb s}a^{}_{2\skb s'})\sigma^{\lambda}_{ss'}\\
h(\Gamma E_{g},\mu,\mu)_{2,n(\lambda),1}=\sum_{\skb}\sum_{ss'}
(a^{\dag}_{1\skb s}a^{}_{3\skb s'}+
a^{\dag}_{3\skb s}a^{}_{1\skb s'})\sigma^{\lambda}_{ss'}
\end{array}
\right.\\[5mm]
\hline \\[-3mm]
\check{G}_0^{(\Gamma E_{g},\mu,1-\mu)}&\left\{
\begin{array}{l}
h(\Gamma E_{g},\mu,1-\mu)_{1,n(\lambda),1}=\sum_{\skb}\sum_{ss'}
i(a^{\dag}_{2\skb s}a^{}_{3\skb s'}-
a^{\dag}_{3\skb s}a^{}_{2\skb s'})\sigma^{\lambda}_{ss'}\\
h(\Gamma E_{g},\mu,1-\mu)_{2,n(\lambda),1}=\sum_{\skb}\sum_{s'}
i(a^{\dag}_{1\skb s}a^{}_{3\skb s'}-
a^{\dag}_{3\skb s}a^{}_{1\skb s'})\sigma^{\lambda}_{ss'}
\end{array}
\right.\\[5mm]
\hline
\hline
\\[-3mm]
\check{G}_0^{(M A_{1g},\mu,\mu)}& 
 h(M A^1_{1g},\mu,\mu)_{1,n(\lambda),1}
=\sum_{\skb}\sum_{ss'}(a^{\dag}_{1(\skb+\sQb_1 s) }a^{}_{1\skb s'}+
a^{\dag}_{2(\skb+\sQb_1 s) }a^{}_{2\skb s'})
\sigma^{\lambda}_{ss'}\\[2mm]
&h(MA^2_{1g},\mu,\mu)_{1,n(\lambda),1}
=\sum_{\skb}\sum_{ss'}a^{\dag}_{3(\skb+\sQb_1 s) }a^{}_{3\skb s'}
\sigma^{\lambda}_{ss'}\\[2mm]
\hline\\[-3mm]
\check{G}_0^{(M A_{2g},\mu,1-\mu)}& 
h(MA_{2g},\mu,1-\mu)_{1,n(\lambda),1}
=\sum_{\skb}\sum_{ss'}i(a^{\dag}_{1(\skb+\sQb_1 s) }
a^{}_{2\skb s'}-
a^{\dag}_{2(\skb+\sQb_1 s) }a^{}_{1\skb s'})
\sigma^{\lambda}_{ss'}\\[2mm]
\hline \\[-3mm]
\check{G}_0^{(M B_{1g},\mu,\mu)}& h(M B_{1g},\mu,\mu)_{1,n(\lambda),1}
=\sum_{\skb}\sum_{ss'}(a^{\dag}_{1(\skb+\sQb_1 s) }a^{}_{1\skb s'}
-a^{\dag}_{2(\skb+\sQb_1 s) }a^{}_{2\skb s'})
\sigma^{\lambda}_{ss'}\\[2mm]
\hline \\[-3mm]
\check{G}_0^{(M B_{2g},\mu,\mu)}& h(M B_{2g},\mu,\mu)_{1,n(\lambda),1}
=\sum_{\skb}\sum_{s s'}(a^{\dag}_{1(\skb+\sQb_1 s) }
a^{}_{2\skb s'}+
a^{\dag}_{2(\skb+\sQb_1 s) }a^{}_{1\skb s'})
\sigma^{\lambda}_{ss'}\\[2mm]
\hline \\[-3mm]
\check{G}_0^{(M E_{g},\mu,\mu)}&\left\{
\begin{array}{l}
h(M E_{g},\mu,\mu)_{1,n(\lambda),1}=\sum_{\skb}\sum_{ss'}
(a^{\dag}_{2(\skb+\sQb_1 s) }a^{}_{3\skb s'}+
a^{\dag}_{3(\skb+\sQb_1 s) }a^{}_{2\skb s'})
\sigma^{\lambda}_{ss'}\\
h(M E_{g},\mu,\mu)_{2,n(\lambda),1}=\sum_{\skb}\sum_{ss'}
(a^{\dag}_{1(\skb+\sQb_1 s) }a^{}_{3\skb s'}+
a^{\dag}_{3(\skb+\sQb_1 s) }a^{}_{1\skb s'})
\sigma^{\lambda}_{ss'}
\end{array}
\right.\\[5mm]
\hline \\[-3mm]
\check{G}_0^{(M E_{g},\mu,1-\mu)}&\left\{
\begin{array}{l}
h(M E_{g},\mu,1-\mu)_{1,n(\lambda),1}=\sum_{\skb}\sum_{ss'}
i(a^{\dag}_{2(\skb+\sQb_1 s) }a^{}_{3\skb s'}-
a^{\dag}_{3(\skb+\sQb_1 s) }a^{}_{2\skb s'})
\sigma^{\lambda}_{ss'}\\
h(M E_{g},\mu,1-\mu)_{2,n(\lambda),1}=\sum_{\skb}\sum_{ss'}
i(a^{\dag}_{1(\skb+\sQb_1 s) }a^{}_{3\skb s'}-
a^{\dag}_{3(\skb+\sQb_1) }a^{}_{1\skb s'})
\sigma^{\lambda}_{ss'}
\end{array}
\right.\\[5mm]
\hline \hline 
\multicolumn{2}{l}{\lambda=0\hspace{2mm} {\rm for }\ \mu=0,\hspace{2mm} 
{\rm and} \hspace{2mm}\lambda=1,2,3\hspace{2mm}
{\rm for }\hspace{2mm} \mu=1.}\\
\multicolumn{2}{l}{n(\lambda)=1\hspace{2mm}{\rm for }\ \lambda=0,\hspace{2mm} 
{\rm and} \hspace{2mm} n(\lambda)=\lambda \hspace{2mm}{\rm for}\hspace{2mm}
\lambda=1,2,3.}\\
\hline 
\hline
\end{array}
$$
}
\end{table}
\begin{table}[!]
{\footnotesize
\caption{Bases of $\check{G}_0^{(X\gamma,\mu,\nu)}$ in $W_{\rm HF}$}
\label{X-mu-nu-base}
$$
\begin{array}{ll}
\hline 
R-rep\ \ & {\rm bases\ in}\ W_{\rm HF}\\
\hline
\\[-3mm]
\check{G}_0^{({ X}A_{g},\mu,\mu)}&\left\{
\begin{array}{l}
h(XA^1_{g},\mu,\mu)_{1,n(\lambda),1}=
 \sum_{\skb}\sum _{ss'}a^{\dag}_{1(\skb+\sQb_2)  s }
a^{}_{1\skb  s'}\sigma^{\lambda}_{ss'}\\[5mm]
h(XA^1_{g},\mu,\mu)_{2,n(\lambda),1}
=\sum_{\skb}\sum _{ss'}a^{\dag}_{2(\skb+\sQb_3)  s }
a^{}_{2\skb s'}\sigma^{\lambda}_{ss'}
\end{array}
\right.\\[8mm]
&\left\{
\begin{array}{l}
h(XA^2_{g},\mu,\mu)_{1,n(\lambda),1}=
\sum_{\skb}\sum _{ss'}a^{\dag}_{2(\skb+\sQb_2)  s }
a^{}_{2\skb  s'}\sigma^{\lambda}_{ss'}\\[5mm]
h(XA^2_{g},\mu,\mu)_{2,n(\lambda),1}
=\sum_{\skb}\sum _{ss'}a^{\dag}_{1(\skb+\sQb_3) s }
a^{}_{1\skb  s'}\sigma^{\lambda}_{ss'}
\end{array}
\right.\\[8mm]
&\left\{
\begin{array}{l}
h(XA^3_{g},\mu,\mu)_{1,n(\lambda),1}=
\sum_{\skb}\sum _{ss'}a^{\dag}_{3(\skb+\sQb_2)  s }
a^{}_{3\skb  s'}\sigma^{\lambda}_{ss'}\\[5mm]
h(XA^3_{g},\mu,\mu)_{2,n(\lambda),1}
=\sum_{\skb}\sum _{ss'}a^{\dag}_{3(\skb+\sQb_3)  s }
a^{}_{3\skb  s'}\sigma^{\lambda}_{ss'}
\end{array}
\right.\\[8mm]
\hline
\\[-3mm]
\check{G}_0^{(XB_{1g},\mu,\mu)}&\left\{
\begin{array}{l}
h({ X}B_{1g},\mu,\mu)_{1,n(\lambda),1}=
\sum_{\skb}\sum _{ss'}\big(a^{\dag}_{1(\skb+\sQb_2)  s }
a^{}_{2\skb  s'}+a^{\dag}_{2(\skb+\sQb_2)  s }
a^{}_{1\skb  s'}\big)\sigma^{\lambda}_{ss'}
\\[5mm]
h({X}B_{1g},\mu,\mu)_{2,n(\lambda),1}=
\sum_{\skb}\sum _{ss'}\big(a^{\dag}_{2(\skb+\sQb_3)  s }
a^{}_{1\skb  s'}+a^{\dag}_{1(\skb+\sQb_3)  s }
a^{}_{2\skb  s'}\big)\sigma^{\lambda}_{ss'}
\end{array}
\right.\\[8mm]
\hline
\\[-3mm]
\check{G}_0^{({\rm X}B_{1g},\mu,1-\mu)}&\left\{
\begin{array}{l}
h({\rm X}B_{1g},\mu,1-\mu)_{1,n(\lambda),1}=
 \sum_{\skb}\sum _{ss'}i\big(a^{\dag}_{1(\skb+\sQb_2)  s }
a^{}_{2\skb  s'}-a^{\dag}_{2(\skb+\sQb_2)  s }
a^{}_{1\skb  s'}\big)\sigma^{\lambda}_{ss'}
\\[5mm]
h({\rm X}B_{1g},\mu,1-\mu)_{2,n(\lambda),1}=
\sum_{\skb}\sum _{ss'}i\big(a^{\dag}_{2(\skb+\sQb_3)  s }
a^{}_{1\skb  s'}-a^{\dag}_{1(\skb+\sQb_3)  s }
a^{}_{2\skb  s'}\big)\sigma^{\lambda}_{ss'}
\end{array}
\right.\\[8mm]
\hline
\\[-3mm]
\check{G}_0^{({\rm X}B_{2g},\mu,\mu)}&\left\{
\begin{array}{l}
h({\rm X}B_{2g},\mu,\mu)_{1,n(\lambda),1}=
\sum_{\skb}\sum _{ss'}\big(a^{\dag}_{1(\skb+\sQb_2)  s }
a^{}_{3\skb  s'}+a^{\dag}_{3(\skb+\sQb_2)  s }
a^{}_{1\skb  s'}\big)\sigma^{\lambda}_{ss'}
\\[5mm]
h({\rm X}B_{2g},\mu,\mu)_{2,n(\lambda),1}=
- \sum_{\skb}\sum _{ss'}\big(a^{\dag}_{2(\skb+\sQb_3)  s }
a^{}_{3\skb  s'}+a^{\dag}_{3(\skb+\sQb_3)  s }
a^{}_{2\skb  s'}\big)\sigma^{\lambda}_{ss'}
\end{array}
\right.\\[8mm]
\hline
\\[-3mm]
\check{G}_0^{({\rm X}B_{2g},\mu,1-\mu)}&\left\{
\begin{array}{l}
h({\rm X}B_{2g},\mu,1-\mu)_{1,n(\lambda),1}=
\sum_{\skb}\sum _{ss'}i\big(a^{\dag}_{1(\skb+\sQb_2)  s }
a^{}_{3\skb  s'}-a^{\dag}_{3(\skb+\sQb_2)  s }
a^{}_{1\skb  s}\big)\sigma^{\lambda}_{ss'}
\\[5mm]
h({\rm X}B_{2g},\mu,1-\mu)_{2,n(\lambda),1}=
-\sum_{\skb}\sum _{ss'}i\big(a^{\dag}_{2(\skb+\sQb_3)  s }
a^{}_{3\skb  s'}-a^{\dag}_{3(\skb+\sQb_3)  s }
a^{}_{2\skb  s'}\big)\sigma^{\lambda}_{ss'}
\end{array}
\right.\\[8mm]
\hline
\\[-3mm]
\check{G}_0^{({\rm X}B_{3g},mu,\mu)}&\left\{
\begin{array}{l}
h({\rm X}B_{3g},\mu,\mu)_{1,n(\lambda),1}=
\sum_{\skb}\sum _{ss'}\big(a^{\dag}_{2(\skb+\sQb_2) s }
a^{}_{3\skb  s}+a^{\dag}_{3(\skb+\sQb_2)  s }
a^{}_{2\skb  s'}\big)\sigma^{\lambda}_{ss'}
\\[5mm]
h({\rm X}B_{3g},\mu,\mu)_{2,n(\lambda),1}=
-\sum_{\skb}\sum _{ss'}\big(a^{\dag}_{1(\skb+\sQb_3)  s }
a^{}_{3\skb s'}+a^{\dag}_{3(\skb+\sQb_3  s }
a^{}_{1\skb  s'}\big)\sigma^{\lambda}_{ss'}
\end{array}
\right.\\[8mm]
\hline
\\[-3mm]
\check{G}_0^{({\rm X}B_{3g},\mu,1-\mu)}&\left\{
\begin{array}{l}
h({\rm X}B_{3g},\mu,1-\mu)_{1,n(\lambda),1}=
\sum_{\skb}\sum _{ss'}i\big(a^{\dag}_{2(\skb+\sQb_2)  s }
a^{}_{3\skb  s'}-a^{\dag}_{3(\skb+\sQb_2)  s }
a^{}_{2\skb  s'}\big)\sigma^{\lambda}_{ss'}
\\[5mm]
h({\rm X}B_{3g},\mu,1-\mu)_{2,n(\lambda),1}=
- \sum_{\skb}\sum _{ss'}i\big(a^{\dag}_{1(\skb+\sQb_3)  s }
a^{}_{3\skb  s'}-a^{\dag}_{3(\skb +\sQb_3 s }
a^{}_{1\skb  s'}\big)\sigma^{\lambda}_{ss'}
\end{array}
\right.\\[8mm]
\hline
\hline
\multicolumn{2}{l}{\lambda=0\hspace{2mm} {\rm for }\ \mu=0,\hspace{2mm} 
{\rm and} \hspace{2mm}\lambda=1,2,3\hspace{2mm}
{\rm for }\hspace{2mm} \mu=1}\\
\multicolumn{2}{l}{n(\lambda)=1\hspace{2mm}{\rm for }\ \lambda=0,\hspace{2mm} 
{\rm and} \hspace{2mm} n(\lambda)=\lambda \hspace{2mm}{\rm for}\hspace{2mm}
\lambda=1,2,3.}\\
\hline
\hline
\end{array}
$$
}
\end{table}
Then bases $h(\Lambda,\mu,\nu)_{m,\lambda,1}$ of 
$\check{\Pb}^{(\Lambda)}\otimes \check{\SSb}^{(\mu)}\otimes 
\check{\Rbb}^{(\nu)}$ transform as follows for 
$p\in \Db_{4h}, T(\mb)\in \Lb_0, u(\nb,\theta)\in \SSb$ 
and $r\in \Rbb$
\begin{align}
p&T(\mb)u(\nb,\theta)r\cdot h(\Lambda,\mu,\nu)_{m,\lambda,1}\nonumber\\
&={\displaystyle \sum_{m'=1}^{\mid \Lambda \mid}
\sum_{\lambda'=1}^{\mid \mu \mid }}
\check{\Pb}^{(\Lambda)}_{m'm}\left(pT(\mb)\right)
\check{\SSb}^{(\mu)}_{\lambda'\lambda}
(u(\nb,\theta))\check{\Rbb}^{(\nu)}(r)
h(\Lambda,\mu,\nu)_{m',\lambda',1},
\end{align}
where $\mid \Lambda \mid $ denotes the dimension of a R-rep $\Lambda $.

In the following sections we show that R-reps 
$\check{G}_0^{(\Lambda,0,0)}(\lambda\neq \Gamma A_{1g},MA_{1g},XA_g)$
derive states with quadrupole moment,
 R-reps $\check{G}_0^{(\Lambda,0,1)}$, in which the 
 time reversal symmetry is broken, 
 derive states with orbital 
 angular momentum,
 R-reps  $\check{G}_0^{(\Lambda,1,1)}$(
 $\Lambda\neq \Gamma A_{1g},MA_{1g},XA_g)$
 derive states with 
 spin quadrupole 
 moment, and R-reps 
$\check{G}_0^{(\Lambda,1,0)}$,  which hold the time reversal symmetry,
derive states with 
spin orbital angular momentum.

\section{Symmetry classes of non magnetic orbital ordered states.}
In this section we consider broken symmetry states derived from 
R-reps $\check{G}_0^{(\Gamma j,0,\nu)}$, $\check{G}_0^{(M j,0,\nu)}$ and 
$\check{G}_0^{(X\gamma,0,\nu)}$, which hold spin rotation symmetry $\SSb$.
In order to list non magnetic ordered states with ordering vectors 
$\Qb_l\ (l=0,1,2,3)$ bifurcating from the normal state,
we present the axial isotropy subgroups of 
the R-reps $\check{G}_0^{(\Gamma j,0,\nu)},\check{G}_0^{(M j,0,\nu)}$ and 
$\check{G}_0^{(X\gamma,0,\nu)}$.
The axial isotropy subgroups of R-reps $\check{G}_0^{(\Gamma j,0,\nu)}$ and 
$\check{G}_0^{(M j,0,\nu)}$ are listed in Table \ref{Isotro-GMj00}.
The axial isotropy subgroups of $\check{G}_0^{(X j,0,\nu)}$ 
are listed in Table \ref{Isotro-Xj0nu}.
\begin{table}[!]
\caption{Axial isotropy subgroup and its Fixed point subspace of 
$\check{G}_0^{(\Gamma j,0,\nu)}$ and $\check{G}_0^{(M j,0,\nu)}$ }
\label{Isotro-GMj00}{\footnotesize
\begin{tabular}{lll}
\hline
\\[-3mm]
{\normalsize R-rep} & {\normalsize axial isotropy subgroup} & {\normalsize Fixed point subspace}
\\[3mm] \hline \\[-3mm]
$\check{G}_0^{(\Gamma A_{1g},0,0)}$ &
$G(\Gamma A_{1g},0,0)=\Db_{4h}\Lb_0\SSb\Rbb$ &
$\{h_{1,1,1}\}_{\mathrm{R}}$
\\[2mm] \hline\\[-4mm]
$\check{G}_0^{(\Gamma A_{2g},0,1)}$ & 
$G(\Gamma A_{2g},0,1)=M_x\Cb_{4h}\Lb_0\SSb$ &
$\{h_{1,1,1}\}_{\mathrm{R}}$
\\[2mm]\hline\\[-4mm]
$\check{G}_0^{(\Gamma B_{1g},0,0)}$ &
$G(\Gamma B_{1g},0,0)=\Db_{2h}\Lb_0\SSb\Rbb$ &
$\{h_{1,1,1}\}_{\mathrm{R}}$
\\[2mm]\hline\\[-4mm]
$\check{G}_0^{(\Gamma B_{2g},0,0)}$ &
$G(\Gamma B_{2g},0,0)=\Db_{2ah}\Lb_0\SSb\Rbb$ &
$\{h_{1,1,1}\}_{\mathrm{R}}$
\\[2mm]\hline\\[-4mm]
$\check{G}_0^{(\Gamma E_{g},0,0)}$ &
$G(\Gamma E_{g},0,0)_1=\Cb_{2xh}\Lb_0\SSb\Rbb$ &
$\{h_{1,1,1}\}_{\mathrm{R}}$
\\[2mm]
& $G(\Gamma E_{g},0,0)_2=\Cb_{2ah}\Lb_0\SSb\Rbb$ &
$\{h_{1,1,1}-h_{2,1,1}\}_{\mathrm{R}}$
\\[2mm]\hline\\[-4mm]
$\check{G}_0^{(\Gamma E_{g},0,1)}$ &
$G(\Gamma E_{g},0,1)_1=M_z\Cb_{2xh}\Lb_0\SSb$ &
$\{h_{1,1,1}\}_{\mathrm{R}}$
\\[2mm]
& $G(\Gamma E_{g},0,1)_2=M_z\Cb_{2ah}\Lb_0\SSb\Rbb$ &
$\{h_{1,1,1}-h_{2,1,1}\}_{\mathrm{R}}$
\\[2mm] \hline \hline\\[-4mm]
$\check{G}_0^{(M A_{1g},0,0)}$ &
$G( MA_{1g},0,0)=\Db_{4h}\Lb_1\SSb\Rbb$ &
$\{h_{1,1,1}\}_{\mathrm{R}}$
\\[2mm]\hline\\[-4mm]
$\check{G}_0^{(M A_{2g},0,1)}$ &
$G(M A_{2g},0,1)=M_xT_x(\eb_1)\Cb_{4h}\Lb_1\SSb$ &
$\{h_{1,1,1}\}_{\mathrm{R}}$
\\[2mm]\hline\\[-4mm]
$\check{G}_0^{(M B_{1g},0,0)}$ &
$G(M B_{1g},0,0)=T_a(\eb_1)\Db_{2h}\Lb_1\SSb\Rbb$ &
$\{h_{1,1,1}\}_{\mathrm{R}}$
\\[2mm]\hline\\[-4mm]
$\check{G}_0^{(M B_{2g},0,0)}$ &
$G(M B_{2g},0,0)=T_x(\eb_1)\Db_{2ah}\Lb_1\SSb\Rbb$ &
$\{h_{1,1,1}\}_{\mathrm{R}}$
\\[2mm]\hline\\[-4mm]
$\check{G}_0^{(M E_{g},0,0)}$ &
$G(M E_{g},0,0)_1=T_z(\eb_1)\Cb_{2xh}\Lb_1\SSb\Rbb$ &
$\{h_{1,1,1}\}_{\mathrm{R}}$%
\\[2mm]
& $G(M E_{g},0,0)_2=T_z(\eb_1)\Cb_{2ah}\Lb_1\SSb\Rbb$ &
$\{h_{1,1,1}-h_{2,1,1}\}_{\mathrm{R}}$
\\[2mm]\hline\\[-4mm]
$\check{G}_0^{(M E_{g},0,1)}$ &
$G(M E_{g},0,1)_1=M_zT_z(\eb_1)\Cb_{2xh}\Lb_1\SSb$ & 
$\{h_{1,1,1}\}_{\mathrm{R}}$
\\[2mm]
& $G(M E_{g},0,1)_2=M_zT_z(\eb_1)\Cb_{2ah}\Lb_1\SSb$ &
$\{h_{1,1,1}-h_{2,1,1}\}_{\mathrm{R}}$
\\[2mm]\hline\\[-4mm]
\multicolumn{3}{l}{$\Lb_1=\Lb(\eb_1+\eb_2,\eb_2-\eb_1)$}\\[2mm]
\multicolumn{3}{l}{$T_i(\eb_1)=\{E,C_{2i}T(\eb_1)\}$}\\[2mm]
\multicolumn{3}{l}{$\Db_{2ah}=\{E,C_{2z},C_{2a},C_{2b},I,
IC_{2z},IC_{2a},IC_{2b}\}$}\\[2mm]
\multicolumn{3}{l}{$\Cb_{2xh}=\{E,C_{2x},I,
IC_{2x}\}$}\\[2mm]
\multicolumn{3}{l}{$\Cb_{2ah}=\{E,C_{2a},I,
IC_{2a}\}$}\\[2mm]
\multicolumn{3}{l}{$M_i=\{E,tC_{2i}\}$}\\[2mm]
\hline \hline
\end{tabular}}
\end{table}

All isotropy subgroups in Table \ref{Isotro-GMj00} 
and \ref{Isotro-Xj0nu} contain $\SSb$. Thus 
from $\SSb$ invariance of $\Rb^{l\mu}$, only 
$\Rb^{l0}$ are non-zero. 
\begin{table}[!]
\caption{Axial isotropy subgroup and its fixed point subspace of
$\check{G}_0^{(\mathrm{X}\gamma ,0,\nu)}$}
\label{Isotro-Xj0nu}{\footnotesize
\begin{tabular}{lll}
\hline \\[-3mm]
{\normalsize R-rep} & {\footnotesize axial isotropy subgroup} & {\normalsize fixed point subspace}
\\[3mm]
\hline \\[-3mm]
$\check{G}_{0}^{(\mathrm{X} A_{g},0,0)}$ &
$G(\mathrm{X} A_{g},0,0)_1=\Db_{2h}\Lb_2 \SSb \Rbb$ &
$\{h_{1,1,1}\}^{}_{\mathrm{R}}$
\\[3mm]
& $G(\mathrm{X} A_{g},0,0)_2=\Db_{4h}\Lb_3 \SSb \Rbb$ &
$\{h_{1,1,1}+h_{2,1,1}\}^{}_{\mathrm{R}}$
\\[3mm]\hline\\[-3mm]
$\check{G}_{0}^{(\mathrm{X} B_{1g},0,0)}$ &
$G(\mathrm{X} B_{1g},0,0)_1=T_x(\eb_1)\Cb_{2zh}\Lb_2 \SSb \Rbb$ &
$\{h_{1,1,1}\}^{}_{\mathrm{R}}$
\\[3mm]
& $G(\mathrm{X} B_{1g},0,0)_2=T_x(\eb_1+\eb_2)\Db_{2ah}\Lb_3 \SSb \Rbb$ &
$\{h_{1,1,1}+h_{2,1,1}\}^{}_{\mathrm{R}}$
\\[3mm]\hline\\[-3mm]
$\check{G}_{0}^{(\mathrm{X} B_{1g},0,1)}$ &
$G(\mathrm{X} B_{1g},0,1)_1=M_yT_x(\eb_1)\Cb_{2zh}\Lb_2 \SSb$ &
$\{h_{1,1,1}\}^{}_{\mathrm{R}}$
\\[3mm]
& $G(\mathrm{X} B_{1g},0,1)_2=M_yT_x(\eb_1+\eb_2)\Db_{2ah}\Lb_3 \SSb$ &
$\{h_{1,1,1}+h_{2,1,1}\}^{}_{\mathrm{R}}$
\\[3mm]\hline\\[-3mm]
$\check{G}_{0}^{(\mathrm{X} B_{2g},0,0)}$ &
$G(\mathrm{X} B_{2g},0,0)_1=T_x(\eb_1)\Cb_{2yh}\Lb_2 \SSb \Rbb$ &
$\{h_{1,1,1}\}^{}_{\mathrm{R}}$
\\[3mm]
& $G(\mathrm{X} B_{2g},0,0)_2=T_x(\eb_1)T_y(\eb_2)\Cb_{2ah}\Lb_3 \SSb \Rbb$ &
$\{h_{1,1,1}+h_{2,1,1}\}^{}_{\mathrm{R}}$
\\[3mm] \hline \\[-3mm]
$\check{G}_{0}^{(\mathrm{X} B_{2g},0,1)}$ &
$G(\mathrm{X} B_{2g},0,1)_1=M_zT_x(\eb_1)\Cb_{2yh}\Lb_2 \SSb$ &
$\{h_{1,1,1}\}^{}_\mathrm{R}$
\\[3mm]
& $G(\mathrm{X} B_{2g},0,1)_2=M_zT_x(\eb_1)T_y(\eb_2)\Cb_{2ah}\Lb_3 \SSb$ &
$\{h_{1,1,1}+h_{2,1,1}\}^{}_\mathrm{R}$
\\[3mm]\hline\\[-3mm]
$\check{G}_{0}^{(\mathrm{X} B_{3g},0,0)}$ &
$G(\mathrm{X} B_{3g},0,0)_1=T_y(\eb_1)\Cb_{2xh}\Lb_2 \SSb \Rbb$ &
$\{h_{1,1,1}\}^{}_\mathrm{R}$
\\[3mm]
& $G(\mathrm{X} B_{3g},0,0)_2=T_x(\eb_2)T_y((\eb_1)\Cb_{2ah}\Lb_3 \SSb \Rbb$ &
$\{h_{1,1,1}+h_{2,1,1}\}^{}_\mathrm{R}$
\\[3mm]
\hline\\[-3mm]
$\check{G}_{0}^{(\mathrm{X} B_{3g},0,1)}$ &
$G(\mathrm{X} B_{3g},0,1)_1=M_zT_y(\eb_1)\Cb_{2xh}\Lb_2 \SSb$ &
$\{h_{1,1,1}\}^{}_\mathrm{R}$
\\[3mm]
& $G(\mathrm{X} B_{3g},0,1)_2=M_zT_x(\eb_2)T_y(\eb_1)\Cb_{2ah}\Lb_3 \SSb$ &
$\{h_{1,1,1}+h_{2,1,1}\}^{}_\mathrm{R}$
\\[3mm]
\hline
\\[-3mm]
\multicolumn{3}{l}{$\Lb_2=\Lb(2\eb_1,\eb_2),\ 
\Lb_3=\Lb(2\eb_1,2\eb_2)$}\\[1mm]
\multicolumn{3}{l}{$T_i(\eb_1+\eb_2)=\big\{E,C_{2i}T(\eb_1+\eb_2)\big\}, 
\ i=z,x,y,a,b$}\\[1mm]
\multicolumn{3}{l}{$T_i(\eb_m)=\big\{E,C_{2i}T(\eb_1)\big\},
\quad i=z,x,y,a,b,\quad m=1,2$}\\[1mm]
\multicolumn{3}{l}{$M_i=\big\{E, tC_{2i}\big\},i=z,x,y,a,b$}\\[1mm]
\multicolumn{3}{l}{$\Db_{2ah}=\big\{E,C_{2z},C_{2a},C_{2b},I,
IC_{2z},IC_{2a},IC_{2b}\big\}$}\\[1mm]
\multicolumn{3}{l}{$\Cb_{2xh}=\big\{E,C_{2x},I,IC_{2x}\big\}$}\\[1mm]
\multicolumn{3}{l}{$\Cb_{2ah}=\big\{E,C_{2a},I,IC_{2a}\big\}$}\\[1mm]
\hline
\hline
\end{tabular}}
\end{table}
From \eqref{Dmupdown} the density matrix is given by
\begin{align}
\Db(\mb)&=\Db^{\uparrow \uparrow}(\mb)=\Db^{\downarrow \downarrow}(\mb)
=\sum_{l=0}^3e^{-i\sQb_l\cdot \smb}\Rb^{l0}.
\end{align}
Since the $\Rb^{l0}$ is an Hermitian matrix, $\Db(\mb)$ is diagonalized 
by a unitary matrix $U$ as 
\begin{align}
U^{\dag}\Db(\mb) U &= \lambdab,\nonumber \\
\lambdab&=\left(
\begin{array}{ccc}
\lambda_1 & 0 & 0\\
0 & \lambda_2 & 0\\
0 & 0 & \lambda_3
\end{array}
\right).
\label{diago-lambda}
\end{align}
We define 
\begin{align}
\Ab^{\dag}_{\smb s}&=(a^{\dag}_{\smb 1 s},a^{\dag}_{\smb 2 s},
a^{\dag}_{\smb 3 s}),\nonumber \\
\alphab^{\dag}_{\smb s}&=(\alpha^{\dag}_{\smb 1 s},
\alpha^{\dag}_{\smb 2 s},
\alpha^{\dag}_{\smb 2 s})\equiv \Ab^{\dag}_{\smb s}U.
\end{align}
These mean 
\begin{align}
\alpha^{\dag}_{\smb j s}&=\sum_{j'=1}^3a^{\dag}_{\smb j's}U_{j'j},\nonumber \\
\alpha^{}_{\smb i s}&=\sum_{i'=1}^3a^{}_{\smb j's}U_{j'j}^*.
\end{align}
Thus we obtain 
\begin{align}
\langle \alpha^{\dag}_{\smb js}\alpha^{}_{\smb is}\rangle &=
U^{\dag}_{ii'}\langle a^{\dag}_{\smb j' s}a^{}_{\smb i' s}\rangle  
U_{j'j}=(U^{\dag}\Db U)_{ij}\nonumber \\
&=(\lambdab)_{ij}=\delta_{ij}\lambda_i.
\end{align}
These represent that the occupation numbers of electron for 
three atomic orbitals $(\psi_1,\psi_2,\psi_3)$  are
$\lambda_1,\lambda_2,\lambda_3$, 
where
\begin{align}
\psi_1&=\phi_1U_{11}+\phi_2U_{21}+\phi_3U_{31},\nonumber\\
\psi_2&=\phi_1U_{12}+\phi_2U_{22}+\phi_3U_{32},\nonumber\\
\psi_3&=\phi_1U_{13}+\phi_2U_{23}+\phi_3U_{33}.
\label{psi123U}
\end{align}

\begin{figure}[!]
\begin{center}
\includegraphics[trim=0mm 0mm 0mm 100mm,width=6cm,clip]{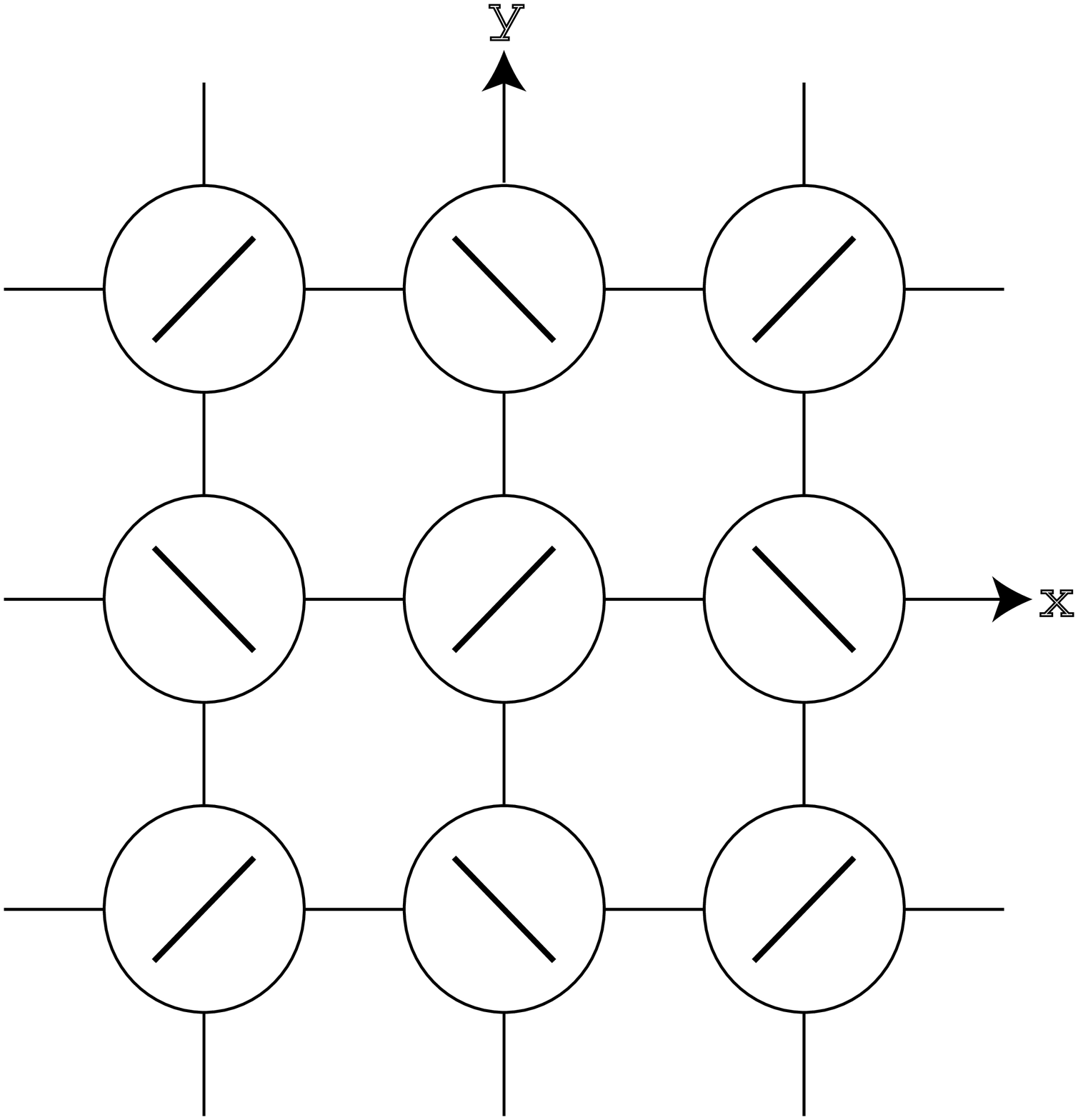}
	\caption{Orbital ordering by 
	$\diagup=\psi_1
	=\frac{1}{\sqrt{2}}(\phi_1+\phi_2)$, and 
	$\diagdown=\psi_2
	=\frac{1}{\sqrt{2}}(\phi_1-\phi_2)$, which have the occupation number 
	$a+c$ in the $G(MB_{2g},0,0)$ state.}
     \label{MB2g-0-0}
	\end{center}
\end{figure}

 We give three examples to show how each axial isotropy subgroup 
 determines the canonical form of the HF Hamiltonian $H_m$, occupied 
 orbitals and their occupation numbers, and the type of the LOP of the state.
\begin{example}
\label{Ex-GA1-0-0}
{\rm 
$G(\Gamma A_{1g},0,0)$ state.\\
In this case the isotropy subgroup $G(H_m)$ of $H_m$ is
\begin{align}
G(\Gamma A_{1g},0,0)&=\Db_{4h}\Lb_0\SSb\Rbb.
\end{align}
From $T(\eb_1)$ and $T(\eb_2)\in \Lb_0$ invariance of $\Rb^{l0}$,
we see that only $\Rb^{00}$ is non-zero.  
$\Rb^{00}$ is invariant under $C_{2z},C_{2x}, C_{2a}\in \Db_{4h}$ 
and $t\in \Rb$. Thus using \eqref{gaction-R} we have   
\begin{align}
C_{2z}\cdot \Rb^{00}&=\left(
\begin{array}{ccc}
R^{00}_{11}&R^{00}_{12}&-R^{00}_{13}\\
R^{00}_{21}&R^{00}_{22}&-R^{00}_{23}\\
-R^{00}_{31}&-R^{00}_{32}&R^{00}_{33}
\end{array}
\right)=\Rb^{00},\nonumber \\[2mm]
C_{2x}\cdot \Rb^{00}&=\left(
\begin{array}{ccc}
R^{00}_{11}&-R^{00}_{12}&-R^{00}_{13}\\
-R^{00}_{21}&R^{00}_{22}&R^{00}_{23}\\
-R^{00}_{31}&R^{00}_{32}&R^{00}_{33}
\end{array}
\right)=\Rb^{00},\nonumber \\[2mm]
C_{2a}\cdot \Rb^{00}&=\left(
\begin{array}{ccc}
R^{00}_{22}&R^{00}_{21}&-R^{00}_{23}\\
R^{00}_{12}&R^{00}_{11}&-R^{00}_{13}\\
-R^{00}_{32}&-R^{00}_{31}&R^{00}_{33}
\end{array}
\right)=\Rb^{00},\nonumber \\[2mm]
t\cdot\Rb^{00}&=(\Rb^{00})^*=\Rb^{00}.
\end{align}
Thus we obtain 
\begin{align}
\Rb^{00}&=\left(
\begin{array}{ccc}
a& 0 & 0\\
0& a & 0\\
0& 0 & b
\end{array}
\right),
\label{Gam-A1-0-0-R}
\end{align}
where $a$ and $b$ are real numbers.
From 
\eqref{WY} and SCF condition \eqref{xlmuconst} we obtain
\begin{align}
x^{00}_{11}&=N(W_{1111}R^{00}_{11}+W_{1212}R^{00}_{22}+W_{1313}R^{00}_{33})
\nonumber\\
&=(U+2U'-J)a+(2U'-J)b\equiv \alpha, \nonumber\\
x^{00}_{22}&=N(W_{2121}R^{00}_{11}+W_{2222}R^{00}_{22}+W_{2323}R^{00}_{33})
\nonumber \\
&=(U+2U'-J)a+(2U'-J)b\equiv \alpha,\nonumber\\
x^{00}_{33}&=N(W_{3131}R^{00}_{11}+W_{3232}R^{00}_{22}+W_{3333}R^{00}_{33})
\nonumber\\
&=2(2U'-J)a+Ub\equiv \beta. 
\end{align}
Thus we obtain HF Hamiltonian $H_m$ as follows:
\begin{align}
H_m&=H_K+\sum_{\skb}\sum_{s}\{\alpha 
(a^{\dag}_{1\skb s}a^{}_{1\skb s}+a^{\dag}_{2\skb s}a^{}_{2\skb s})
+\beta a^{\dag}_{3\skb s}a^{}_{3\skb s}
\}\nonumber\\
&=H_K+\alpha h(\Gamma A_{1g}^1,0,0)_{1,1,1}
+\beta h(\Gamma A_{1g}^2,0,0)_{1,1,1}.
\end{align}
Here the bases $h(\Gamma A_{1g}^j,0,0)_{1,1,1} (j=1,2)$ are  given in
Table \ref{basis-GamM-mu-nu}. This state corresponds to the normal 
paramagnetic  state.
}
\end{example}
Now we consider M point non magnetic states.
All isotropy subgroups of M point non magnetic states contain 
$\Lb_1=\Lb(\eb_1+\eb_2,\eb_2-\eb_1)$ and $\SSb$. Thus from $\SSb$ and $\Lb_1$ 
invariance of 
$\Rb^{l\lambda}$, only $\Rb^{00}$ and $\Rb^{10}$ are non-zero.
From \eqref{Dmupdown}, 
for $\mb$ such that $T(\mb)\in \Lb_1$, we have 
\begin{align}
\Db(\mb)&=\Db^{\uparrow\uparrow}(\mb)=\Db^{\downarrow\downarrow}(\mb)=
\Rb^{00}+\Rb^{10},\nonumber\\
\Db(\mb+\eb_i)&=
\Db^{\uparrow\uparrow}(\mb+\eb_i)=\Db^{\downarrow\downarrow}(\mb+\eb_i)
=\Rb^{00}-\Rb^{10},
\end{align}
where $i=1,2$. Thus
diagonalizing $\Db(\mb)$ and $\Db(\mb+\eb_i),$
we obtain 
occupied atomic orbitals and their occupation numbers at 
sites $\mb$ and $\mb+\eb_i$. 
We consider the $G(MB_{2g},0,0)$ state.
\begin{example}
\label{Ex-MB2-0-0}{\rm 
$G(MB_{2g},0,0)$ state.\\
Since
\begin{align}
G(MB_{2g},0,0)&=T_x(\eb_1)\Db_{2ah}\Lb_1\SSb\Rbb,
\end{align}
$\Rb^{00}$ and $\Rb^{10}$ are invariant under  $C_{2z},C_{2a}\in \Db_{2ah}$ 
and $t\in \Rb$.
Thus using \eqref{gaction-R} we have for $l=0,1$  
\begin{align}
C_{2z}\cdot \Rb^{l0}&=\left(
\begin{array}{ccc}
R^{l0}_{11}&R^{l0}_{12}&-R^{l0}_{13}\\
R^{l0}_{21}&R^{l0}_{22}&-R^{l0}_{23}\\
-R^{l0}_{31}&-R^{l0}_{32}&R^{l0}_{33}
\end{array}
\right)=\Rb^{l0},\nonumber \\[2mm]
C_{2a}\cdot \Rb^{l0}&=\left(
\begin{array}{ccc}
R^{l0}_{22}&R^{l0}_{21}&-R^{l0}_{23}\\
R^{l0}_{12}&R^{l0}_{11}&-R^{l0}_{13}\\
-R^{l0}_{32}&-R^{l0}_{31}&R^{l0}_{33}
\end{array}
\right)=\Rb^{l0},\nonumber \\[2mm]
t\cdot\Rb^{l0}&=(\Rb^{l0})^*=\Rb^{l0}.
\end{align}
Then $\Rb^{l0}$ have  forms 
\begin{align}
\Rb^{l0}&=\left(
\begin{array}{ccc}
R^{l0}_{11} & R^{l0}_{12}& 0\\
R^{l0}_{12} & R^{l0}_{11} & 0\\
0 & 0 & R^{l0}_{33}
\end{array}
\right).
\end{align}
From $C_{2x}T(\eb_1)$ invariance, 
we obtain
\begin{align}
C_{2x}T(\eb_1)\cdot \Rb^{00}&=\left(
\begin{array}{ccc}
R^{00}_{11}& -R^{00}_{12}&0\\
-R^{00}_{21} & R^{00}_{11} & 0\\
0 & 0 & R^{00}_{33}
\end{array}
\right)=\Rb^{00},\nonumber \\
C_{2x}T(\eb_1)\cdot \Rb^{10}&=\left(
\begin{array}{ccc}
-R^{10}_{11}& R^{10}_{12}&0\\
R^{10}_{21} & -R^{10}_{22} & 0\\
0 & 0 & -R^{10}_{33}
\end{array}
\right)=\Rb^{10}.
\end{align}
 Thus we obtain 
\begin{align}
\Rb^{00}&=\left(
\begin{array}{ccc}
a& 0 & 0\\
0& a & 0\\
0& 0 & b
\end{array}
\right),\ \ 
\Rb^{10}=\left(
\begin{array}{ccc}
0& c & 0\\
c& 0 & 0\\
0& 0 & 0
\end{array}
\right),
\label{MB2-0-0-R}
\end{align}
where $a$, $b$ and $c$ are real numbers.
From 
\eqref{WY} and SCF condition \eqref{xlmuconst} 
we obtain
\begin{align}
x^{00}_{11}&=N(W_{1111}R^{00}_{11}+W_{1212}R^{00}_{22}+W_{1313}R^{00}_{33})
\nonumber\\
&=(U+2U'-J)a+(2U'-J)b\equiv \alpha, \nonumber \\
x^{00}_{22}&=N(W_{2121}R^{00}_{11}+W_{2222}R^{00}_{22}+W_{2323}R^{00}_{33})
\nonumber \\
&=(U+2U'-J)a+(2U'-J)b\equiv \alpha,\nonumber \\
x^{00}_{33}&=N(W_{3131}R^{00}_{11}+W_{3232}R^{00}_{22}+W_{3333}R^{00}_{33})
\nonumber \\
&=2(2U'-J)a+Ub\equiv \beta, \nonumber \\
x^{10}_{12}&=N(W_{1122}R^{10}_{21}+W_{1221}R^{10}_{12})\nonumber \\
&=(J'+2J-U')c\equiv \gamma, \nonumber \\
x^{10}_{21}&=N(W_{2211}R^{10}_{12}+W_{2112}R^{10}_{21})\nonumber \\
&=(J'+2J-U')c\equiv \gamma. 
\end{align}
Thus we obtain the HF Hamiltonian $H_m$ as 
\begin{align}
H_m&=H_K+\sum_{\skb}\sum_{s}\{\alpha 
(a^{\dag}_{1\skb s}a^{}_{1\skb s}+a^{\dag}_{2\skb s}a^{}_{2\skb s})
+\beta a^{\dag}_{3\skb s}a^{}_{3\skb s}\nonumber \\
&\ +\gamma(
a^{\dag}_{1(\skb+\sQb_1) s}a^{}_{2\skb s}+a^{\dag}_{2(\skb+\sQb_1) s}
a^{}_{1\skb s})
\}\nonumber \\
&=H_K+\alpha h(\Gamma A^1_{1g},0,0)_{1,1,1}+
\beta h(\Gamma A^2_{1g},0,0)_{1,1,1}+
\gamma h(M B_{2g},0,0)_{1,1,1}.
\label{Hm-MB2g00}
\end{align}
The fourth term of \eqref{Hm-MB2g00} is the primary part which leads 
to the transition to the $G(MB_{2g},0,0)$ state. 
From \eqref{HF-energy} we obtain 
the HF energy $E_\mathrm{HF}$ as
\begin{align}
E_\mathrm{HF} &= \langle H_\mathrm{K} \rangle' \nonumber \\
&\ +N\{2(U+2U'-J)a^2+Ub^2+4(2U'-J)ab
+2(2J+J'-U')c^2\}.
\end{align} 
For $\mb$ such that $T(\mb)\in \Lb_1$ we obtain 
\begin{align}
\Db(\mb)&=\Rb^{00}+\Rb^{10}=\left(
\begin{array}{ccc}
a & c & 0 \\
c & a & 0 \\
0 & 0 & b
\end{array}
\right),\nonumber \\
\Db(\mb+\eb_j)&=\Rb^{00}-\Rb^{10}=\left(
\begin{array}{ccc}
a & -c & 0 \\
-c & a & 0 \\
0 & 0 & b
\end{array}
\right) (j=1,2).
\end{align}
Diagonalization of $\Db(\mb)$ and $\Db(\mb+\eb_j)$ 
are written as 
\begin{align}
U^{\dag}\Db(\mb) U&=\left(
\begin{array}{ccc}
a+c & 0 & 0\\
0 & a-c & 0\\
0 & 0 & b
\end{array}
\right),\nonumber \\[1mm]
U^{\dag}\Db(\mb+\eb_j) U&=\left(
\begin{array}{ccc}
a-c & 0 & 0\\
0 & a+c & 0\\
0 & 0 & b
\end{array}
\right),
\end{align}
where 
\begin{align}
U=\left(
\begin{array}{ccc}
\frac{1}{\sqrt{2}} & \frac{1}{\sqrt{2}} &0\\[2mm]
\frac{1}{\sqrt{2}} & -\frac{1}{\sqrt{2}} &0\\[2mm]
0& 0 & 1
\end{array}
\right).
\end{align}
From \eqref{psi123U}
we obtain the occupied atomic orbitals and their occupation numbers
for $G(MB_{2g},0,0)$ state as shown in Table \ref{MB2g-0-0-occupation}, 
\begin{table}[!]
\caption{Occupied atomic orbitals and their occupation numbers 
for $G(M B_{2g},0,0)$ state}
\label{MB2g-0-0-occupation}
\begin{center}
\begin{tabular}{lllc}
\hline
site & spin & atomic orbital & occupation number\\
\hline\\[-3mm] 
$\mb$ & $\uparrow \downarrow$ & $\psi_1=\frac{1}{\sqrt{2}}(\phi_1+\phi_2)$ & $a+c$
\\[2mm]
$\mb$ & $\uparrow \downarrow$ & $\psi_2=\frac{1}{\sqrt{2}}(\phi_1-\phi_2)$ & $a-c$
\\[2mm]
$\mb$ & $\uparrow \downarrow$ & $\psi_3=\phi_3$ & $b$
\\[2mm]
$\mb+\eb_j$ & $\uparrow \downarrow$ & $\psi_1=\frac{1}{\sqrt{2}}(\phi_1+\phi_2)$ & $a-c$
\\[2mm]
$\mb+\eb_j$ & $\uparrow \downarrow$ & $\psi_2=\frac{1}{\sqrt{2}}(\phi_1-\phi_2)$ & $a+c$
\\[2mm]
$\mb+\eb_j$ & $\uparrow \downarrow$ & $\psi_3=\phi_3$ & $b$
\\[2mm]
\hline\\[-3mm]
\multicolumn{4}{l}{$\mb: T(\mb)\in \Lb_1,\ j=1,2,\ a=R^{00}_{11}=R^{00}_{22},\ b=R^{00}_{33},\ 
c=R^{00}_{12}=R^{00}_{21}$}
\\[2mm]
\hline \hline
\end{tabular}
\end{center}
\end{table}
}
\end{example}
In Fig.\ref{MB2g-0-0} we show the pattern of orbital order by
 $\frac{1}{\sqrt{2}}(\phi_1+\phi_2)$
and $\frac{1}{\sqrt{2}}(\phi_1-\phi_2)$ with the occupation number $a+c$.
Note that this pattern has symmetries of 
$C_{2a},C_{2b},C_{2z},C_{2x}T(\eb_1)$ and $C_{2x}T(\eb_2)\in G(MB_{2g}0,0)$. 
From \eqref{OP-R} we see that this state has alternating quadrupole moments,  
for $\mb$ such that $T(\mb)\in \Lb_1$,
as follows 
\begin{align}
Q_{12}(\mb)&=4I_2c,\  Q_{12}(\mb+\eb_j)=-4I_2c,\ j=1,2.
\end{align}

Finally we consider X point non magnetic state. 
As shown in Table \ref{Isotro-Xj0nu}, 
in cases of $G(X\gamma,0,\nu)_1$, isotropy subgroups contain 
$\Lb_2=\Lb(2\eb_1,\eb_2)$ and $\SSb$.  
Thus from $\Lb_2=\Lb(2\eb_1,\eb_2)$ and  $\SSb$
invariance of $\Rb^{l\lambda}$, only $\Rb^{l0}\ (l=0,2)$ are non-zero.
From \eqref{Dmupdown}, for $\mb$ such that $T(\mb)\in \Lb_2$, we have
\begin{align}
\Db(\mb)&=\Db^{\uparrow\uparrow}(\mb)=\Db^{\downarrow\downarrow}(\mb)
=\Rb^{00}+\Rb^{20},\nonumber \\
\Db(\mb+\eb_1)&=\Db^{\uparrow\uparrow}(\mb+\eb_1)
=\Db^{\downarrow\downarrow}(\mb+\eb_1)
=\Rb^{00}-\Rb^{20}.
\label{XB3-0-1-Dm2}
\end{align}
In the cases of $G(X\gamma,0,\nu)_2$, isotropy subgroups contain 
$\Lb_3=\Lb(2\eb_1,2\eb_2)$ and $\SSb$.  
Thus from $\Lb_3=\Lb(2\eb_1,2\eb_2)$ and  $\SSb$
invariance of $\Rb^{l\lambda}$, only $\Rb^{l0}\ (l=0,1,2,3)$ are non-zero.
From \eqref{Dmupdown}, for $\mb$ such that $T(\mb)\in \Lb_2$, we have
{\small
\begin{align}
\Db(\mb)&=\Db^{\uparrow\uparrow}(\mb)=\Db^{\downarrow\downarrow}(\mb)
=\Rb^{00}+\Rb^{10}+\Rb^{20}+\Rb^{30},\nonumber \\
\Db(\mb+\eb_1)&=\Db^{\uparrow\uparrow}(\mb+\eb_1)
=\Db^{\downarrow\downarrow}(\mb+\eb_1)
=\Rb^{00}-\Rb^{10}-\Rb^{20}+\Rb^{30},\nonumber \\
\Db(\mb+\eb_2)&=\Db^{\uparrow\uparrow}(\mb+\eb_2)
=\Db^{\downarrow\downarrow}(\mb+\eb_2)
=\Rb^{00}-\Rb^{10}+\Rb^{20}-\Rb^{30},\nonumber \\
\Db(\mb+\eb_1+\eb_2)&=\Db^{\uparrow\uparrow}(\mb+\eb_1+\eb_2)
=\Db^{\downarrow\downarrow}(\mb+\eb_1+\eb_2)\nonumber \\
&=\Rb^{00}+\Rb^{10}-\Rb^{20}-\Rb^{{30}}.
\label{XB3-0-1-Dm}
\end{align}
}
We consider the  $G(XB_{3g},0,1)_2$ state which breaks time reversal symmetry
as an example.
\begin{example}{\rm 
\label{Ex-XB3-0-1-2}
{$G(XB_3,0,1)_2$ state.}\\
Since
\begin{align}
G(XB_{3g},0,1)_2&=M_z T_x(\eb_2)T_y(\eb_1)
\Cb_{2ah}\Lb_3\SSb,
\end{align}
${\Rb^{l0}}\ (l=0,1,2,3)$ are invariant under 
$C_{2a},C_{2x}T(\eb_2),C_{2y}T(\eb_1),tC_{2z}$. Thus we obtain 
\begin{align}
\Rb^{00}&=\left(
\begin{array}{ccc}
a & 0 & 0\\
0 & a & 0\\
0 & 0 & b
\end{array}
\right),&
\Rb^{10}&=\left(
\begin{array}{ccc}
0 & c & 0\\
c & 0 & 0\\
0 & 0 & 0
\end{array}
\right),\nonumber \\[2mm]
\Rb^{20}&=\left(
\begin{array}{ccc}
0 & 0 & 0\\
0 & 0 & id\\
0 & -id & 0
\end{array}
\right),& 
\Rb^{30}&=\left(
\begin{array}{ccc}
0 & 0 & -id\\
0 & 0 & 0\\
id & 0& 0
\end{array}
\right),
\end{align}
where $a$, $b$, $c$ and $d$ are real numbers. Using \eqref{xlmuconst} we obtain 
$H_{m}$ as 
\begin{align}
H_m&=H_\mathrm{K}+\sum_{\skb}\sum_{s}\{\alpha(a^{\dag}_{1\skb s} 
a^{}_{1\skb s}+a^{\dag}_{2\skb s} 
a^{}_{2\skb s})+\beta a^{\dag}_{3\skb s} 
a^{}_{3\skb s}\nonumber \\
&\ +\gamma (a^{\dag}_{1(\skb+\sQb_1)s}a^{}_{2\skb s}+
a^{\dag}_{2(\skb+\sQb_1)s}a^{}_{1\skb s})\nonumber \\
&\ i\delta(a^{\dag}_{2(\skb+\sQb_2)s}a^{}_{3\skb s}-
a^{\dag}_{3(\skb+\sQb_2)s}a^{}_{2\skb s}\nonumber \\
&\ -a^{\dag}_{1(\skb+\sQb_3)s}a^{}_{3\skb s}+
a^{\dag}_{3(\skb+\sQb_3)s}a^{}_{1\skb s})\}\nonumber \\
&=H_\mathrm{K}+\alpha h(\Gamma A^1_{1g},0,0)_{1,0,1}+
\beta h(\Gamma A^2_{1g},0,0)_{1,0,1}+\gamma h(M B_{2g},0,0)_{1,0,1}
\nonumber \\
&\ \delta \{ h(XB_{3g},0,1)_{1,0,1}+h(XB_{3g},0,1)_{2,0,1} \},
\label{Hm-XB3g01}
\end{align}
where $\alpha$, $\beta$, $\delta$ and $\gamma$ are determined by SCF conditions
\begin{align}
\alpha &=(U+2U'-J)a+(2U'-J)b,\nonumber \\
\beta&=2(2U'-J)a+Ub,\nonumber \\
\gamma&=(2J+J'-U')c,\nonumber \\
\delta &=(2J-J'-U')d.
\end{align}
The fifth term of \eqref{Hm-XB3g01} is the primary part which leads to the 
transition to the $G(XB_{3g},0,1)_2$ state. The fourth term of 
\eqref{Hm-XB3g01} is the secondary part induced by the transition.

The HF energy is written as 
\begin{align}
E_\mathrm{HF}&=\langle H_\mathrm{K} \rangle' +
\dfrac{N}{2}\{2(U+2U'-J)a^2+Ub^2+4(2U'-J)ab\}\nonumber \\
&\ \ +\dfrac{N}{2}\{2(2J+J'-U')c^2+4(2J-U'-J')d^2\}.
\end{align}
From \eqref{XB3-0-1-Dm} we obtain, for $\mb$ such that $T(\mb)\in \Lb_3$, 
{\small
\begin{align}
\Db(\mb)&=\left(
\begin{array}{ccc}
a & c & -id\\
c & a & id\\
id & -id & b
\end{array}
\right), & 
\Db(\mb+\eb_2)&=\left(
\begin{array}{ccc}
a & -c & id\\
-c & a & id\\
-id & -id & b
\end{array}
\right), \nonumber \\[2mm]
\Db(\mb+\eb_1)&=\left(
\begin{array}{ccc}
a & -c & -id\\
-c & a & -id\\
id & id & b
\end{array}
\right), & 
\Db(\mb+\eb_1+\eb_2)&=\left(
\begin{array}{ccc}
a & c & id\\
c & a & -id\\
-id & id & b
\end{array}
\right).
\end{align}
}
\begin{table}[!]
\caption{Occupied atomic orbitals and their occupation numbers for 
$G(XB_{3g},0,1)_2$ state}
\label{XB3-0-1-2-occ}
\begin{tabular}{lllc}
\hline
site & spin & atomic orbital & occupation number \\
\hline\\[-3mm] 
$\mb$ & $\uparrow \downarrow$ &
$\psi_1=\frac{1}{\sqrt{2}}\mathrm{sgn}(d)u\phi_1-\frac{1}{\sqrt{2}}\mathrm{sgn}(d)u\phi_2+iw\phi_3$ &
$\lambda_1$ \\
$\mb$ & $\uparrow \downarrow$ &
$\psi_2=-\frac{1}{\sqrt{2}}\mathrm{sgn}(d)w\phi_1+\frac{1}{\sqrt{2}}\mathrm{sgn}(d)w\phi_2+iu\phi_3$ &
$\lambda_2$ \\
$\mb$ & $\uparrow \downarrow$ &
$\psi_3=\frac{1}{\sqrt{2}}(\phi_1+\phi_2)$
& $\lambda_3$
\\[2mm]\hline\\[-3mm]
$\mb+\eb_1$ & $\uparrow \downarrow$ &
$\psi_1=-\frac{1}{\sqrt{2}}\mathrm{sgn}(d)u\phi_1-\frac{1}{\sqrt{2}}\mathrm{sgn}(d)u\phi_2-iw\phi_3$ &
$\lambda_1$ \\
$\mb+\eb_1$ & $\uparrow \downarrow$ &
$\psi_2=-\frac{1}{\sqrt{2}}\mathrm{sgn}(d)w\phi_1-\frac{1}{\sqrt{2}}\mathrm{sgn}(d)w\phi_2+iu\phi_3$ &
$\lambda_2$ \\
$\mb+\eb_1$ & $\uparrow \downarrow$ &
$\psi_3=\frac{1}{\sqrt{2}}(\phi_1-\phi_2)$
$\lambda_3$
\\[2mm]\hline\\[-3mm]
$\mb+\eb_2$ & $\uparrow \downarrow$ &
$\psi_1=\frac{1}{\sqrt{2}}\mathrm{sgn}(d)u\phi_1+\frac{1}{\sqrt{2}}\mathrm{sgn}(d)u\phi_2-iw\phi_3$ &
$\lambda_1$ \\
$\mb+\eb_2$ & $\uparrow \downarrow$ &
$\psi_2=\frac{1}{\sqrt{2}}\mathrm{sgn}(d)w\phi_1+\frac{1}{\sqrt{2}}\mathrm{sgn}(d)w\phi_2+iu\phi_3$ &
$\lambda_2$ \\
$\mb+\eb_2$ & $\uparrow \downarrow$ &
$\psi_3=\frac{1}{\sqrt{2}}(\phi_1-\phi_2)$ &
$\lambda_3$
\\[2mm]\hline\\[-3mm]
$\mb+\eb_1+\eb_2$ & $\uparrow \downarrow$ &
$\psi_1=-\frac{1}{\sqrt{2}}\mathrm{sgn}(d)u\phi_1+\frac{1}{\sqrt{2}}\mathrm{sgn}(d)u\phi_2+iw\phi_3$ &
$\lambda_1$ \\
$\mb+\eb_1+\eb_2$ & $\uparrow \downarrow$ &
$\psi_2=\frac{1}{\sqrt{2}}\mathrm{sgn}(d)w\phi_1-\frac{1}{\sqrt{2}}\mathrm{sgn}(d)w\phi_2+iu\phi_3$ &
$\lambda_2$ \\
$\mb+\eb_1+\eb_2$ & $\uparrow \downarrow$ &
$\psi_3=\frac{1}{\sqrt{2}}(\phi_1+\phi_2)$ &
$\lambda_3$
\\[2mm]\hline\\[-3mm]
\multicolumn{4}{l}{$T(\mb)\in \Lb_3=
(2\eb_1,2\eb_2)$}
\\[1mm]
\multicolumn{4}{l}{$a=R^{00}_{11}=R^{00}_{22},\ b=R^{00}_{33},\ 
c=R^{10}_{12}=R^{10}_{21},\  
id=R^{20}_{23}=-R^{20}_{32}=-R^{30}_{13}=R^{30}_{31}$}
\\[1mm]
\multicolumn{4}{l}{$u=\frac{1}{\sqrt{2}}
(1+\frac{a-b-c}{\sqrt{{(a-b-c)^2+8d^2}}})^{\frac{1}{2}},
w=\frac{1}{\sqrt{2}}
(1-\frac{a-b-c}{\sqrt{{(a-b-c)^2+8d^2}}})^{\frac{1}{2}}$}
\\[3mm]
\multicolumn{4}{l}{$\lambda_1=\frac{a+b-c+\sqrt{(a-b-c)^2+8d^2}}{2},
\lambda_2=\frac{a+b-c-\sqrt{(a-b-c)^2+8d^2}}{2},
\lambda_3=a+c$}
\\[1mm]
\hline \hline
\end{tabular}
\end{table}
\begin{figure}[!]
\begin{center}
\includegraphics[trim=0mm 0mm 0mm 0mm,width=7cm,clip]{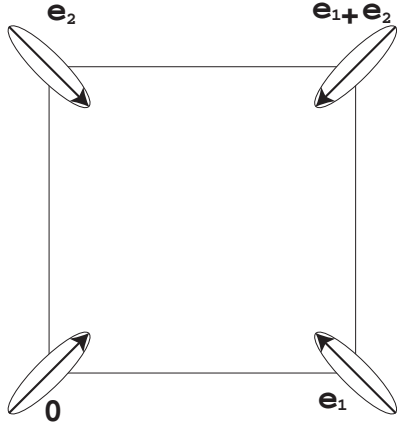}
	\caption{The ordering pattern for quadrupole moments:\hspace{1mm}$Q_{12}$ 
	and 
	orbital angular momenta:\hspace{2cm} $\lb=(l_1,l_2)$ in the
	$G(XB_{3g}0,1)_2$ state. The ovals 
	and arrows represent $Q_{12}$ and orbital angular momentum $\lb$, respectively.} 
	\label{GB3g-0-1-3-fig}
	\end{center}
\end{figure}
Diagonalizing $\Db(\mb),\Db(\mb+\eb_1),\Db(\mb+\eb_2),\Db(\mb+\eb_1+\eb_2)$, 
we obtain occupied atomic orbitals and their occupation numbers at the sites 
$\mb,\mb+\eb_1,\mb+\eb_2,\mb+\eb_1+\eb_2$ for the $G(XB_{3g},0,1)_2$ state 
as shown in Table \ref{XB3-0-1-2-occ}. Note that 
occupied atomic orbitals have complex coefficients. 

From \eqref{OP-R}, we see that this state has the following orbital 
angular momentum(the primary LOP)  and quadrupole 
moment(the secondary LOP). 
 For $\mb$ such that $T(\mb)\in \Lb_3$, 
we obtain
\begin{align}
l_1(\mb)&=4d,\ \ \ \  l_2(\mb)=4d,\nonumber \\
l_1(\mb+\eb_1)&=-4d,\ \ \ l_2(\mb+\eb_1)=4d,\nonumber \\
l_1(\mb+\eb_2)&=4d,\ \ \  \ l_2(\mb+\eb_2)=-4d,\nonumber \\
l_1(\mb+\eb_1+\eb_2)&=-4d,\ \ \ l_2(\mb+\eb_1+\eb_2)=-4d\nonumber \\
Q_{12}(\mb)&=4I_2c,\hspace{7mm} Q_{12}(\mb+\eb_1)=-4I_2c,\nonumber \\
Q_{12}(\mb+\eb_2)&=-4I_2c,\ 
\ Q_{12}(\mb+\eb_1+\eb_2)=4I_2c
\end{align}
The ordering pattern of $Q_{12}$ and $\lb =(l_1,l_2)$ 
for the $G(XB_{3g},0,1)_3$ state is shown in 
Fig. \ref{GB3g-0-1-3-fig}. 
 Note that the onset of quadrupole moment $Q_{12}$ is induced by the transition
to the $G(XB_{3g},0,1)_2$ with the orbital angular momentum.
}
\end{example}
By a similar manner to the case of $G(XB_{3g},0,1)_2$,  
we can see that all states having isotropy subgroups 
such as $G(\Lambda, 0,1)$ or $G(\Lambda, 0,1)_j~(j=1,2)$, which break 
the time reversal symmetry, have 
complex occupied orbitals and orbital angular momentum as LOP.

\section{Symmetry classes of magnetic orbital ordered states}
 In order to list magnetic orbital ordered states with ordering vectors 
$\Qb_l\ (l=0,1,2,3)$, bifurcating from the normal state,
we present the axial isotropy subgroups of 
the R-reps $\check{G}_0^{(\Gamma j,1,\nu)},\check{G}_0^{(M j,1,\nu)}$ and 
$\check{G}_0^{(X\gamma,1,\nu)}$.
The axial isotropy subgroups of R-reps $\check{G}_0^{(\Gamma j,1,\nu)}$ and 
$\check{G}_0^{(M j,1,\nu)}$ are listed in Table \ref{Isotro-GMj1nu}.
\begin{table}[!]
\caption{Axial isotropy subgroup and its Fixed point subspace of 
$\check{G}_0^{(\Gamma j,1,\nu)}$ and $\check{G}_0^{(M j,1,\nu)}$}
\label{Isotro-GMj1nu}
{\footnotesize
\begin{tabular}{lll}
\hline
\\[-3mm]
{\normalsize R-rep} & {\normalsize axial isotropy subgroup} & {\normalsize Fixed point subspace}
\\[3mm] \hline \\[-3mm]
$\check{G}_0^{(\Gamma A_{1g},1,1)}$ &
$G(\Gamma A_{1g},1,1)=M(\eb_2)\Db_{4h}\Lb_0\Ab(\eb_3)$ &
$\{h_{1,1,1}\}_{\mathrm{R}}$
\\[2mm]\hline\\[-4mm]
$\check{G}_0^{(\Gamma A_{2g},1,0)}$ &
$G(\Gamma A_{2g},1,0)=\Db_{4h}(E,u_{2x})\Lb_0\Ab(\eb_3)\Rbb$ &
$\{h_{1,3,1}\}_\mathrm{R}$
\\[2mm]\hline\\[-4mm]
$\check{G}_0^{(\Gamma B_{1g},1,1)}$ &
$G(\Gamma B_{1g},1,1)=M(\eb_2)\Db_{4h}(u_{2x},E)\Lb_0\Ab(\eb_3)$ &
$\{h_{1,3,1}\}_\mathrm{R}$
\\[2mm]\hline\\[-4mm]
$\check{G}_0^{(\Gamma B_{2g},1,1)}$ &
$G(\Gamma B_{2g},1,1)=M(\eb_2)\Db_{4h}(u_{2x},u_{2x})\Lb_0\Ab(\eb_3)$ &
$\{h_{1,3,1}\}_\mathrm{R}$
\\[2mm]\hline\\[-4mm]
$\check{G}_0^{(\Gamma E_{g},1,1)}$ &
$G(\Gamma E_{g},1,1)_1=M(\eb_2)\Db_{2h}(u_{2x},E)\Lb_0\Ab(\eb_3)$ &
$\{h_{1,3,1}\}_\mathrm{R}$\\[2mm]
& $G(\Gamma E_{g},1,1)_2=M(\eb_2)\Db_{2ah}(u_{2x},E)\Lb_0\Ab(\eb_3)$ &
$\{h_{1,3,1}+h_{2,3,1}\}_\mathrm{R}$
\\[2mm]
& $G(\Gamma E_{g},1,1)_3=M(\eb_3)\Db_{4h}(u_{4z}^+,u_{2x})\Lb_0$ &
$\{h_{1,1,1}+h_{2,2,1}\}_\mathrm{R}$
\\[2mm]\hline\\[-4mm]
$\check{G}_0^{(\Gamma E_{g},1,0)}$ &
$G(\Gamma E_{g},1,0)_1=\Db_{2h}(u_{2x},E)\Lb_0\Ab(\eb_3)\Rbb$ &
$\{h_{1,3,1}\}_\mathrm{R}$
\\[2mm]
& $G(\Gamma E_{g},1,0)_2=\Db_{2ah}(u_{2x},E)\Lb_0\Ab(\eb_3)\Rbb$ &
$\{h_{1,3,1}+h_{2,3,1}\}_\mathrm{R}$
\\[2mm]
& $G(\Gamma E_{g},1,0)_3=\Db_{4h}(u_{4z}^+,u_{2x})\Lb_0\Rbb$ & 
$\{h_{1,3,1}+h_{2,2,1}\}_\mathrm{R}$
\\[2mm] \hline \hline\\[-4mm]
$\check{G}_0^{(M A_{1g},1,1)}$ &
$G(M A_{1g},1,1)=M(\eb_2)T^x(\eb_1)\Db_{4h}\Lb_1\Ab(\eb_3)$ &
$\{h_{1,1,1}\}_\mathrm{R}$
\\[2mm]\hline\\[-4mm]
$\check{G}_0^{(M A_{2g},1,0)}$ &
$G(M A_{2g},1,0)=T_x(\eb_1)T^x(\eb_1)\Db_{4h}(E,u_{2x})\Lb_1\Ab(\eb_3)\Rbb$ &
$\{h_{1,3,1}\}_\mathrm{R}$
\\[2mm]\hline\\[-4mm]
$\check{G}_0^{(M B_{1g},1,1)}$ &
$G(M B_{1g},1,1)=M(\eb_2)T^x(\eb_1)T_a(\eb_1)\Db_{4h}(u_{2x},E)\Lb_1\Ab(\eb_3)$ & 
$\{h_{1,3,1}\}_\mathrm{R}$
\\[2mm]\hline\\[-4mm]
$\check{G}_0^{(M B_{2g},1,1)}$ &
$G(M B_{2g},1,1)=M(\eb_2)T^x(\eb_1)T_x(\eb_1)\Db_{4h}(u_{2x},u_{2x})\Lb_1\Ab(\eb_3)$ &
$\{h_{1,3,1}\}_\mathrm{R}$
\\[2mm]\hline\\[-4mm]
$\check{G}_0^{(M E_{g},1,1)}$ &
$G(M E_{g},1,1)_1=M(\eb_2)T^x(\eb_1)T_y(\eb_1)\Db_{2h}(u_{2x},E)\Lb_1\Ab(\eb_3)$ &
$\{h_{1,3,1}\}_\mathrm{R}$
\\[2mm]
& $G(M E_{g},1,1)_2=M(\eb_2)T^x(\eb_1)T_b(\eb_1)\Db_{2ah}(u_{2x},E)\Lb_1\Ab(\eb_3)$ &
$\{h_{1,3,1}+h_{2,3,1}\}_\mathrm{R}$
\\[2mm]
& $G(M E_{g},1,1)_3=M(\eb_3)T_z(\eb_1)\Db_{4h}(u_{4z}^+,u_{2x})\Lb_1$ &
$\{h_{1,1,1}+h_{2,2,1}\}_\mathrm{R}$
\\[2mm]\hline\\[-4mm]
$\check{G}_0^{(M E_{g},1,0)}$ &
$G(M E_{g},1,0)_1=T^x(\eb_1)T_y(\eb_1)\Db_{2h}(u_{2x},E)\Lb_1\Ab(\eb_3)\Rbb$ &
$\{h_{1,3,1}\}_\mathrm{R}$
\\[2mm]
& $G(M E_{g},1,0)_2=T^x(\eb_1)T_y(\eb_1)\Db_{2ah}(u_{2x},E)\Lb_1\Ab(\eb_3)\Rbb$ &
$\{h_{1,3,1}+h_{2,3,1}\}_\mathrm{R}$
\\[2mm]
& $G(M E_{g},1,0)_3=T^z(\eb_1)T_z(\eb_1)\Db_{4h}(u_{4z}^+,u_{2x})\Lb_1\Rbb$ &
$\{h_{1,3,1}+h_{2,2,1}\}_\mathrm{R}$
\\[2mm]\hline\\[-4mm]
\multicolumn{3}{l}{$\Lb_0=\Lb(\eb_1,\eb_2),\ 
\Lb_1=\Lb(\eb_1+\eb_2,\eb_2-\eb_1)$}\\[2mm]
\multicolumn{3}{l}{$u_{2i}=u(\eb_i,\pi)\in \SSb,\ 
M(\eb_i)=\{E,tu_{2i}\}$}\\[2mm]
\multicolumn{3}{l}{$\Ab(\eb_i)=\{u(\eb_3,\theta)
\mid 0 \leq \theta \leq 2\pi\}$}\\[2mm]
\multicolumn{3}{l}{$\Db_{4h}(\alpha,\beta)=
\{(E,C_{4z}^+\alpha,C_{2z}\alpha ^2,C_{4z}^-\alpha ^{-})+
C_{2x}\beta(E,C_{4z}^+\alpha,C_{2z}\alpha ^2,C_{4z}^-\alpha ^{-})\}
\times \Cb_I$}\\[2mm]
\multicolumn{3}{l}{$\Db_{2h}(\alpha,\beta)=\{(E,C_{2z}\alpha)
+C_{2x}\beta (E,C_{2z}\alpha)\}\times \Cb_I$}\\[2mm]
\multicolumn{3}{l}{$\Db_{2ah}(\alpha,\beta)=\{(E,C_{2z}\alpha)
+C_{2a}\beta (E,C_{2z}\alpha)\}\times \Cb_I$}\\[2mm]
\multicolumn{3}{l}{$T^i(\eb_m)=\{E,u_{2i}T(\eb_m)\}$}\\[2mm]
\multicolumn{3}{l}{$T_i(\eb_m)=\{E,C_{2i}T(\eb_m)\}$}\\[2mm]
\hline \hline
\end{tabular}}
\end{table}
Those of $\check{G}_0^{(Xj,1,\nu)}$ 
are listed in Table \ref{Isotro-Xj1nu}. 
\begin{table}[!]
\caption{Axial isotropy subgroup and its Fixed point subspace of 
$\check{G}_0^{(Xj,1,\nu)}$}
\label{Isotro-Xj1nu}
{\footnotesize
\begin{tabular}{lll}
\hline
\\[-3mm]
{\normalsize R-rep} & {\normalsize axial isotropy subgroup} & {\normalsize Fixed point subspace} \\[3mm]
\hline \\[-3mm]
$\check{G}_0^{(XA_{1g},1,1)}$ &
$G(X A_{1g},1,1)_1= M(\eb_2)T^x(\eb_1)\Db_{2h}\Lb_2\Ab(\eb_3)$ &
$\{h_{1,3,1}\}_\mathrm{R}$
\\[2mm]\\[-4mm]
& $G(X A_{1g},1,1)_2=M(\eb_2)T^x(\eb_1+\eb_2)\Db_{4h}\Lb_3\Ab(\eb_3)$ &
$\{h_{1,3,1}+h_{2,3,1}\}_\mathrm{R}$
\\[2mm]\\[-4mm]
& $G(X A_{1g},1,1)_3=M(\eb_3)T^x(\eb_2)T^y(\eb_1)\Db_{4h}(u_{2a},E)\Lb_3$ &
$\{h_{1,1,1}+h_{2,2,1}\}_\mathrm{R}$
\\[2mm]\hline\\[-4mm]
$\check{G}_0^{(XB_{1g},1,1)}$ &
$G(X B_{1g},1,1)_1=M(\eb_2)T^x(\eb_1)\Db_{2h}(E,u_{2x})\Lb_2\Ab(\eb_3)$ &
$\{h_{1,3,1}\}_\mathrm{R}$
\\[2mm]\\[-4mm]
& $G(X B_{1g},1,1)_2=M(\eb_2)T^x(\eb_1+\eb_2)\Db_{4h}(u_{2x},u_{2x})\Lb_3\Ab(\eb_3)$ &
$\{h_{1,3,1}+h_{2,3,1}\}_\mathrm{R}$
\\[2mm]\\[-4mm]
& $G(X B_{1g},1,1)_3=M(\eb_3)T^x(\eb_2)T^y(\eb_1)\Db_{4h}(u_{2b},u_{2x})\Lb_3$ &
$\{h_{1,1,1}+h_{2,2,1}\}_\mathrm{R}$
\\[2mm]\hline\\[-4mm]
$\check{G}_0^{(XB_{2g},1,1)}$ &
$G(X B_{2g},1,1)_1=M(\eb_2)T^x(\eb_1)\Db_{2h}(u_{2x},u_{2x})\Lb_2\Ab(\eb_3)$ &
$\{h_{1,3,1}\}_\mathrm{R}$
\\[2mm]\\[-4mm]
& $G(X B_{2g},1,1)_2=M(\eb_2)T_x(\eb_1)T_y(\eb_2)\Db_{2ah}(u_{2x},E)\Lb_3\Ab(\eb_3)$ &
$\{h_{1,3,1}+h_{2,3,1}\}_\mathrm{R}$
\\[2mm]\\[-4mm]
& $G(X B_{2g},1,1)_3=M(\eb_3)T_x(\eb_1)T_y(\eb_2)\Db_{4h}(u_{4z}^+,u_{2x})\Lb_3$ &
$\{h_{1,1,1}+h_{2,2,1}\}_\mathrm{R}$
\\[2mm]\hline\\[-4mm]
$\check{G}_0^{(XB_{3g},1,1)}$ &
$G(X B_{3g},1,1)_1=M(\eb_2)T_y(\eb_1)\Db_{2h}(u_{2x},E)\Lb_2\Ab(\eb_3)$ & 
$\{h_{1,3,1}\}_\mathrm{R}$
\\[2mm]\\[-4mm]
& $G(X B_{3g},1,1)_2=M(\eb_2)T_x(\eb_2)T_y(\eb_1)\Db_{2ah}(u_{2x},E)\Lb_3\Ab(\eb_3)$ &
$\{h_{1,3,1}+h_{2,3,1}\}_\mathrm{R}$
\\[2mm]\\[-4mm]
& $G(X B_{3g},1,1)_3=M(\eb_3)T_x(\eb_2)T_y(\eb_1)\Db_{4h}(u_{4z}^+,u_{2x})\Lb_3$ &
$\{h_{1,1,1}+h_{2,2,1}\}_\mathrm{R}$
\\[2mm]\hline \hline\\[-4mm]
$\check{G}_0^{(XB_{1g},1,0)}$ &
$G(X B_{1g},1,0)_1=T^x(\eb_1)\Db_{2h}(E,u_{2x})\Lb_2\Ab(\eb_3)\Rbb$ &
$\{h_{1,3,1}\}_\mathrm{R}$
\\[2mm]\\[-4mm]
& $G(X B_{1g},1,0)_2=T^x(\eb_1+\eb_2)\Db_{4h}(u_{2x},u_{2x})\Lb_3\Ab(\eb_3)\Rbb$ &
$\{h_{1,3,1}+h_{2,3,1}\}_\mathrm{R}$
\\[2mm]\\[-4mm]
& $G(X B_{1g},1,0)_3=T^x(\eb_2)T^y(\eb_1)\Db_{4h}(u_{2b},u_{2x})\Lb_3\Rbb$ &
$\{h_{1,1,1}+h_{2,2,1}\}_\mathrm{R}$
\\[2mm]\hline\\[-4mm]
$\check{G}_0^{(XB_{2g},1,0)}$ &
$G(X B_{2g},1,0)_1=T^x(\eb_1)\Db_{2h}(u_{2x},u_{2x})\Lb_2\Ab(\eb_3)\Rbb$ &
$\{h_{1,3,1}\}_\mathrm{R}$
\\[2mm]\\[-4mm]
& $G(X B_{2g},1,0)_2=T_x(\eb_1)T_y(\eb_2)\Db_{2ah}(u_{2x},E)\Lb_3\Ab(\eb_3)\Rbb$ &
$\{h_{1,3,1}+h_{2,3,1}\}_\mathrm{R}$
\\[2mm]\\[-4mm]
& $G(X B_{2g},1,0)_3=T_x(\eb_1)T_y(\eb_2)\Db_{4h}(u_{4z}^+,u_{2x})\Lb_3\Rbb$ &
$\{h_{1,1,1}+h_{2,2,1}\}_\mathrm{R}$
\\[2mm]\hline\\[-4mm]
$\check{G}_0^{(XB_{3g},1,0)}$ &
$G(X B_{3g},1,0)_1=T_y(\eb_1)\Db_{2h}(u_{2x},E)\Lb_2\Ab(\eb_3)\Rbb$ & 
$\{h_{1,3,1}\}_\mathrm{R}$
\\[2mm]\\[-4mm]
& $G(X B_{3g},1,0)_2=T_x(\eb_2)T_y(\eb_1)\Db_{2ah}(u_{2x},E)\Lb_3\Ab(\eb_3)\Rbb$ &
$\{h_{1,3,1}+h_{2,3,1}\}_\mathrm{R}$
\\[2mm]\\[-4mm]
& $G(X B_{2g},1,0)_3=T_x(\eb_2)T_y(\eb_1)\Db_{4h}(u_{4z}^+,u_{2x})\Lb_3\Rbb$ & 
$\{h_{1,1,1}+h_{2,2,1}\}_\mathrm{R}$
\\[2mm]\hline\\[-4mm]
\multicolumn{3}{l}{$\Lb_2=\Lb(2\eb_1,\eb_2),\ 
\Lb_3=\Lb(2\eb_1,2\eb_2)$}\\[2mm]
\multicolumn{3}{l}{$u_{2i}=u(\eb_i,\pi),u_{2a}=u(\eb_1+\eb_2,\pi),
u_{2b}=u(-\eb_1+\eb_2,\pi)\in \SSb$}\\[2mm]
\multicolumn{3}{l}{$M(\eb_i)=\{E,tu_{2i}\},
\ T^i(\eb_m)=\{E,u_{2i}T(\eb_m)\}
, \ T_i(\eb_m)=\{E,C_{2i}T(\eb_m)\}$}\\[2mm]
\multicolumn{3}{l}{$\Db_{2h}(\alpha,\beta)=
\{(E,C_{2z}\alpha )+C_{2x}\beta (E,C_{2z}\alpha )\}\times \Cb_I$}\\[2mm]
\multicolumn{3}{l}{$\Db_{2ah}(\alpha,\beta)=
\{(E,C_{2z}\alpha )+C_{2a}\beta (E,C_{2z}\alpha )\}\times \Cb_I$}\\[2mm]
\multicolumn{3}{l}{$\Db_{4h}(\alpha,\beta)=
\{(E,C_{4z}^+\alpha, C_{2z}\alpha^2,C_{4z}^-\alpha^{-1} )
+C_{2x}\beta (E,C_{4z}^+\alpha, C_{2z}\alpha^2,
C_{4z}^-\alpha^{-1} )\}\times \Cb_I$}\\[2mm]
\hline \hline
\end{tabular}}
\end{table}

Spin magnetic states are classified into two groups: collinear and 
non-collinear magnetic states. 
The collinear magnetic state has the isotropy subgroup containing the 
subgroup $\Ab(\eb_3)$:
the spin rotation around $z$ axis.
The isotropy subgroup of the non-collinear magnetic state does not 
contain $\Ab(\eb_3)$. 

Non-collinear magnetic states are derived from the R-reps: 
$G_0^{(\Lambda E_g,1,1)}$, $G_0^{(\Lambda E_g,1,0)}$, 
$G_0^{(XA_{1g},1,1)}$, $G_0^{(XB_{jg},1,1)}$ and $G_0^{(XB_{jg},1,0)}$
where $\Lambda=\Gamma, M$ and $j=1,2,3$.
The corresponding non-collinear magnetic states are 
$G(\Gamma E_g,1,1)_3,
G(\Gamma E_g,1,0)_3,$\\
$G(M E_g,1,1)_3,
G(M E_g,1,0)_3,
G(XA_g,1,1)_3,G(XB_{ig},1,1)_3$
 and $G(XB_{ig},1,0)_3\ (i=1,2,3)$.
All states in  Table \ref{Isotro-GMj1nu} and 
\ref{Isotro-Xj1nu} except these eleven states are 
collinear magnetic states. We consider collinear and 
non-collinear magnetic states seperately.
\subsection{Collinear magnetic state}
All axial isotropy subgroups of collinear magnetic states 
contain $A(\eb_3)$, then $\Rb^{l1}=\Rb^{l2}=0$ for $l=0,1,2,3$.
Thus in these cases  we obtain from \eqref{Dmupdown}
\begin{align}
\Db^{\uparrow \downarrow}(\mb)&=\Db^{\downarrow \uparrow }(\mb)=0.
\end{align}
We consider two examples.  
\begin{example}{\rm 
\label{Ex-GA1-1-1}
$G(\Gamma A_{1g},1,1)$ state.\\
In this case, the isotropy subgroup $G(H_m)$ of $H_m$ is 
\begin{align}
G(\Gamma A_{1g},1,1)&=M(\eb_2)\Db_{4h}\Lb_0\Ab(\eb_3).
\end{align}
From $\Lb_0$ invariance of $\Rb^{l\lambda}$ we can see that only 
$\Rb^{00}$ and $\Rb^{03}$ are non-zero.\\ 
From $C_{2x}, C_{2z},C_{2a}(\in \Db_{4h})$ invariance of $\Rb^{00}$ and
$\Rb^{03}$, we obtain 
\begin{align}
\Rb^{00}&=\left(
\begin{array}{ccc}
a& 0 & 0 \\
0 & a & 0 \\
0 & 0 & b\\
\end{array}
\right),\ \ 
\Rb^{03}=\left(
\begin{array}{ccc}
c& 0 & 0 \\
0 & c & 0 \\
0 & 0 & d\\
\end{array}
\right),
\label{GA-1-1-abcd}
\end{align}
where $a,b,c$ and $d$ are real numbers. 
From \eqref{Dmupdown}
we obtain 
\begin{align}
\Db^{\uparrow \uparrow}(\mb)&=\Rb^{00}+\Rb^{03}=\left(
\begin{array}{ccc}
a+c & 0 & 0\\
0 & a+c & 0\\
0 & 0 & b+d
\end{array}
\right),\nonumber \\
\Db^{\downarrow\downarrow}(\mb)&=\Rb^{00}-\Rb^{03}=\left(
\begin{array}{ccc}
a-c & 0 & 0\\
0 & a-c & 0\\
0 & 0 & b-d
\end{array}
\right).
\end{align}
In Table \ref{GA11-occupation}, we list the occupied atomic orbitals and their occupation numbers for $G(\Gamma A_{1g},1,1)$ state.
\begin{table}[!]
\begin{center}
\caption{Occupied atomic orbitals and their occupation numbers 
for $G(\Gamma A_{1g},1,1)$ state}
\label{GA11-occupation}
\begin{tabular}{lllc}
\hline
{site} & { spin }& { atomic\  orbital}& { occupation\ number} \\
\hline\\[-3mm] 
$\mb$ & $\uparrow$    & $\psi_1=\phi_1$ & $a+c$ \\[2mm]
$\mb$ & $\uparrow$    & $\psi_2=\phi_2$ & $a+c$ \\[2mm]
$\mb$ & $\uparrow$    & $\psi_3=\phi_3$ & $b+d$ \\[2mm]
$\mb$ & $\downarrow$  & $\psi_1=\phi_1$ & $a-c$ \\[2mm]
$\mb$ & $\downarrow$  & $\psi_2=\phi_2$ & $a-c$ \\[2mm]
$\mb$ & $\downarrow$  & $\psi_3=\phi_3$ & $b-d$ \\[2mm]
\hline\\[-3mm]
\multicolumn{4}{l}{$a=R^{00}_{11}=R^{00}_{22},\ b=R^{00}_{33},\ 
c=R^{03}_{11}=R^{03}_{22},\ d=R^{03}_{33}$}
\\[2mm]
\hline \hline
\end{tabular}
\end{center}
\end{table}
 From \eqref{OP-R} the spin density at the site $\mb$
is given by
\begin{align}
s^3(\mb)&=\sum_{j=1}^3R^{03}_{jj}=2c+d
\end{align}
This state corresponds to the usual ferromagnetic state without
orbital order.
}
\end{example}
Next we consider M point collinear magnetic state. 
From $\Lb_1$ invariance of $\Rb^{l\lambda}$, only  $\Rb^{l\lambda}(l=1,0,\lambda=0,3)$ are 
non-zero. 
\begin{example}{\rm 
\label{Ex-MB1g-1-1}
$G(MB_{1g},1,1)$ state.\\
In this case, the isotropy subgroup $G(H_m)$ of $H_m$
is 
\begin{align}
G(MB_{1g},1,1)&=M(\eb_2)T^x(\eb_1)T_a(\eb_1)
\Db_{4h}(u_{2x},E)\Lb_1\Ab(\eb_3).
\end{align}
From $C_{2x},C_{2z}, C_{2a}T(\eb_1)\ \in G(MB_{1g},1,1)$ invariance, 
we can see that 
only $\Rb^{00}$ and 
$\Rb^{13}$ are non-zero and 
\begin{align}
\Rb^{00}&=\left(
\begin{array}{ccc}
a & 0 & 0\\
0 & a & 0 \\
0 & 0 & b
\end{array}
\right),\ 
\Rb^{13}=\left(
\begin{array}{ccc}
c & 0 & 0\\
0 &-c  & 0 \\
0 & 0 & 0
\end{array}
\right),
\end{align}
where $a$, $b$ and $c$ are real numbers.
From \eqref{Dmupdown},
we obtain for $\mb$ such that $T(\mb)\in \Lb_1$
\begin{align}
\Db^{\uparrow\uparrow}(\mb)&=\left(
\begin{array}{ccc}
a+c & 0 &0\\
0 & a-c & 0\\
0 & 0 & b
\end{array}
\right),\ 
\Db^{\downarrow \downarrow}(\mb)=\left(
\begin{array}{ccc}
a-c & 0 &0\\
0 & a+c & 0\\
0 & 0 & b
\end{array}
\right),\nonumber \\[2mm]
\Db^{\uparrow\uparrow}(\mb+\eb_j)&=\left(
\begin{array}{ccc}
a-c & 0 &0\\
0 & a+c & 0\\
0 & 0 & b
\end{array}
\right),\ 
\Db^{\downarrow \downarrow}(\mb)=\left(
\begin{array}{ccc}
a+c & 0 &0\\
0 & a-c & 0\\
0 & 0 & b
\end{array}
\right).
\end{align}
In Table \ref{MB1-1-1-occ} we list the occupied atomic orbitals
and their occupation numbers for $G(MB_{1g},1,1)$ state.
\begin{table}[!]
\begin{center}
{\footnotesize 
\caption{Occupied atomic orbitals and their occupation numbers 
for $G(MB_{1g},1,1)$ state}
\label{MB1-1-1-occ}
\begin{tabular}{lllc}
\hline
{\normalsize site} & {\normalsize spin }& {\normalsize atomic orbital} 
& {\normalsize occupation\ number} \\
\hline\\[-3mm] 
$\mb$ & $\uparrow$   & $\psi_1=\phi_1$ & $a+c$ \\[2mm]
$\mb$ & $\uparrow$   & $\psi_2=\phi_2$ & $a-c$ \\[2mm]
$\mb$ & $\uparrow$   & $\psi_3=\phi_3$ & $b$   \\[2mm]
$\mb$ & $\downarrow$ & $\psi_1=\phi_1$ & $a-c$ \\[2mm]
$\mb$ & $\downarrow$ & $\psi_2=\phi_2$ & $a+c$ \\[2mm]
$\mb$ & $\downarrow$ & $\psi_3=\phi_3$ & $b$   \\[2mm]
\hline 
\hline\\[-3mm] 
$\mb+\eb_j$ & $\uparrow$   & $\psi_1=\phi_1$ & $a-c$ \\[2mm]
$\mb+\eb_j$ & $\uparrow$   & $\psi_2=\phi_2$ & $a+c$ \\[2mm]
$\mb+\eb_j$ & $\uparrow$   & $\psi_3=\phi_3$ & $b$   \\[2mm]
$\mb+\eb_j$ & $\downarrow$ & $\psi_1=\phi_1$ & $a+c$ \\[2mm]
$\mb+\eb_j$ & $\downarrow$ & $\psi_2=\phi_2$ & $a-c$ \\[2mm]
$\mb+\eb_j$ & $\downarrow$ & $\psi_3=\phi_3$ & $b$   \\[2mm]
\hline\\[-3mm]
\multicolumn{4}{l}{j=1,2,\ $a=R^{00}_{11}=R^{00}_{22},\ b=R^{00}_{33},\ 
c=R^{13}_{11}=-R^{13}_{22}$}
\\[2mm]
\hline \hline
\end{tabular}
}
\end{center}
\end{table}
From \eqref{OP-R},  we see that
this state has the spin quadrupole moment $Q_{2}^{23}(\mb)$, for 
$\mb$ such that 
$T(\mb)\in \Lb_1$, as follows:
\begin{align}
Q^{23}_{2}(\mb)&=2cI_1, Q^{23}_{2}(\mb+\eb_j)=-2cI_1, j=1,2.
\end{align}
 From \eqref{OP-R} we obtain
\begin{align}
Q_2^2(\mb)&=I_1(R^{00}_{11}-R^{00}_{22})=c-c=0,\nonumber \\
s^3(\mb)&=e^{-i\sQb_1\cdot \smb}\sum_{j=1}^3R^{13}_{jj}
=e^{-i\sQb_1\cdot \smb}(c-c)=0.
\end{align}
Thus $G(MB_{1g},1,1)$ state, which has the spin quadrupole 
moment $Q_{2}^{23}$, 
is not the coexisting state of 
quadrupole moment $Q_2^2(\mb)$ and spin density $s^3(\mb)$.
 }
\end{example}
\subsection{Non-collinear magnetic state}
First we consider M point non-collinear magnetic state. 
\begin{example}{\rm 
\label{Ex-MEg-1-1-3}
$G(ME_{g},1,1)_3$ state.\\
In this case, the isotropy subgroup $G(H_m)$ of $H_m$
is 
\begin{align}
G(M E_{g},1,1)_3&=M(\eb_3)T_z(\eb_1)\Db_{4h}(u_{4z}^+,u_{2x})\Lb_1.
\end{align}
From $T(\eb_1+\eb_2)\in \Lb_1$ invariance of $\Rb^{l\lambda}$, we can see that 
$\Rb^{2\lambda}=\Rb^{3\lambda}=0$. From $tu_{2z}\in M(\eb_3)$ invariance, 
we see that $\Rb^{l0},\Rb^{l1}$ and $\Rb^{l2} (l=0,1)$ are real matrices and 
$\Rb^{03}$ and $\Rb^{13}$ are pure imaginary matrices.
From $C_{2z}T(\eb_1)$ and $C_{2z}u_{2z},C_{2x}u_{2x},\\
C_{2a}u_{2a}\in 
\Db_{4h}(u_{4z}^+,u_{2x})$ invariance
we can see that only $\Rb^{00},\Rb^{03},\Rb^{11}$ and $\Rb^{12}$ are 
non-zero and have following forms.
\begin{align}
\Rb^{00}&=\left(
\begin{array}{ccc}
a & 0 &0 \\
0 & a & 0\\
0 & 0 & b
\end{array}
\right),&
\Rb^{03}&=\left(
\begin{array}{ccc}
0& ic &0 \\
-ic & 0 & 0\\
0 & 0 & 0
\end{array}
\right),\nonumber \\[2mm]
\Rb^{11}&=\left(
\begin{array}{ccc}
0 &0 & 0\\
0& 0 & d\\
0 & d& 0
\end{array}
\right),& 
\Rb^{12}&=\left(
\begin{array}{ccc}
0& 0 &-d\\
0 & 0 & 0\\
-d & 0 & 0
\end{array}
\right),
\label{ME-1-1-3-abcdf}
\end{align}
where $a$, $b$, $c$ and $d$ are real numbers, and $i$ is imaginary unit.
From \eqref{Dmupdown} and \eqref{calDb-def},
we obtain, for $\mb$ such that $T(\mb)\in \Lb_1$,
\begin{align}
\calDb(\mb)&=\left(
\begin{array}{cccccc}
a   & ic & 0  & 0  & 0  & id\\
-ic &  a & 0  & 0  & 0  & d\\
0   &  0 & b  & id & d  & 0\\
0   & 0  & -id& a  &-ic & 0\\
0   & 0  &  d & ic & a  & 0 \\
-id & d  &  0 & 0  & 0  & b
\end{array}
\right),\nonumber \\[2mm]
\calDb(\mb+\eb_j)&=\left(
\begin{array}{cccccc}
a   & ic & 0  & 0  & 0  & -id\\
-ic &  a & 0  & 0  & 0  &-d\\
0   &  0 & b  & -id & -d  & 0\\
0   & 0  & id& a  &-ic & 0\\
0   & 0  &  -d & ic & a  & 0 \\
id & -d  &  0 & 0  & 0  & b
\end{array}
\right)\ (j=1,2).
\end{align}       
Diagonalizing $\calDb$,
we obtain occupied general spin  orbitals and their occupation 
numbers as in 
Table \ref{MEg-1-1-3-occupation}. 
\begin{table}[!]
\caption{Occupied general spin orbitals and 
their occupation numbers 
for $G(ME_g,1,1)_3$ state}
\label{MEg-1-1-3-occupation}
\begin{tabular}{llc}
\hline
{\normalsize site} & {\normalsize general\ spin orbital}& {\normalsize occupation number} \\
\hline\\[-3mm] 
$\mb$ & $\psi_1=\frac{\mathrm{sgn}(d)}{\sqrt{2}}u(i\phi_1+\phi_2)
\mid \uparrow \rangle+w\phi_3\mid \downarrow \rangle$ & $\lambda_1$
\\[3mm]
$\mb$ & $\psi_2=\frac{\mathrm{sgn}(d)}{\sqrt{2}}u(-i\phi_1+\phi_2)
\mid \downarrow \rangle+w\phi_3\mid \uparrow \rangle$ & $\lambda_1$
\\[3mm]
$\mb$ & $\psi_3=\frac{\mathrm{sgn}(d)}{\sqrt{2}}w(-i\phi_1-\phi_2)
\mid \uparrow \rangle+u\phi_3\mid \downarrow \rangle$ & $\lambda_2$
\\[3mm]
$\mb$ & $\psi_4=\frac{\mathrm{sgn}(d)}{\sqrt{2}}w(i\phi_1-\phi_2)
\mid \downarrow \rangle+u\phi_3\mid \uparrow \rangle$ & $\lambda_2$
\\[3mm]
$\mb$ & $\psi_5=\frac{1}{\sqrt{2}}(\phi_1+i\phi_2)
\mid \uparrow \rangle$ & $\lambda_3$
\\[3mm]
$\mb$ & $\psi_6=\frac{1}{\sqrt{2}}(\phi_1-i\phi_2)
\mid \downarrow \rangle$ & $\lambda_3$
\\[3mm]
\hline
\\[-3mm]
$\mb+\eb_j$ & $\psi_1=\frac{\mathrm{sgn}(d)}{\sqrt{2}}u(i\phi_1+\phi_2)
\mid \uparrow \rangle-w\phi_3\mid \downarrow \rangle$ & $\lambda_1$
\\[3mm]
$\mb+\eb_j$ & $\psi_2=\frac{\mathrm{sgn}(d)}{\sqrt{2}}u(-i\phi_1+\phi_2)
\mid \downarrow \rangle-w\phi_3\mid \uparrow \rangle$ & $\lambda_1$
\\[3mm]
$\mb+\eb_j$ & $\psi_3=\frac{\mathrm{sgn}(d)}{\sqrt{2}}w(-i\phi_1-\phi_2)
\mid \uparrow \rangle-u\phi_3\mid \downarrow \rangle$ & $\lambda_2$
\\[3mm]
$\mb+\eb_j$ & $\psi_4=\frac{\mathrm{sgn}(d)}{\sqrt{2}}w(i\phi_1-\phi_2)
\mid \downarrow \rangle-u\phi_3\mid \uparrow \rangle$ & $\lambda_2$
\\[3mm]
$\mb+\eb_j$ & $\psi_5=\frac{1}{\sqrt{2}}(\phi_1+i\phi_2)
\mid \uparrow \rangle$ & $\lambda_3$
\\[3mm]
$\mb+\eb_j$ & $\psi_6=\frac{1}{\sqrt{2}}(\phi_1-i\phi_2)
\mid \downarrow \rangle$ & $\lambda_3$
\\[1mm]
\hline\\[-2mm]
\multicolumn{3}{l}{$T(\mb)\in \Lb_1$}\\
\multicolumn{3}{l}{$j=1,2,\ u=\frac{1}{\sqrt{2}}(1+\frac{a-b+c}
{\sqrt{(a-b+c)^2+8d^2}})^{\frac{1}{2}},\ 
w=\frac{1}{\sqrt{2}}(1-\frac{a-b+c}
{\sqrt{(a-b+c)^2+8d^2}})^{\frac{1}{2}}$}\\[3mm]
\multicolumn{3}{l}{$
a=R^{00}_{11}=R^{00}_{22},\ b=R^{00}_{33},\ 
ic=R^{03}_{12}=-R^{03}_{21},d=R^{11}_{23}=R^{11}_{32}=-R^{12}_{13}
=-R^{12}_{31}$}\\[2mm]
\multicolumn{3}{l}{$
\lambda_1=\frac{a+b+c+\sqrt{(a-b+c)^2+8d^2}}{2},
\ \lambda_2=\frac{a+b+c-\sqrt{(a-b+c)^2+8d^2}}{2},\ 
\lambda_3=a-c$}
\\[2mm]
\hline \hline
\end{tabular}
\end{table}
From \eqref{OP-R} we see that this state has following 
local order parameters. For $\mb$ such that 
$T(\mb)\in \Lb_1$ and $j=1,2$, we obtain 
\begin{align}
Q_{23}^1(\mb)&=2I_2d,\ 
Q_{23}^1(\mb+\eb_j)=-2I_2d,\nonumber \\
Q_{31}^2(\mb)&=-2I_2d,\ \  Q_{31}^2(\mb+\eb_j)=2I_2d,\nonumber \\
l_3^3(\mb)&=l_3^3(\mb+\eb_j)=2c.
\end{align}
Note that the onset of ferro spin obital angular momentum:$l_3^3$ is induced by the 
transition to the $G(ME_g,1,1)_3$ state having non-collinear spin 
quadrupole moment:$\{Q_{23}^1,Q_{31}^2\}$.

From \eqref{OP-R} we can see that for all site $\mb$
\begin{align}
s^3(\mb) &= s^1(\mb)=s^2(\mb)=0,\nonumber \\
l_3(\mb) &= 0,\nonumber \\
Q_{23}(\mb) &= Q_{31}(\mb)=0.
\end{align}
The important point to note is that 
the existence of order parameters $l_3^3(\mb), Q_{23}^1(\mb)$ and 
$Q_{31}^2(\mb)$ does not mean the coexistence  of 
spin density $s^3(\mb)$ and orbital angular momentum $l_3(\mb)$ nor 
spin densities $s^1(\mb), s^2(\mb)$ and quadrupole moment 
$Q_{23}(\mb), Q_{31}(\mb)$.
}
\end{example}
Next we consider X point non-collinear magnetic state. 
We consider two examples.
\begin{example}{\rm 
\label{Ex-XA-1-1-3}
$G(XA_g,1,1)_3$ state.\\
The isotropy subgroup of the $G(XA_g,1,1)_3$ state is
\begin{align}
G(XA_g,1,1)_3&=M(\eb_3)T^x(\eb_2)T^y(\eb_1)\Db_{4h}(u_{2a},E)\Lb_3.
\end{align}
From $u_{2x}T(\eb_2)\ (\in T^x(\eb_2))$ and $u_{2y}T(\eb_1)\ (\in T^y(\eb_1))$
invariance of $\Rb^{l\lambda}$, we see that  
only $\Rb^{00},\Rb^{13},\Rb^{21}$ and $\Rb^{32}$ are non-zero.
From $\Db_{2h}$, $C_{4z}^+u_{2a}\ (\in \Db_{4h}(u_{2a},E))$ 
and $M(\eb_3)$ invariance
of $\Rb^{l\lambda}$ we see that 
only $\Rb^{00},\Rb^{21}$ and $\Rb^{32}$ are non-zero and  have following 
forms,
\begin{align}
\Rb^{00}&=\left(
\begin{array}{ccc}
a & 0 & 0 \\
0 & a & 0\\
0 & 0 & b
\end{array}
\right),\ \ 
\Rb^{21}=\left(
\begin{array}{ccc}
c & 0 & 0 \\
0 & d & 0\\
0 & 0 & f
\end{array}
\right),\ \ 
\Rb^{32}=\left(
\begin{array}{ccc}
d& 0 & 0 \\
0 & c & 0\\
0 & 0 & f
\end{array}
\right),
\label{XA-1-1-3-abcdf}
\end{align}
where $a,b,c,d$ and $f$ are real numbers.
From \eqref{Dmupdown} we obtain 
\begin{align}
\Db^{\uparrow \uparrow}(\mb)&=\Db^{\downarrow \downarrow}(\mb)
=\Rb^{00},\nonumber \\
\Db^{\uparrow \downarrow}(\mb)&=e^{-i\sQb_2\cdot \mb}\Rb^{21}-
ie^{-i\sQb_3\cdot \mb}\Rb^{32},\nonumber \\
\Db^{\downarrow\uparrow }(\mb)&=e^{-i\sQb_2\cdot \mb}\Rb^{21}+
ie^{-i\sQb_3\cdot \mb}\Rb^{32}.
\end{align}
\begin{table}[!]
\caption{Occupied general spin orbitals and 
their occupation numbers 
for $G(XA_g,1,1)_3$ state}
\label{XAg-1-1-3-occupation}
{\footnotesize 
\begin{tabular}{llc}
\hline
{\normalsize site} & {\normalsize general spin orbital}& {\normalsize occupation number} \\
\hline\\[-3mm] 
$\mb$ & $\psi_1=\frac{\phi_1}{\sqrt{2}}(\mid \uparrow \rangle  +
e^{i\gamma}\mid \downarrow \rangle )$ & $\lambda_1$
\\[3mm]
$\mb$ & $\psi_2=\frac{\phi_2}{\sqrt{2}}(\mid \uparrow \rangle  +
e^{i(\frac{\pi}{2}-\gamma)}\mid \downarrow \rangle )$ & $\lambda_1$
\\[3mm]
$\mb$ & $\psi_3=\frac{\phi_1}{\sqrt{2}}(\mid \uparrow \rangle 
+e^{i(\pi+\gamma )}\mid \downarrow \rangle )$ & $\lambda_2$
\\[3mm]
$\mb$ & $\psi_4=\frac{\phi_2}{\sqrt{2}}(\mid \uparrow \rangle 
+e^{i(\frac{3\pi}{2}-\gamma)}\mid \downarrow \rangle )$ & $\lambda_2$
\\[3mm]
$\mb$ & $\psi_5=\frac{\phi_3}{\sqrt{2}}(\mid \uparrow \rangle
+e^{\frac{i\pi}{4}}\mid \downarrow \rangle )$ & $\lambda_3$
\\[3mm]
$\mb$ & $\psi_6=\frac{\phi_3}{\sqrt{2}}(\mid \uparrow \rangle
+e^{i(\pi+\frac{\pi}{4})}\mid \downarrow \rangle)$ & $\lambda_4$
\\[3mm]
\hline\\[-3mm]
$\mb+\eb_1$ & $\psi_1=\frac{\phi_1}{\sqrt{2}}(\mid \uparrow \rangle  +
e^{i(\pi-\gamma)}\mid \downarrow \rangle )$ & $\lambda_1$
\\[3mm]
$\mb+\eb_1$ & $\psi_2=\frac{\phi_2}{\sqrt{2}}(\mid \uparrow \rangle  +
e^{i(\frac{\pi}{2}+\gamma)}\mid \downarrow \rangle )$ & $\lambda_1$
\\[3mm]
$\mb+\eb_1$ & $\psi_3=\frac{\phi_1}{\sqrt{2}}(\mid \uparrow \rangle 
+e^{-i\gamma }\mid \downarrow \rangle )$ & $\lambda_2$
\\[3mm]
$\mb+\eb_1$ & $\psi_4=\frac{\phi_2}{\sqrt{2}}(\mid \uparrow \rangle 
+e^{i(-\frac{\pi}{2}+\gamma)}\mid \downarrow \rangle )$ & $\lambda_2$
\\[3mm]
$\mb+\eb_1$ & $\psi_5=\frac{\phi_3}{\sqrt{2}}(\mid \uparrow \rangle
+e^{i(\pi-\frac{\pi}{4})}\mid \downarrow \rangle )$ & $\lambda_3$
\\[3mm]
$\mb+\eb_1$ & $\psi_6=\frac{\phi_3}{\sqrt{2}}(\mid \uparrow \rangle
+e^{i(-\frac{\pi}{4})}\mid \downarrow \rangle )$ & $\lambda_4$
\\[3mm]
\hline\\[-3mm]
$\mb+\eb_2$ & $\psi_1=\frac{\phi_1}{\sqrt{2}}(\mid \uparrow \rangle  +
e^{i(-\gamma)}\mid \downarrow \rangle )$ & $\lambda_1$
\\[3mm]
$\mb+\eb_2$ & $\psi_2=\frac{\phi_2}{\sqrt{2}}(\mid \uparrow \rangle  +
e^{i(-\frac{\pi}{2}+\gamma)}\mid \downarrow \rangle )$ & $\lambda_1$
\\[3mm]
$\mb+\eb_2$ & $\psi_3=\frac{\phi_1}{\sqrt{2}}(\mid \uparrow \rangle 
+e^{i(\pi-\gamma) }\mid \downarrow \rangle )$ & $\lambda_2$
\\[3mm]
$\mb+\eb_2$ & $\psi_4=\frac{\phi_2}{\sqrt{2}}(\mid \uparrow \rangle 
+e^{i(\frac{\pi}{2}+\gamma)}\mid \downarrow \rangle )$ & $\lambda_2$
\\[3mm]
$\mb+\eb_2$ & $\psi_5=\frac{\phi_3}{\sqrt{2}}(\mid \uparrow \rangle
+e^{i(-\frac{\pi}{4})}\mid \downarrow \rangle )$ & $\lambda_3$
\\[3mm]
$\mb+\eb_2$ & $\psi_6=\frac{\phi_3}{\sqrt{2}}(\mid \uparrow \rangle
+e^{i(\pi-\frac{\pi}{4})}\mid \downarrow \rangle )$ & $\lambda_4$
\\[3mm]
\hline\\[-3mm]
$\mb+\eb_1+\eb_2$ & $\psi_1=\frac{\phi_1}{\sqrt{2}}(\mid \uparrow \rangle  +
e^{i(\pi+\gamma)}\mid \downarrow \rangle )$ & $\lambda_1$
\\[3mm]
$\mb+\eb_1+\eb_2$ & $\psi_2=\frac{\phi_2}{\sqrt{2}}(\mid \uparrow \rangle  +
e^{i(-\frac{\pi}{2}-\gamma)}\mid \downarrow \rangle )$ & $\lambda_1$
\\[3mm]
$\mb+\eb_1+\eb_2$ & $\psi_3=\frac{\phi_1}{\sqrt{2}}(\mid \uparrow \rangle 
+e^{i\gamma }\mid \downarrow \rangle )$ & $\lambda_2$
\\[3mm]
$\mb+\eb_1+\eb_2$ & $\psi_4=\frac{\phi_2}{\sqrt{2}}(\mid \uparrow \rangle 
+e^{i(\frac{\pi}{2}-\gamma)}\mid \downarrow \rangle )$ & $\lambda_2$
\\[3mm]
$\mb+\eb_1+\eb_2$ & $\psi_5=\frac{\phi_3}{\sqrt{2}}(\mid \uparrow \rangle
+e^{i(\pi+\frac{\pi}{4})}\mid \downarrow \rangle )$ & $\lambda_3$
\\[3mm]
$\mb+\eb_1+\eb_2$ & $\psi_6=\frac{\phi_3}{\sqrt{2}}(\mid \uparrow \rangle
+e^{i(\frac{\pi}{4})}\mid \downarrow \rangle )$ & $\lambda_4$
\\[3mm]
\hline\\[-3mm]
\multicolumn{3}{l}{$T(\mb)\in \Lb_3,\ \lambda_1=a+\sqrt{c^2+d^2},\ 
\lambda_2=a-\sqrt{c^2+d^2},\ \lambda_3=b+\sqrt{2}f,\ 
\lambda_4=b-\sqrt{2}f$}\\
\multicolumn{3}{l}{$e^{i\gamma}=\dfrac{c+id}{\sqrt{c^2+d^2}},\ 
a=R^{00}_{11}=R^{00}_{22},\ b=R^{00}_{33},\ 
c=R^{21}_{11}=R^{32}_{22},d=R^{32}_{11}=R^{21}_{22}, f=R^{21}_{33}=R^{32}_{33}$}
\\[2mm]
\hline \hline
\end{tabular}
}
\end{table}
Then from \eqref{calDb-def} we obtain, for $\mb$ such that $T(\mb)\in \Lb_3$,
{\footnotesize
\begin{align}
\calDb(\mb)&=\left(
\begin{array}{cccccc}
a & 0 & 0 & c-id & 0 & 0 \\
0 & a & 0 & 0 & d-ic & 0 \\
0 & 0 & b & 0 & 0 & f-if\\
c+id & 0 & 0 &a  & 0 & 0 \\
0 & d+ic & 0 & 0 & a & 0 \\
0 & 0 & f+if & 0 & 0 & b 
\end{array}
\right),\nonumber \\[3mm]
\calDb(\mb+\eb_1)&=\left(
\begin{array}{cccccc}
a & 0 & 0 & -(c+id) & 0 & 0 \\
0 & a & 0 & 0 & -(d+ic) & 0 \\
0 & 0 & b & 0 & 0 & -(f+if)\\
-(c-id) & 0 & 0 &a  & 0 & 0 \\
0 & -(d-ic) & 0 & 0 & a & 0 \\
0 & 0 & -(f-if) & 0 & 0 & b 
\end{array}
\right),\nonumber \\[3mm]
\calDb(\mb+\eb_2)&=\left(
\begin{array}{cccccc}
a & 0 & 0 & c+id & 0 & 0 \\
0 & a & 0 & 0 & d+ic & 0 \\
0 & 0 & b & 0 & 0 & f+if\\
c-id & 0 & 0 &a  & 0 & 0 \\
0 & d-ic & 0 & 0 & a & 0 \\
0 & 0 & f-if & 0 & 0 & b 
\end{array}
\right),\nonumber \\[3mm]
\calDb(\mb+\eb_1+\eb_2)&=\left(
\begin{array}{cccccc}
a & 0 & 0 & -(c-id) & 0 & 0 \\
0 & a & 0 & 0 & -(d-ic) & 0 \\
0 & 0 & b & 0 & 0 & -(f-if)\\
-(c+id) & 0 & 0 &a  & 0 & 0 \\
0 & -(d+ic) & 0 & 0 & a & 0 \\
0 & 0 & -(f+if) & 0 & 0 & b 
\end{array}
\right).
\end{align}
}
Diagonalizing  $\calDb$, 
we obtain occupied general spin orbitals and their occupation 
numbers as in 
Table \ref{XAg-1-1-3-occupation}. 

In Fig.\ref{XA-1-1-1-3-spin} we show the spin density pattern of 
$\ssb(\mb)=(s^1(\mb),s^2(\mb))$ for four sites in the unit cell.
Note that this spin density pattern has $G(XA_g,1,1)_3$ symmetry.
\begin{figure}[!]
\begin{center}
\includegraphics[width=5.5cm,clip]{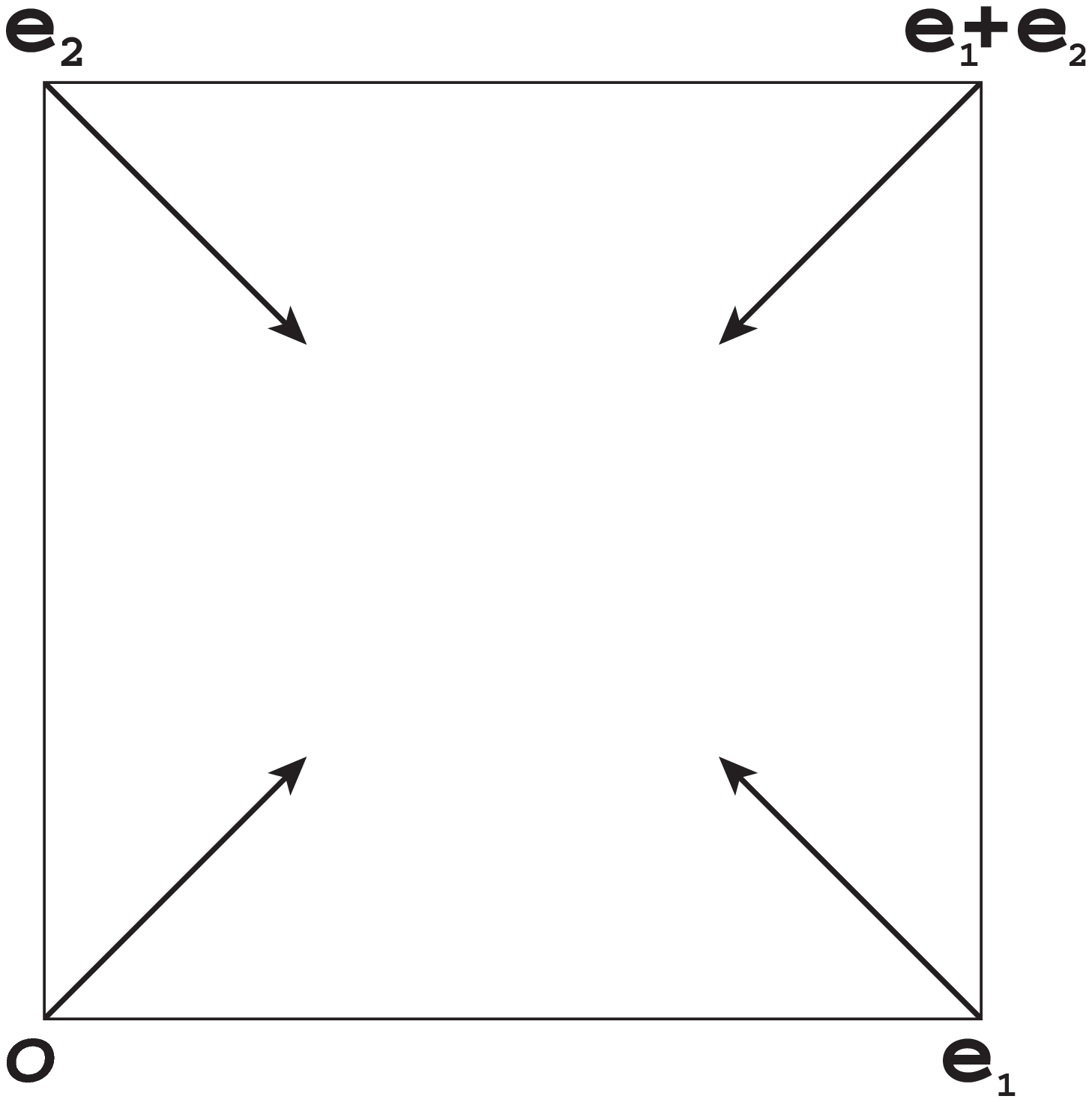}
	\caption{Spin density $\ssb(\mb)=\left(s^1(\mb),s^2(\mb)\right)$ of the
	$G(XA_g,1,1)_3$ state. $|\ssb(\mb)|=c+d+f$. }
	\label{XA-1-1-1-3-spin}
\end{center}
\end{figure}
}
\end{example}
\begin{example}{\rm 
\label{Ex-XB3-1-1-3}
$G(XB_{3g},1,1)_3$ state.\\
The isotropy subgroup of the $G(XB_{3g},1,1)_3$ state is 
\begin{align}
G(XB_{3g},1,1)_3&=M(\eb_3)T_{x}(\eb_2)T_y(\eb_1)
\Db_{4h}(u_{4z}^+,u_{2x})\Lb_3.
\end{align}
From $G(XB_{3g},1,1)_3$ invariance, we obtain  non-zero 
$\Rb^{00},\Rb^{13},\Rb^{21}$ and $\Rb^{32}$ as follows
\begin{align}
\Rb^{00}&=\left(
\begin{array}{ccc}
a & 0 & 0\\
0 & a & 0\\
0 & 0 & b
\end{array}
\right),&
\Rb^{13}&=\left(
\begin{array}{ccc}
0 & ic & 0\\
-ic & 0 & 0\\
0 & 0 & 0
\end{array}
\right),\\[1mm]
\Rb^{21}&=\left(
\begin{array}{ccc}
0 & 0 & 0\\
0 & 0 & d\\
0 & d & 0
\end{array}
\right),&
\Rb^{32}&=\left(
\begin{array}{ccc}
0 & 0 & -d\\
0 & 0 & 0\\
-d & 0 & 0
\end{array}
\right),
\label{XB3-1-1-3-abcdf}
\end{align}
where $a$, $b$, $c$ and $d$ are real numbers, and $i$ is imaginary unit.
From \eqref{Dmupdown} and \eqref{calDb-def}
we obtain, for $\mb$ such that $T(\mb)\in \Lb_3$,
\begin{align}
\calDb(\mb)&=\left(
\begin{array}{cccccc}
a & ic &0 &      0 & 0 & id\\
-ic & a & 0&     0 & 0 & d\\
0  & 0 & b &     id &d& 0\\
0 & 0 & -id     & a & -ic & 0 \\
0 & 0 & d      & ic & a & 0\\
-id & d & 0     & 0 & 0 & b 
\end{array}
\right),\nonumber \\[2mm]
\calDb(\mb+\eb_1)&=\left(
\begin{array}{cccccc}
a & -ic &0 &      0 & 0 & id\\
ic & a & 0&     0 & 0 & -d\\
0  & 0 & b &     id &-d& 0\\
0 & 0 & -id     & a & ic & 0 \\
0 & 0 & -d      & -ic & a & 0\\
-id & -d & 0     & 0 & 0 & b 
\end{array}
\right),\nonumber \\[2mm]
\calDb(\mb+\eb_2)&=\left(
\begin{array}{cccccc}
a & -ic &0 &      0 & 0 & -id\\
ic & a & 0&     0 & 0 & d\\
0  & 0 & b &     -id &d& 0\\
0 & 0 & id     & a & ic & 0 \\
0 & 0 & d      & -ic & a & 0\\
id & d & 0     & 0 & 0 & b 
\end{array}
\right),\nonumber \\[2mm]
\calDb(\mb+\eb_1+\eb_2)&=\left(
\begin{array}{cccccc}
a & ic &0 &      0 & 0 & -id\\
-ic & a & 0&     0 & 0 & -d\\
0  & 0 & b &     -id &-d& 0\\
0 & 0 & id     & a & -ic & 0 \\
0 & 0 & -d      & ic & a & 0\\
id & -d & 0     & 0 & 0 & b 
\end{array}
\right).
\end{align}
\begin{table}[!]
\caption{Occupied general spin orbitals and 
their occupation numbers 
for $G(XB_{3g},1,1)_3$ state}
\label{XB3g-1-1-3-occupation}
{\footnotesize
\begin{tabular}{llc}
\hline
{\normalsize site} & {\normalsize general spin orbital}& {\normalsize occupation number} \\
\hline\\[-3mm] 
$\mb$ & $\psi_1=i\frac{\mathrm{sgn}(d)}{\sqrt{2}}u(\phi_1-i\phi_2)
\mid \uparrow \rangle+w\phi_3\mid \downarrow \rangle$ & $\lambda_1$
\\[3mm]
$\mb$ & $\psi_2=-i\frac{\mathrm{sgn}(d)}{\sqrt{2}}u(\phi_1+i\phi_2)
\mid \downarrow \rangle+w\phi_3\mid \uparrow \rangle$ & $\lambda_1$
\\[3mm]
$\mb$ & $\psi_3=-i\frac{\mathrm{sgn}(d)}{\sqrt{2}}w(\phi_1-i\phi_2)
\mid \uparrow \rangle+u\phi_3\mid \downarrow \rangle$ & $\lambda_2$
\\[3mm]
$\mb$ & $\psi_4=i\frac{\mathrm{sgn}(d)}{\sqrt{2}}w(\phi_1+i\phi_2)
\mid \downarrow \rangle+u\phi_3\mid \uparrow \rangle$ & $\lambda_2$
\\[3mm]
$\mb$ & $\psi_5=\frac{1}{\sqrt{2}}(\phi_1+i\phi_2)\mid \uparrow \rangle$ &
$\lambda_3$
\\[3mm]
$\mb$ & $\psi_6=\frac{1}{\sqrt{2}}(\phi_1-i\phi_2)\mid \downarrow \rangle$ &
$\lambda_3$
\\[3mm]
\hline
$\mb+\eb_1$ & $\psi_1=i\frac{\mathrm{sgn}(d)}{\sqrt{2}}u(\phi_1+i\phi_2)
\mid \uparrow \rangle+w\phi_3\mid \downarrow \rangle$ & $\lambda_1$
\\[3mm]
$\mb+\eb_1$ & $\psi_2=-i\frac{\mathrm{sgn}(d)}{\sqrt{2}}u(\phi_1-i\phi_2)
\mid \downarrow \rangle+w\phi_3\mid \uparrow \rangle$ & $\lambda_1$
\\[3mm]
$\mb+\eb_1$ & $\psi_3=-i\frac{\mathrm{sgn}(d)}{\sqrt{2}}w(\phi_1+i\phi_2)
\mid \uparrow \rangle+u\phi_3\mid \downarrow \rangle$ & $\lambda_2$
\\[3mm]
$\mb+\eb_1$ & $\psi_4=i\frac{\mathrm{sgn}(d)}{\sqrt{2}}w(\phi_1-i\phi_2)
\mid \downarrow \rangle+u\phi_3\mid \uparrow \rangle$ & $\lambda_2$
\\[3mm]
$\mb+\eb_1$ & $\psi_5=\frac{1}{\sqrt{2}}(\phi_1-i\phi_2)\mid \uparrow \rangle$ &
$\lambda_3$
\\[3mm]
$\mb+\eb_1$ & $\psi_6=\frac{1}{\sqrt{2}}(\phi_1+i\phi_2)\mid \downarrow \rangle$ &
$\lambda_3$
\\[3mm]
\hline
$\mb+\eb_2$ & $\psi_1=-i\frac{\mathrm{sgn}(d)}{\sqrt{2}}u(\phi_1+i\phi_2)
\mid \uparrow \rangle+w\phi_3\mid \downarrow \rangle$ & $\lambda_1$
\\[3mm]
$\mb+\eb_2$ & $\psi_2=i\frac{\mathrm{sgn}(d)}{\sqrt{2}}u(\phi_1-i\phi_2)
\mid \downarrow \rangle+w\phi_3\mid \uparrow \rangle$ & $\lambda_1$
\\[3mm]
$\mb+\eb_2$ & $\psi_3=i\frac{\mathrm{sgn}(d)}{\sqrt{2}}w(\phi_1+i\phi_2)
\mid \uparrow \rangle+u\phi_3\mid \downarrow \rangle$ & $\lambda_2$
\\[3mm]
$\mb+\eb_2$ & $\psi_4=-i\frac{\mathrm{sgn}(d)}{\sqrt{2}}w(\phi_1-i\phi_2)
\mid \downarrow \rangle+u\phi_3\mid \uparrow \rangle$ & $\lambda_2$
\\[3mm]
$\mb+\eb_2$ & $\psi_5=\frac{1}{\sqrt{2}}(\phi_1-i\phi_2)\mid \uparrow \rangle$ &
$\lambda_3$
\\[3mm]
$\mb+\eb_2$ & $\psi_6=\frac{1}{\sqrt{2}}(\phi_1+i\phi_2)\mid \downarrow \rangle$ &
$\lambda_3$
\\[3mm]
\hline
$\mb+\eb_1+\eb_2$ & $\psi_1=-i\frac{\mathrm{sgn}(d)}{\sqrt{2}}u(\phi_1-i\phi_2)
\mid \uparrow \rangle+w\phi_3\mid \downarrow \rangle$ & $\lambda_1$
\\[3mm]
$\mb+\eb_1+\eb_2$ & $\psi_2=i\frac{\mathrm{sgn}(d)}{\sqrt{2}}u(\phi_1+i\phi_2)
\mid \downarrow \rangle+w\phi_3\mid \uparrow \rangle$ & $\lambda_1$
\\[3mm]
$\mb+\eb_1+\eb_2$ & $\psi_3=i\frac{\mathrm{sgn}(d)}{\sqrt{2}}w(\phi_1-i\phi_2)
\mid \uparrow \rangle+u\phi_3\mid \downarrow \rangle$ & $\lambda_2$
\\[3mm]
$\mb+\eb_1+\eb_2$ & $\psi_4=-i\frac{\mathrm{sgn}(d)}{\sqrt{2}}w(\phi_1+i\phi_2)
\mid \downarrow \rangle+u\phi_3\mid \uparrow \rangle$ & $\lambda_2$
\\[3mm]
$\mb+\eb_1+\eb_2$ & $\psi_5=\frac{1}{\sqrt{2}}
(\phi_1+i\phi_2)\mid \uparrow \rangle$ & $\lambda_3$
\\[3mm]
$\mb+\eb_1+\eb_2$ & $\psi_6=\frac{1}{\sqrt{2}}(\phi_1-i\phi_2)
\mid \downarrow \rangle$ & $\lambda_3$
\\[3mm]
\hline\\[-3mm]
\multicolumn{3}{l}{$\mb: T(\mb)\in \Lb_3$}\\
\multicolumn{3}{l}{$
a=R^{00}_{11}=R^{00}_{22},\ b=R^{00}_{33},\ 
ic=R^{13}_{12}=-R^{13}_{21},d=R^{21}_{23}=R^{21}_{32}=-R^{32}_{13}
=-R^{32}_{31}$}\\[2mm]
\multicolumn{3}{l}{$
\lambda_1=\frac{a+b+c+\sqrt{(a-b+c)^2+8d^2}}{2},
\ \lambda_2=\frac{a+b+c-\sqrt{(a-b+c)^2+8d^2}}{2},\ 
\lambda_3=a-c$}
\\[2mm]
\multicolumn{3}{l}{$u=\frac{1}{\sqrt{2}}(1+\frac{a-b+c}
{\sqrt{(a-b+c)^2+8d^2}})^{\frac{1}{2}},\ 
w=\frac{1}{\sqrt{2}}(1-\frac{a-b+c}
{\sqrt{(a-b+c)^2+8d^2}})^{\frac{1}{2}}$}\\[3mm]
\hline \hline
\end{tabular}
}
\end{table}
Diagonalizing $\calDb$, we obtain 
the occupied general spin orbital and their occupation numbers as shown in
Table \ref{XB3g-1-1-3-occupation}.
From \eqref{OP-R} we see that this state has spin orbital angular momenta
$l^3_z(\mb)$ and spin quadrupole momenta $Q^1_{23}(\mb),Q^2_{31}(\mb)$
as follows:
\begin{align}
Q^1_{23}(\mb)&=2I_2d,\ \ \ Q^2_{31}(\mb)=-2I_2d,\nonumber \\
Q^1_{23}(\mb+\eb_1)&=-2I_2d,\ \ \ Q^2_{31}(\mb+\eb_1)=-2I_2d,\nonumber \\
Q^1_{23}(\mb+\eb_2)&=2I_2d,\ \ \ Q^2_{31}(\mb+\eb_2)=-2I_2d,\nonumber \\
Q^1_{23}(\mb+\eb_1+\eb_2)&=-2I_2d,\ \ \ Q^2_{31}(\mb+\eb_1+\eb_2)=2I_2d,
\nonumber \\
l_3^3(\mb)&=l^3_3(\mb+\eb_1+\eb_2)=2c,\nonumber \\
l_3^3(\mb+\eb_1)&=l^3_3(\mb+\eb_2)=-2c,
\end{align}
for $\mb$ such taht $T(\mb)\in \Lb_3$.
Here we note that the secondary LOP:$l_3^3$  is induced by the appearance
of the primary LOP:$\{Q_{23}^1,Q_{31}^2\}$.
From \eqref{OP-R} we can see that for all site $\mb$
\begin{align}
s^1(\mb) &=s^2(\mb)=s^3(\mb)=0,\nonumber \\
l_3(\mb) &=0,\nonumber \\
Q_{23}(\mb) &=Q_{31}(\mb)=0.
\end{align}
Thus the $G(XB_{3g},1,1)_3$ state is not a coexistent state of  
spin density and quadrupole moment. 
}
\end{example}
Finally we note that many states in Table \ref{Isotro-GMj1nu} and 
\ref{Isotro-Xj1nu}
have  spin orbital angular momentum and/or  spin quadrupole moment,
however they are not coexistent states of \{spin density\}, and
\{orbital angular momentum or quadrupole moment\} 
except $G(XA_g,1,1)_1$ state.

\section{Some numerical results}
In this section we present some calculated results of HF equations for 
the states listed in  Table \ref{Isotro-GMj00}, Table \ref{Isotro-Xj0nu},
Table \ref{Isotro-GMj1nu} and Table \ref{Isotro-Xj1nu}.
We solve the Hartree-Fock equation of a state (characterized by 
an isotropy subgroup $G_j$) self-consistently 
by starting initial values of $\Rb^{l\lambda}$ with  $G_j$ symmetry. 
From the SCF condition \eqref{xlmuconst} we obtain the initial values of 
$\xb^{l\lambda}$. After diagonalizing $H_m$ of \eqref{Hmdef2} with these 
$\xb^{l\lambda}$, we obtain new $\Rb^{l\lambda}$. The obtained $\Rb^{l\lambda}$ are
substituted into \eqref{xlmuconst} to compute new $\xb^{l\lambda}$. We use 
them as inputs to repeat the above process until the relative errors 
in $\Rb^{l\lambda}$  between successive iterations are less than the desired 
accuracy. After obtaining  converged $\Rb^{l\lambda}$ , we obtain the HF energy $E_{\rm HF}$  from \eqref{HF-energy}. 

We consider the cases of parameter
 sets listed in Table \ref{Parameter set}.
Since it is known that Ca$_{2-x}$Sr$_{x}$RuO$_{4}$ has four 4d electrons in the $\mathrm{t_{2g}}$ orbitals
\cite{Kubota,Hotta}, we use number of electrons per site $n_e^0= 4.0$ for all parameter sets. The eighth column denotes the  most stable state among  all states listed 
in listed in  Table \ref{Isotro-GMj00}, Table \ref{Isotro-Xj0nu},
Table \ref{Isotro-GMj1nu} and Table \ref{Isotro-Xj1nu}.
\begin{table}[!]
\caption{Parameter sets for model calculations}
\label{Parameter set}
\begin{tabular}{c|llllll|l}
\hline
N.O & $\delta$  & $t_1$ & $t_2$ & $t_3$ & $U$ & $J=J'$ & most stable state\\
\hline
(1)&0.0& 1.0 & 0.0 & 1.0 & 9.0& 2.25   & $G(\Gamma A_{1g},1,1)$ \\[1mm]
(2)& 0.4 & 1.0 & 0.0 & 1.0 & 9.0 & 0.4 & $G(MB_{1g},1,1)$\\[1mm]
(3) & 0.0 & 1.0 & 0.0& 1.0 & 9.0 & 0.7 & $G(ME_{g},1,1)_3, G(XB_{3g},1,1)_3$
\\[1mm]
(4)& -0.14& 1.0&  0.8& 0.8 & 8.0& 1.0& $G(XA_{g},1,1)_3$\\[1mm]
\hline
\multicolumn{8}{l}{number of electrons per site = $n_e^0=4.0,\  U'=U-2J$}\\
\hline \hline
\end{tabular}
\end{table}
\subsection{Parameter set {\rm (1)}}
In the parameter set (1), where $J$ has rather large values, 
the $G(\Gamma A_{1g},1,1)$ state is most stable. The calculated values of 
$a, b, c$ and $d$ in \eqref{GA-1-1-abcd} are 
\begin{align}
a&=0.6679,\ \ b=0.6643,\ \ c=0.3321,\ \ d=0.3357.
\end{align}
Thus we obtain $a+c\approx 1.0, a-c\approx \frac{1}{3}, b+d\approx 1.0$ and 
$b-d\approx \frac{1}{3}$. 
Then from Table \ref{GA11-occupation}, the occupation numbers for 
$\phi_i\mid \uparrow \rangle$ and 
$\phi_i\mid \downarrow \rangle$ are 
\begin{align}
n(\phi_1\mid \uparrow \rangle)&=n(\phi_2\mid \uparrow \rangle)
=n(\phi_3\mid \uparrow \rangle)=1,
 \nonumber\\
n(\phi_1\mid \downarrow \rangle)&=n(\phi_2\mid \downarrow \rangle)
=n(\phi_3\mid \downarrow \rangle)\approx \frac{1}{3}.
\end{align}
\subsection{Parameter set {\rm (2)}}
\begin{figure}[!]
\begin{center}
\includegraphics[trim=40mm 5mm 0mm 100mm,width=9cm,clip]{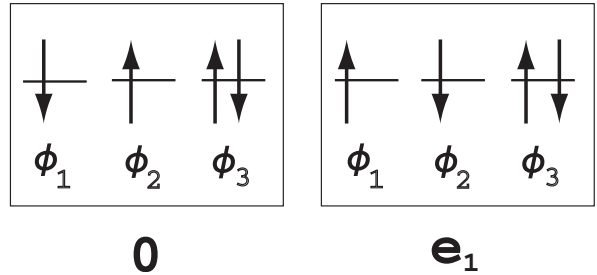}
	\caption{Schematic pattern of the electron occupation in 
	$G(MB_{1g},1,1)$ state. $\uparrow (\downarrow)$ indicates one electron with
	up(down) spin.}
     \label{MB1g-1-1}
	\end{center}
\end{figure}

In the parameter set (2), the $G(MB_{1g},1,1)$ state is most 
stable. The calculated  values of $a,b$ and $c$ are 
\begin{align}
a&=0.5, \ \ b=1,\ \ c=-0.473. 
\end{align} 
Thus we have
\begin{align}
a+c&=0.027\approx 0.0,\ \  a-c=0.974\approx 1.0, \ \  b=1.0.
\end{align}
From Table \ref{MB1-1-1-occ}, we obtain a
 qualitative pattern  
of the electron occupations as shown in Fig.\ref{MB1g-1-1}.
\subsection{Parameter set {\rm (3)}}
In the parameter set (3),  
the $G(ME_{g},1,1)_3$ and the $G(XB_{3g},1,1)_3$ states 
have the same energy and are  most stable. The calculated values of 
$a, b, c$ and $d$ in \eqref{ME-1-1-3-abcdf} and \eqref{XB3-1-1-3-abcdf}
 are 
\begin{align}
a&=0.7113,\ \ b=0.5775,\ \ c=-0.2775,\ \ d=0.3240.
\end{align}
Then the occupation numbers for general spin orbitals
$\psi_j\ (j=1, \dots, 6)$ in Table \ref{MEg-1-1-3-occupation} and 
\ref{XB3g-1-1-3-occupation} are
\begin{align}
n(\psi_1)&=n(\psi_2)=\lambda_1
=\frac{a+b+c+\sqrt{(a-b+c)^2+8d^2}}{2}\approx 1.0,\nonumber \\
n(\psi_3)&=n(\psi_4)=\lambda_2
=\frac{a+b+c-\sqrt{(a-b+c)^2+8d^2}}{2}\approx 0.0,\nonumber \\
n(\psi_5)&=n(\psi_6)=\lambda_3
=a-c\approx 1.0.
\end{align}
\subsection{Parameter set {\rm (4)}}
In the parameter set (4), the $G(XA_g,1,1)_3$ state is most 
stable. The calculated values of $a,b,c,d$ and $f$ in   
\eqref{XA-1-1-3-abcdf} are
\begin{align}
a&=0.75,\ \ b=0.5,\ \ c=-0.1802,\ \ d=-0.1408,\ \ f=0.3416.
\end{align}
 Then the occupation numbers for general spin orbitals
$\psi_j\ (j=1,\cdots, 6)$ in Table 
\ref{XAg-1-1-3-occupation} are
\begin{align}
n(\psi_1)&=n(\psi_2)=\lambda_1
=a+\sqrt{c^2+d^2}=0.9786\approx 1.0,\nonumber \\
n(\psi_3)&=n(\psi_4)=\lambda_2
=a-\sqrt{c^2+d^2}=0.521\approx 0.5, \nonumber \\
n(\psi_5)&=\lambda_3=b+\sqrt{2}f=0.017\approx 0.0,\nonumber \\
n(\psi_6)&=\lambda_4=b-\sqrt{2}f=0.983\approx 1.0.
\end{align}
The magnitude of spin density $|\ssb(\mb)|$ is 
$c+d+f=-0.1802-0.1408+0.3416=0.0206$.

\section{Summary and discussion}
We applied the group theoretical bifurcation theory of the HF equation
to the $\mathrm{t_{2g}}$-Hubbard model on a two-dimensional square lattice.
By enumerating the 
axial isotropy subgroups of the R-reps of the group  $G_0$
in the HF Hamiltonian space $W_{\rm HF}$ in the cases of ordering vectors 
$\Qb_i,i=0,1,2,3$, we have 
obtained many types of broken symmetry states which bifurcate from the 
normal state through one step transition. 

It is shown that 
these states have various types of local order parameters (LOP):
spin density $\ssb=(s^1,s^2,s^3)$, quadrupole moment 
$\Qb=(Q_{2}^2,Q_{12},Q_{23},Q_{31})$, orbital angular momentum 
$\lb=(l_1,l_2,l_3)$, 
spin quadrupole moment $\Qb^{\lambda}=(Q_2^{2\lambda},
 Q_{12}^{\lambda},Q_{23}^{\lambda},Q_{31}^{\lambda})$ and spin orbital angular 
 momentum $\lb^{\lambda}=(l_1^{\lambda},l_2^{\lambda},l_3^{\lambda})$, where 
 $\lambda=1,2,3$. 
 We have illustrated how the isotropy subgroup of a state
can determine the canonical form of the HF Hamiltonian, 
the standard forms of the occupied spin orbitals and 
their occupation number, and types of LOP.

We performed numerical calculations for all states discussed in \S 5 and \S 6
with some parameter sets. We found that various types of broken symmetric
solutions can be the most stable state depending on parameter sets.
Through these calculations we found that  non-collinear magnetic 
states:$G(ME_g,1,1)_3, G(XB_{3g},1,1)_3$ and 
$G(XA_g,1,1)_3$ can be the ground state for some parameter sets.

In \S 5 and \S 6 we considered broken symmetry states 
bifurcating through a single phase transition from 
the normal state.
However as shown in \S 7 of I, there are many other states derived 
through two step  transitions from the normal state. 
Here we consider two states bifurcatig from the $G(MB_{2g},0,0)$ treated 
in the Example 5.2. The isotropy subgroup of the $G(MB_{2g},0,0)$ is 
\begin{align}
G(MB_{2g},0,0)=\left(E+T(\eb_1)C_{2x}\right)\Db_{2a}\Lb_1\SSb\Rb
\end{align}
By a similar method to that in \S 7 of the previous paper, we obtain 
the following two types of isotropic subgroups
\begin{align}
G(MB_{2g},0,0)_a&= (E+tu_{2y})(E+T(\eb_1)C_{2x})\Db_{2ah}\Lb_1\Ab(\eb_3),
\nonumber \\
G(MB_{2g},0,0)_b&= (E+tu_{2y})(E+T(\eb_1)C_{2x}u_{2x})\Db_{2ah}\Lb_1\Ab(\eb_3).
\end{align}
We consider these states seperately.\\
{\bf (a) } $G(MB_{2g},0,0)_a$ state.\\
This state has following non zero $\Rb^{l\nu}$
\begin{align}
\Rb^{00}&=\left(
\begin{array}{ccc}
a & 0 & 0\\
0 & a & 0 \\
0 & 0 & b 
\end{array}
\right),
\ \ 
\Rb^{10}=\left(
\begin{array}{ccc}
0 & c & 0\\
c & 0 & 0 \\
0 & 0 & 0 
\end{array}
\right),\nonumber \\
\Rb^{03}&=\left(
\begin{array}{ccc}
d & 0 & 0\\
0 & d & 0 \\
0 & 0 & f
\end{array}
\right),
\ \ 
\Rb^{13}=\left(
\begin{array}{ccc}
0 & e & 0\\
e & 0 & 0 \\
0 & 0 & 0 
\end{array}
\right).
\end{align}
Thus from \eqref{OP-R} we obtain for $\mb$ such that $T(\mb)\in \Lb_1$
\begin{align}
Q_{12}(\mb)&=4I_2c, \ \ \ Q_{12}(\mb+\eb_j)=-4I_2c, \nonumber \\
s^3(\mb)&=2d+f,\ \ s^3(\mb+\eb_j)=2d+f,\nonumber \\
Q_{12}^3(\mb)&=2I_2e,\ \ \ Q_{12}^3(\mb+\eb_j)=-2I_2e,
\end{align}
where $j=1,2$. Thus we see that through the ferromagnetic 
transition of the
$G(MB_{2g},0,0)$ state which has the anti-ferro quadrupole moment $Q_{12}$,
there appears anti-ferro spin quadrupole moment $Q_{12}^3$ as well 
as ferro spin density $s^3$.

\noindent
{\bf (b) } $G(MB_{2g},0,0)_b$ state.\\
This state has following non zero $\Rb^{l\nu}$
\begin{align}
\Rb^{00}&=\left(
\begin{array}{ccc}
a & 0 & 0\\
0 & a & 0 \\
0 & 0 & b 
\end{array}
\right)
\ \ 
\Rb^{10}=\left(
\begin{array}{ccc}
0 & c & 0\\
c & 0 & 0 \\
0 & 0 & 0 
\end{array}
\right)\nonumber \\
\Rb^{03}&=\left(
\begin{array}{ccc}
0 & e & 0\\
e & 0 & 0 \\
0 & 0 & f
\end{array}
\right)
\ \ 
\Rb^{13}=\left(
\begin{array}{ccc}
d& 0 & 0\\
0 & d & 0 \\
0 & 0 & f 
\end{array}
\right)
\end{align}
Thus we obtain, for $\mb$ such that $T(\mb)\in \Lb_1$ and $j=1,2$,
\begin{align}
Q_{12}(\mb)&=4I_2c, \ \ \ Q_{12}(\mb+\eb_j)=-4I_2c, \nonumber \\
s^3(\mb)&=2d+f,\ \ s^3(\mb+\eb_j)=-(2d+f),\nonumber \\
Q_{12}^3(\mb)&=2I_2e,\ \ \ Q_{12}^3(\mb+\eb_j)=2I_2e.
\end{align}
Thus we see that through the anti-ferromagnetic 
transition of the
$G(MB_{2g},0,0)$ state,
there appears ferro spin quadrupole moment $Q_{12}^3$ as well
 as anti-ferro spin density $s^3$.

By similar manner to that of above cases, we can see 
that states, which are  derived through two step transition 
from the normal state, are coexisting states of 
\{spin sensity ,  quadrupole moment and spin quadrupole moment
\}, or \{spin sensity, orbital angular momentum  and 
spin orbital angular momentum\}.

\section*{Acknowledgements}
The authors would like to thank Professor M. Aihara and A. Takahashi
for valuable discussions and comments.

\appendix 
\section{Derivation of quadrupole moment}
Here we derive  \eqref{quadrupole-1} for the quadrupole moment
at the site $\mb$.
The quadrupole moment operator $\hat{Q}_{ij}(\mb)$ at a site $\mb$ is 
defined by\cite{Uimin} 
\begin{align}
\hat{\calQ}_{ij}(\mb)&=\sum_{p,q=1}^3\sum_{s,s'=1}^2
\langle \phi_{\smb p s}| (3x_ix_j-\delta_{ij}r^2)| 
\phi_{\smb q s'}\rangle
a^{\dag}_{\smb p s}a^{}_{\smb q s'}\nonumber \\
&=\sum_{s=1}^2\sum_{p,q=1}^3(\calQb_{ij})_{pq}
a^{\dag}_{\mb p s}a^{}_{\mb q s},
\end{align}
where $\phi_{\smb p s}=\phi_p(\rb-\mb)\chi_s$, $\chi_s$  are 
the spin function: $\chi_1=| \uparrow \rangle, 
\chi_2=| \downarrow \rangle$, and  
$\calQb_{ij}$ is a $3\times 3$ symmetric matrix 
whose ($p,q$) component is defined by
\begin{align}
\left(\calQb_{ij}\right)_{pq}&=
\int d\rb \phi_p(\rb)(3x_ix_j-r^2\delta_{ij})\phi_q(\rb).
\end{align}
Thus the expectation value of the quadrupole moment at a site $\mb$ 
given by
\begin{align}
Q_{ij}(\mb)&=\sum_{s=1}^2\sum_{p,q=1}^3(\calQb_{ij})_{pq}
\langle a^{\dag}_{\mb p s}a^{}_{\mb q s}\rangle. 
\label{calQ-2}
\end{align}
The explicit forms of $\calQb_{ij}$ are given by
\begin{align}
\calQb_{11}&=\left(
\begin{array}{ccc}
2I_1 & 0 & 0 \\
0 & -I_1 & 0 \\
0 & 0 & -I_1 
\end{array}
\right),\quad
\calQb_{12}=\Qb_{21}=
\left(
\begin{array}{ccc}
0 & I_2 & 0 \\
I_2 & 0 & 0 \\
0 & 0 & 0
\end{array}
\right),\nonumber \\[2mm]
\calQb_{22}&=\left(
\begin{array}{ccc}
-I_1 & 0 & 0 \\
0 & 2I_1 & 0 \\
0 & 0 & -I_1 
\end{array}
\right), \quad 
\calQb_{23}=\Qb_{32}=
\left(
\begin{array}{ccc}
0 & 0 & 0 \\
0 & 0 & I_2 \\
0 & I_2 & 0
\end{array}
\right),\nonumber \\[2mm]
\calQb_{33}&=\left(
\begin{array}{ccc}
-I_1 & 0 & 0 \\
0 & -I_1 & 0 \\
0 & 0 & 2I_1
\end{array}
\right),\quad 
\calQb_{31}=\Qb_{31}=\left(
\begin{array}{ccc}
0 & 0 & I_2 \\
0 & 0 & 0 \\
I_2 & 0 &0 
\end{array}
\right),
\label{calQI1I2}
\end{align}
where 
\begin{align}
I_1&=\int d\rb \phi_1(\rb)(x^2-y^2)\phi_1(\rb),\nonumber \\
I_2&=3\int d\rb \phi_1(\rb)xy \phi_2(\rb).
\end{align}
Thus from\eqref{calQ-2} and \eqref{calQI1I2} we obtain 
\begin{align}
Q_{11}(\mb)&=I_1\sum_{s,s'=1}^2\left(2\langle a^{\dag}_{\smb 1 s}
a^{}_{\smb 1 s'}\rangle -\langle a^{\dag}_{\smb 2 s}
a^{}_{\smb 2 s'}\rangle -\langle a^{\dag}_{\smb 3 s}
a^{}_{\smb 3 s'}\rangle \right)\sigma^{0}_{ss'},\nonumber \\
Q_{22}(\mb)&=I_1\sum_{s,s'=1}^2\left(-\langle a^{\dag}_{\smb 1 s}
a^{}_{\smb 1 s'}\rangle +2\langle a^{\dag}_{\smb 2 s}
a^{}_{\smb 2 s'}\rangle -\langle a^{\dag}_{\smb 3 s}
a^{}_{\smb 3 s'}\rangle \right)\sigma^{0}_{ss'},\nonumber \\
Q_{33}(\mb)&=I_1\sum_{s,s'=1}^2\left(-\langle a^{\dag}_{\smb 1 s}
a^{}_{\smb 1 s'}\rangle -\langle a^{\dag}_{\smb 2 s}
a^{}_{\smb 2 s'}\rangle +2\langle a^{\dag}_{\smb 3 s}
a^{}_{\smb 3 s'}\rangle \right)\sigma^{0}_{ss'},\nonumber \\
Q_{12}(\mb)&=Q_{21}(\mb)=I_2\sum_{s,s'=1}^2\left(\langle a^{\dag}_{\smb 1 s}
a^{}_{\smb 2 s'}\rangle +\langle a^{\dag}_{\smb 2 s}
a^{}_{\smb 1 s'}\rangle \right)\sigma^{0}_{ss'},\nonumber \\
Q_{23}(\mb)&=Q_{32}(\mb)=I_2\sum_{s,s'=1}^2\left(\langle a^{\dag}_{\smb 2 s}
a^{}_{\smb 3 s'}\rangle +\langle a^{\dag}_{\smb 3 s}
a^{}_{\smb 2 s'}\rangle \right)\sigma^{0}_{ss'},\nonumber \\
Q_{31}(\mb)&=Q_{13}(\mb)=I_2\sum_{s,s'=1}^2\left(\langle a^{\dag}_{\smb 3 s}
a^{}_{\smb 1 s'}\rangle +\langle a^{\dag}_{\smb 1 s}
a^{}_{\smb 3 s'}\rangle \right)\sigma^{0}_{ss'}.
\end{align}
The spin quadrupole moment  is defined by
\begin{align}
\hat{\calQ}^{\lambda}_{ij}(\mb)&=\dfrac{1}{2}\sum_{p,q=1}^3\sum_{s,s'=1}^2
\langle \phi_{\smb p s}| (3x_ix_j-\delta_{ij}r^2)\sigma^{\lambda}_{ss'}
| \phi_{\smb q s'}\rangle
a^{\dag}_{\smb p s}a^{}_{\smb q s'}.
\end{align}
Then we obtain \eqref{spin-Quadru-1} for the expectation values of 
spin quadrupole moment.

\section{Derivation of the orbital angular momentum}
Here we derive \eqref{orbital-angular-1} for the orbital angular 
momentum at the site $\mb$. 
Let $\hat{l}_j ~(j=1,2,3)$ be the orbital angular momentum. 
Then operators of the orbital angular  mormentum at the site $\mb$ 
are expressed by
\begin{align}
\hat{l}_i(\mb)&=\sum_{p,q=1}^3\sum_{s,s'=1}^2
\langle \phi_{\smb ps} | \hat{l}_i | \phi_{\smb q s'}\rangle
a^{\dag}_{\smb p s}a^{}_{\smb q s'}\nonumber \\
&=\sum_{p,q=1}^3\sum_{s=1}^2
\langle \phi_{\smb ps} | \hat{l}_i | \phi_{\smb q s}\rangle
a^{\dag}_{\smb p s}a^{}_{\smb q s}\nonumber \\
&=\sum_{p,q=1}^3\sum_{s=1}^2
(L_i)_{pq}a^{\dag}_{\smb p s}a^{}_{\smb q s}.
\label{lim}
\end{align}
where 
\begin{align}
(L_i)_{pq}&=
\langle \phi_{\smb ps} | \hat{l}_i | \phi_{\smb q s}\rangle.
\end{align}
We denote a $3\times 3$ matrix, whose $(p,q)$ component is $(L_i)_{pq}$, 
by $L_i$.
Within the $\mathrm{t_{2g}}$ subspace  we have \cite{Kamimura}
\begin{align}
L_1&=\left(
\begin{array}{ccc}
0 & 0 & 0\\
0 & 0 & i \\
0 & -i & 0
\end{array}
\right),\ 
L_2=\left(
\begin{array}{ccc}
0 & 0 & -i\\
0 & 0 & 0 \\
i & 0 & 0
\end{array}
\right),\ 
L_3=\left(
\begin{array}{ccc}
0 & i & 0\\
-i& 0 & 0 \\
0 & 0 & 0
\end{array}
\right).
\label{lbxyz}
\end{align}
Thus form \eqref{lim} and \eqref{lbxyz} the expectation 
values $l_i(\mb)$
of the orbital angular momentum $\hat{l}_i(\mb)$ are given by
\begin{align}
l_1(\mb)&=\sum_{ss'}(i)(\langle a^{\dag}_{\mb 2 s}a^{}_{\smb 3 s'}\rangle 
-\langle a^{\dag}_{\mb 3 s}a^{}_{\smb 2 s'}\rangle )\sigma^{0}_{ss'},
\nonumber \\
l_2(\mb)&=\sum_{ss'}(i)(\langle a^{\dag}_{\mb 3 s}a^{}_{\smb 1 s'}\rangle 
-\langle a^{\dag}_{\mb 1 s}a^{}_{\smb 3 s'}\rangle )\sigma^{0}_{ss'}
,\nonumber \\
l_3(\mb)&=\sum_{ss'}(i)(\langle a^{\dag}_{\mb 1 s}a^{}_{\smb 2 s'}\rangle 
-\langle a^{\dag}_{\mb 2 s}a^{}_{\smb 1 s'}\rangle )\sigma^{0}_{ss'}.
\end{align}
The spin orbital angular momentum is defined by
\begin{align}
\hat{l}^{\lambda}_i(\mb)&=\dfrac{1}{2}\sum_{p,q=1}^3\sum_{s,s'=1}^2
\langle \phi_{\smb ps} | \hat{l}_i \sigma^{\lambda}_{ss'}| \phi_{\smb q s'}\rangle
a^{\dag}_{\smb p s}a^{}_{\smb q s'}.
\end{align}
Then we obtain \eqref{spin-orbital-ang} for the spin orbital angular 
momenta.


\end{document}